\documentclass[11pt]{article}

\usepackage{latexsym,multirow}
\usepackage{amssymb,amsmath, bm}
\usepackage{graphicx}
\usepackage{booktabs}

\usepackage{enumerate}

\usepackage{natbib}
\usepackage[pdftex, bookmarksopen=true, bookmarksnumbered=true,
pdfstartview=FitH, breaklinks=true, urlbordercolor={0 1 0}, citebordercolor={0 0 1}]{hyperref}
\usepackage{colortbl}
\usepackage{subcaption}

\usepackage{dcolumn}
\newcolumntype{.}{D{.}{.}{-1}}
\newcolumntype{d}[1]{D{.}{.}{#1}}
\usepackage{theorem}
\theoremstyle{plain}
\theoremheaderfont{\scshape}
\newtheorem{assumption}{Assumption}

\newtheorem{proposition}{Proposition}
\newtheorem{theorem}{Theorem}

\newtheorem{lemma}{Lemma}

\newcommand{\ind}{\mbox{$\perp\!\!\!\perp$}}

\newcommand{\argmax}{\operatornamewithlimits{argmax}}

\usepackage{rotating}

\usepackage[compact]{titlesec}

\allowdisplaybreaks

\newcommand\spacingset[1]{\renewcommand{\baselinestretch}%
  {#1}\small\normalsize}

\let\S\relax 
\DeclareRobustCommand{\S}{%
  \ifmmode
    \mathsection
  \else
    \textsection~%
  \fi
}
\usepackage{versions}
\excludeversion{synth}
\includeversion{real}
\graphicspath{{./}}

\newcommand{\blind}{0}

\newcommand*{\QEDB}{\hfill\ensuremath{\square}}

\newcommand{\bone}{\mathbf{1}}

\newcommand{\pr}{\text{pr}}

\newcommand{\bW}{\bm{W}}

\newcommand{\E}{\mathbb{E}}
\newcommand{\bx}{\mathbf{x}}
\newcommand{\bX}{\mathbf{X}}

\newcommand{\cX}{\mathcal{X}}
\newcommand{\bI}{\mathbf{I}}

\newcommand{\ACEP}{\textsf{APCEp}}
\newcommand{\ACER}{\textsf{APCEr}}
\newcommand{\ACES}{\textsf{APCEs}}
\newcommand{\ACE}{\textsf{APCE}}

\begin{document} 

\newcommand{\tit}{Experimental Evaluation of Algorithm-Assisted Human
  Decision-Making: Application to Pretrial Public Safety Assessment}
%

\spacingset{0.95}

\if0\blind

{\title{{\bf\tit}\thanks{We thank Dean Knox and many participants at
      various seminars for helpful comments.  We acknowledge partial
      financial support from Arnold Ventures, Cisco Systems, Inc. (CG\#
      2370386), the National Science Foundation (SES--2051196), and
      the Sloan Foundation (Economics Program; 2020-13946).  The
      experiment analyzed in this paper has been approved by Harvard
      University's Institutional Review Board (\#16--1258).}}

  \author{Kosuke Imai\thanks{Corresponding author. Professor,
      Department of Government and Department of Statistics, Harvard
      University.  1737 Cambridge Street, Institute for Quantitative
      Social Science, Cambridge MA 02138.  Email:
      \href{mailto:imai@harvard.edu}{imai@harvard.edu} URL:
      \href{https://imai.fas.harvard.edu}{https://imai.fas.harvard.edu}}
    \and Zhichao Jiang\thanks{Assistant Professor, Department of
      Biostatistics and Epidemiology, University of Massachusetts,
      Amherst MA 01003. Email:
      \href{mailto:zhichaojiang@umass.edu}{zhichaojiang@umass.edu} }
    \and D. James Greiner\thanks{Honorable S.  William Green Professor
      of Public Law, Harvard Law School, 1525 Massachusetts Avenue,
      Griswold 504, Cambridge, MA 02138.} \and Ryan Halen\thanks{Data
      Analyst, Access to Justice Lab at Harvard Law School, 1607
      Massachusetts Avenue, Third Floor, Cambridge, MA 02138.} \and
    Sooahn Shin\thanks{Ph.D. student, Department of Government,
      Harvard University, Cambridge, MA 02138. Email:
      \href{mailto:sooahnshin@g.harvard.edu}{sooahnshin@g.harvard.edu}
      URL: \href{https://sooahnshin.com}{https://sooahnshin.com}}}

    \date{
      [To be read before The Royal Statistical Society at an online meeting organized by the Statistics and the Law Section, the Data Ethics and Governance Section, and the Discussion Meetings Committee, on Tuesday, 8th February 2022, Dr. J. Mortera in the Chair]
}

\maketitle

}\fi

\if1\blind
\title{\bf \tit}

\maketitle
\fi

\pdfbookmark[1]{Title Page}{Title Page}

\thispagestyle{empty}
\setcounter{page}{0}
         
\begin{abstract}
  Despite an increasing reliance on fully-automated algorithmic
  decision-making in our day-to-day lives, human beings still make
  highly consequential decisions. As frequently seen in business,
  healthcare, and public policy, recommendations produced by
  algorithms are provided to human decision-makers to guide their
  decisions. While there exists a fast-growing literature evaluating
  the bias and fairness of such algorithmic recommendations, an
  overlooked question is whether they help humans make better
  decisions. We develop a general statistical methodology for
  experimentally evaluating the causal impacts of algorithmic
  recommendations on human decisions. We also show how to examine
  whether algorithmic recommendations improve the fairness of human
  decisions and derive the optimal decision rules under various
  settings. We apply the proposed methodology to preliminary data from
  the first-ever randomized controlled trial that evaluates the
  pretrial Public Safety Assessment (PSA) in the criminal justice
  system. A goal of the PSA is to help judges decide which arrested
  individuals should be released. On the basis of the preliminary data
  available, we find that providing the PSA to the judge has little
  overall impact on the judge's decisions and subsequent arrestee
  behavior.  Our analysis, however, yields some potentially suggestive
  evidence that the PSA may help avoid unnecessarily harsh decisions
  for female arrestees regardless of their risk levels while it
  encourages the judge to make stricter decisions for male arrestees
  who are deemed to be risky. In terms of fairness, the PSA appears to
  increase an existing gender difference while having little effect on
  any racial differences in judges' decision. Finally, we find that
  the PSA's recommendations might be unnecessarily severe unless the
  cost of a new crime is sufficiently high.
  


\end{abstract}


\clearpage
\spacingset{1.5}

\section{Introduction}

A growing body of literature has suggested the potential superiority
of algorithmic decision-making over purely human choices across a
variety of tasks \citep[e.g.,][]{hans:hasa:15,he:etal:15}. Although
some of this evidence is decades old
\citep[e.g.,][]{dawe:faus:meeh:89}, it has recently gained significant
public attention by the spectacular defeats of humanity's best in
cerebral games \citep[e.g.,][]{silv:etal:18}. Yet, even in contexts
where research has warned of human frailties, we humans still make
many consequential decisions for a variety of reasons including the
preservation of human agency and accountability.

The desire for a human decision-maker as well as the precision and
efficiency of algorithms have led to the adoption of hybrid systems
involving both. By far the most popular system uses algorithmic
recommendations to inform human decision-making. Such
algorithm-assisted human decision-making has been deployed in many
aspects of our daily lives, including medicine, hiring, credit
lending, investment decisions, and online shopping. And of particular
interest, algorithmic recommendations are increasingly of use in the
realm of evidence-based public policy making. A prominent example,
studied in this paper, is the use of risk assessment instruments in
the criminal justice system that are designed to improve incarceration
rulings and other decisions made by judges.

While there exists a fast-growing literature in computer science that
studies the bias and fairness of algorithms \citep[see][for a review
and many references therein]{chou:roth:20}, an overlooked question is
whether such algorithms help humans make better decisions \citep[see
e.g.,][for an exception]{gree:chen:19}. In this paper, we develop a
general methodological framework for experimentally evaluating the
impacts of algorithmic recommendations on human decision-making. We
conducted the first-ever real-world field experiment by providing, for
a randomly selected cases, information from a system consisting of
Public Safety Assessment (PSA) risk scores and a recommendation from a
Decision Making Framework (DMF) to a judge who makes an initial
release decision. We evaluate whether the PSA-DMF system (which for
brevity we refer to as the PSA hereafter) helps judges achieve their
goal of preventing arrestees from committing a new crime or failing to
appear in court while avoiding an unnecessarily harsh decision.

Using the concept of principal stratification from the causal
inference literature \citep[e.g.,][]{fran:rubi:02,ding:lu:17}, we
propose the evaluation quantities of interest, identification
assumptions, and estimation strategies. We also develop sensitivity
analyses to assess the robustness of empirical findings to the
potential violation of a key identification assumption \citep[see
also][]{hira:imbe:rubi:zhou:00,
  schwartz2011bayesian,mattei2013exploiting,jiang2016principal}. In
addition, we examine whether algorithmic recommendations improve the
fairness of human decisions, using the concept of principal fairness
that, unlike other fairness criteria, accounts for how the decision in
question affects individuals \citep{imai:jian:20}.  Finally, we
consider how the data from an experimental evaluation can be used to
inform an optimal decision rule and assess the optimality of
algorithmic recommendations and human decisions \citep[see][for a
methodological framework for learning an optimal algorithmic
recommendation]{benm:etal:21}. Although we describe and apply the
proposed methodology in the context of evaluating the PSA, it is
directly applicable or extendable to many other settings of
algorithm-assisted human decision-making.

The use of risk assessment scores, which serves as the main
application of the current paper, has played a prominent role in the
literature on algorithmic fairness since the controversy over the
potential racial bias of COMPAS risk assessment score used in the
United States (US) criminal justice system \citep[see
e.g.,][]{angw:etal:16,diet:mend:bren:16,flor:bech:lowe:16,dres:fari:18}.
With few exceptions, however, much of this debate focused upon the
accuracy and fairness properties of risk assessment scores itself
rather than how they affect judges' decisions \citep[see e.g.,][and
references therein]{berk:etal:18,klei:etal:18,
  rudi:wang:coke:20}. Even studies that directly estimate the impacts
of risk assessment scores on judges' decisions are based on either
observational data or hypothetical vignettes in surveys
\citep[e.g.,][]{mill:malo:13,berk:17,stev:18, albr:19,gree:chen:19,
  garr:mona:20,skee:scru:mona:20,stev:dole:21}.

We contribute to this literature by demonstrating how to evaluate the
use of risk assessment scores experimentally when humans are ultimate
decision makers. To the best of our knowledge, this is the first
real-world randomized controlled trial (RCT) that evaluates the
impacts of algorithmic risk assessment scores on judges' decisions in
the criminal justice system (see also the Manhattan Bail Project and
Philadelphia Bail Experiment that evaluated the effects of bail
guidelines on judges' decisions several decades ago
\citep{ares:rank:stur:63,gold:gott:84,gold:gott:85}). Using the
concept of principal stratification from causal inference literature,
the proposed methodology allows us to evaluate the effects of the PSA
on judges' decisions separately for the subgroups of arrestees with
different levels of risks.

Based on the preliminary data from our experiment (complete data will
not be available for some time), we find that the provision of the PSA
has little overall impact on the judge's decisions across three
outcomes we examine: failure to appear (FTA), new criminal activity
(NCA), and new violent criminal activity (NVCA).  Our analysis,
however, provides some suggestive evidence that the PSA may make the
judge's decisions more lenient for female arrestees regardless of
their risk levels, while it encourages the judge to make stricter
decisions for male arrestees who are deemed to be risky. In terms of
fairness, the PSA appears to increase an existing gender difference
while having no substantial impact on any racial differences in
judges' decisions. Finally, we use the experimental data to learn
about the optimal decision rule that minimizes the prevalence of
negative outcomes (FTA, NCA, and NVCA) while avoiding unnecessarily
harsh decisions. Our analysis suggests that the PSA's recommendations
may be unnecessarily severe unless a jurisdiction considers the costs
of FTA, NCA, and NVCA to be sufficiently high. This might suggest that
incarceration decisions themselves, whether PSA-informed or otherwise,
are also unnecessarily severe.

\section{Experimental Evaluation of Pretrial Public Safety Assessment}
\label{sec:RCT}

In this section, we briefly describe our field experiment after
providing some background about the use of the PSA in the US criminal
justice system. Additional details about our experiment are given in
\citep{grei:etal:20}.

\subsection{Background}

The US criminal justice apparatus consists of thousands of diverse
systems. Some are similar in the decision points they feature as an
individual suspected of a crime travels from investigation to
sentencing. Common decision points include whether to stop and frisk
an individual in a public place, whether to arrest or issue a citation
to an individual suspect of committing a crime, whether to release the
arrestee while they await the disposition of any charges against them
(the subject of this paper), what charge(s) to be filed against the
individual, whether to find the defendant guilty of those charges, and
what sentence to impose on a defendant found guilty.

At present, human judges make all of these decisions. In theory,
algorithms could inform any of them, and could even make some of these
decisions without human involvement. To date, algorithmic outputs have
appeared most frequently in two settings: (i) at the ``first
appearance'' hearing, during which a judge decides whether to release
an arrestee pending disposition of any criminal charges, and (ii) at
sentencing, in which the judge imposes a punishment on a defendant
found guilty. The first of these two motivates the present paper, but
the proposed methodology is applicable or extendable to other
settings.

We describe a typical first appearance hearing. The key decision the
judge must make at a first appearance hearing is whether to release
the arrestee pending disposition of any criminal charges and, if the
arrestee is to be released, what conditions to impose. Almost all
jurisdictions allow the judge to release the arrestee with only a
promise to reappear at subsequent court dates. In addition, because
arrestees have not yet been adjudicated guilty of any charge at the
time of a pretrial hearing, there exists a consensus that pretrial
incarceration is to be avoided unless the risks associated with
release are sufficiently high.

Judges deciding whether to release arrestees ordinarily consider two
risk factors among a variety of other concerns; the risk that the
arrestee will fail to appear at subsequent court dates, and the risk
that the arrestee will engage in new criminal activity (NCA) before
the case is resolved (e.g., 18 U.S.C. \S 3142(e)(1)). Jurisdiction
laws vary regarding how these two risks are to be weighed. Some
jurisdictions direct judges to consider both simultaneously along with
other factors (e.g., Ariz. Const. art. II, \S 22, Iowa Code \S
811.2(1)(a)), while others focus on only FTA risk (e.g.,
N.Y. Crim. Proc. Law \S 510.30(2)(a)). Despite these variations, NCA
and FTA are constant and prominent in the debate over the first
appearance decisions.

Concerns about the consequential nature of the first appearance
decision have led to the development of the PSA, which is ordinarily
offered as an input to first appearance judges. Predisposition risk
assessment instruments take various forms, but most focus on
classifying arrestees according to FTA and NCA risks. They are
generally constructed by fitting a statistical model to a training
dataset based on past first appearance hearings and the subsequent
incidences (or lack thereof) of FTA and NCA. The hope is that
providing such instruments will improve the assessment of FTA and NCA
risks and thereby lead to better decisions. The goal of this paper is
to develop a general methodological framework for evaluating the
impact of providing the PSA to judges at first appearance hearings
using an RCT, to which we now turn.

\subsection{The Experiment}
\label{subsec:experiment}

We conducted a field RCT in Dane county, Wisconsin, to evaluate the
impacts of PSA provision on judges' decisions. The PSA consists of
three scores --- two six-point scores separately summarizing FTA and
NCA risks as well as a binary score for the risk of NVCA. These scores
are based on the weighted indices of nine factors drawn from criminal
history information, primarily prior convictions and FTA, and a single
demographic factor, age. Notably, gender and race are not used to
compute the PSA. The weights are calculated using past data.  A
Decision Making Framework (DMF) combines information from the three
PSA scores with other considerations to produce an overall
recommendation to the judge, which the judge may accept or modify or
ignore as they see fit.  The details about the construction of the PSA
and other relevant information are available at
\url{https://advancingpretrial.org/psa/factors/}

The field operation was straightforward. In this county, a court
employee assigned each matter a case number sequentially as it entered
the system. No one but this clerk was aware of the pending matter
numbers, so manipulation of the number by charging assistant district
attorneys was not possible. Employees of the Clerk’s office scanned
online record systems to calculate the PSA for all cases. If the last
digit of the case number was even, these employees made the PSA
(specifically, a printout of the PSA scores, the DMF recommendation,
and the supporting criminal history and age information) available to
the judge. Otherwise, no PSA was made available. Thus, the provision
of the PSA to judges was essentially randomized. Indeed, the
comparison of the observed covariate distributions suggests that this
scheme produced groups comparable on background variables
\citep{grei:etal:20}.

The judge presiding over the first appearance hearing by law was to
consider the risk of FTA and NCA, along with other factors including
ties to the community as prescribed by statute. The judge could order
the arrestee released with or without bail of varying amount. The
judge could also condition release on compliance with certain
conditions such as monitoring, but for the sake of simplicity, we
focus on bail decisions and ignore other conditions in this paper.

When making decisions, the judge also had information other than the
PSA and its inputs. In all cases, the judge had a copy of an affidavit
sworn to by a police officer recounting the circumstances of the
incident that led to the arrest. The defense attorney sometimes
informed the judge of the following regarding the arrestee’s
connections to the community: length of time lived there, employment
there, and family living there. When available, this information
ordinarily stemmed from an arrestee interview conducted earlier by a
paralegal. The assistant district attorney sometimes provided
additional information regarding the circumstances of the arrest or
criminal history. Given the lack of access to this additional
information, we develop a sensitivity analysis to address a potential
unobserved confounding bias.

\subsection{The Data}
\label{subsec:data}

The field operation design called for approximately a 30-month
treatment assignment period (from the middle of 2017 until the end of
2019) followed by the collection of data on FTA, NCA, NVCA, and other
outcomes for a period of two years after randomization. At the time of
this writing, we have outcome data from a 12-month follow-up of each
first appearance event that occurred in the first 12 months of
randomization.  The 30-month randomization period has expired, and we
will report the results of our comprehensive analysis of a full data
set in the future. Furthermore, although some arrestees had multiple
cases during the study period, this paper focuses only on the first of
the first appearance hearings for any individual arrestee. This leads
to a total of 1891 cases for our analysis, of which 40.0\% (38.8\%)
are white male arrestees and 13.0 are white female arrestees
(non-white male and female arrestees account for 38.8\% and 8.1\%,
respectively).

\begin{table}[!t] \centering \spacingset{1} \setlength{\tabcolsep}{3.5pt}
	\begin{tabular}{@{\extracolsep{5pt}} rrrrrrrr}
		\\[-1.8ex]\hline
		\hline \\[-1.8ex]
		& \multicolumn{3}{c}{\textit{no} PSA (Control Group)} &
       \multicolumn{3}{c}{PSA
       (Treatment Group)}
		\\\cmidrule(lr){2-4}\cmidrule(lr){5-7}
		& \multicolumn{1}{c}{Signature} & \multicolumn{2}{c}{Cash bond} & \multicolumn{1}{c}{Signature} & \multicolumn{2}{c}{Cash bond} \\
		& \multicolumn{1}{c}{bond} & $\leq$\$1000 & $>$\$1000 & \multicolumn{1}{c}{bond} & $\leq$\$1000 & $>$\$1000 & Total (\%) \\
		\hline \\[-1.8ex]
		Non-white Female & 64 & 11 & 6 & 67 & 6 & 0 & 154 \\
		& (3.4) & (0.6) & (0.3) & (3.5) & (0.3) & (0.0) & (8.1) \\
		White Female & 91 & 17 & 7 & 104 & 17 & 10 & 246 \\
		& (4.8) & (0.9) & (0.4) & (5.5) & (0.9) & (0.5) & (13.0) \\
		Non-white Male & 261 & 56 & 49 & 258 & 53 & 57 & 734 \\
		& (13.8) & (3.0) & (2.6) & (13.6) & (2.8) & (3.0) & (38.8) \\
		White Male & 289 & 48 & 44 & 276 & 54 & 46 & 757 \\
		& (15.3) & (2.5) & (2.3) & (14.6) & (2.9) & (2.4) & (40.0) \\ \hline
		FTA committed & 218 & 42 & 16 & 221 & 45 & 16 & 558 \\
		& (11.5) & (2.2) & (0.8) & (11.7) & (2.4) & (0.8) & (29.4) \\
		\textit{not} committed & 487 & 90 & 90 & 484 & 85 & 97 & 1333 \\
		& (25.8) & (4.8) & (4.8) & (25.6) & (4.5) & (5.1) & (70.6) \\ \hline
		NCA committed & 211 & 39 & 14 & 202 & 40 & 17 & 523 \\
		& (11.2) & (2.1) & (0.7) & (10.7) & (2.1) & (0.9) & (27.7) \\
		\textit{not} committed & 494 & 93 & 92 & 503 & 90 & 96 & 1368 \\
		& (26.1) & (4.9) & (4.9) & (26.6) & (4.8) & (5.1) & (72.4) \\ \hline
		NVCA committed & 36 & 10 & 3 & 44 & 10 & 6 & 109 \\
		& (1.9) & (0.5) & (0.2) & (2.3) & (0.5) & (0.3) & (5.7) \\
		\textit{not} committed & 669 & 122 & 103 & 661 & 120 & 107 & 1782 \\
		& (35.4) & (6.5) & (5.4) & (35.0) & (6.3) & (5.7) & (94.3) \\ \hline
		Total & 705 & 132 & 106 & 705 & 130 & 113 & 1891\\
		& (37.3) & (7.0) & (5.6) & (37.3) & (6.9) & (6.0) & (100) \\ \hline \\[-1.8ex]
	\end{tabular}
	\caption{The Joint Distribution of Treatment Assignment,
     Judge's Decisions, and Outcomes. The table shows the number
     of cases in each category with the corresponding percentage
     in parentheses. Only about 20\% of all arrestees are
     female. Few cases result in NVCA (new violent criminal
     activity), while FTA (failure to appear in court) and NCA
     (new criminal activity) occur in slightly above 25\% each. A
     majority of decisions are signature bonds rather than cash
     bonds.}
	\label{tab:prop}
\end{table}

Based on the empirical distribution of bail amounts and expert's
opinion, we categorize the judge's decisions into three ordinal
categories: signature bond, small cash bond (less than \$1,000), and
large cash bond (greater than or equal to \$1,000). A signature bond
requires an arrestee to sign a promise to return to the court for
trial, but does not require any payment to be released. Cash bonds
require an arrestee to deposit money with the court to obtain release.
Table~\ref{tab:prop} summarizes the joint distribution of treatment
assignment (PSA provision), the judge's decisions (three ordinal
categories), and three binary outcomes. We observe that in about three
quarters of cases the judge imposed signature bonds, while in the
remaining cases the judge imposed bail. For the outcome variables,
slightly less than 30\% of arrestees commit FTA or NCA whereas the
proportion of those who commit NVCA is only about 6\%.

\subsection{The Overall Impact of PSA Provision on Judge's Decisions}

\begin{figure}[t!]
\vspace{-.2in}
 \centering \spacingset{1}
 \begin{subfigure}[t]{\textwidth}
 \subcaption{Treatment Group}
 \includegraphics[width = \textwidth]{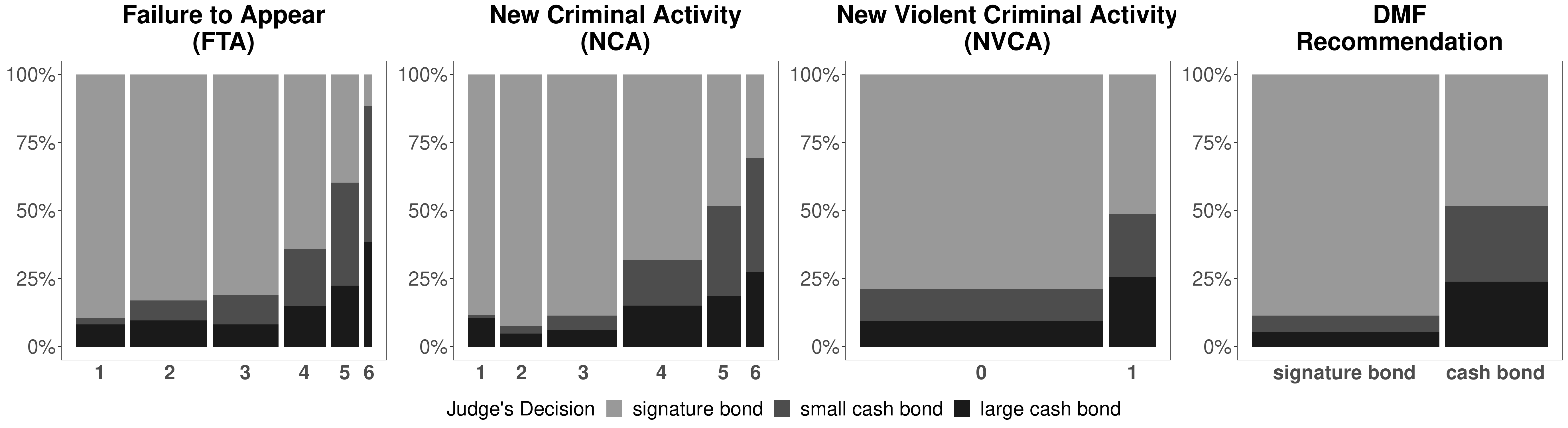}
 \end{subfigure} \\
 \vspace{.25in}
 \begin{subfigure}[t]{\textwidth}
 \subcaption{Control Group}
 \includegraphics[width = \textwidth, trim = 0 0 0 57, clip]{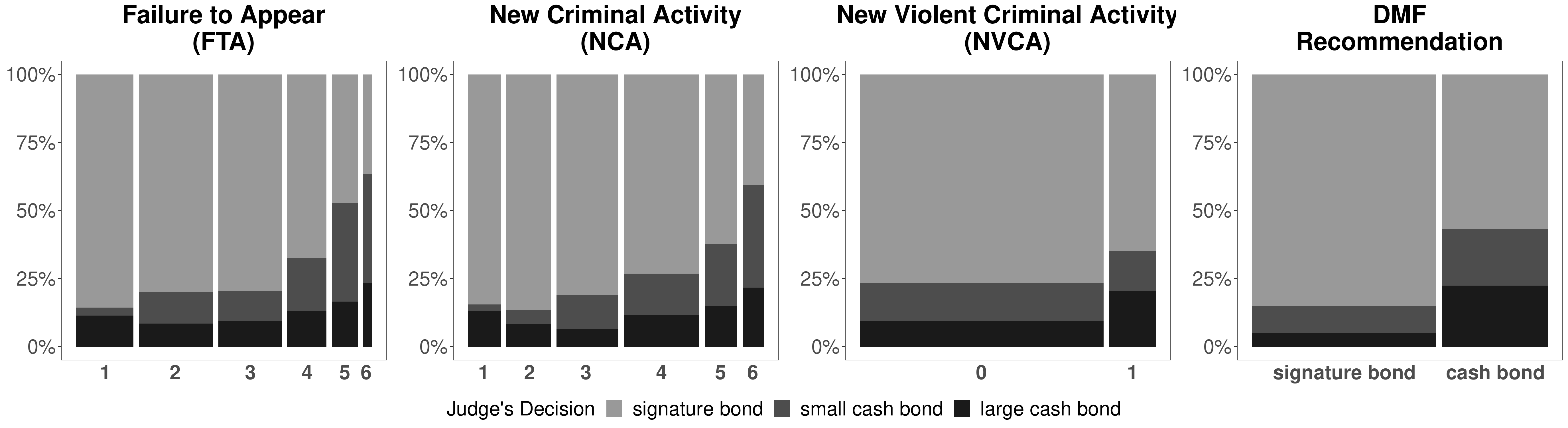}
 \end{subfigure}
 \caption{The Distribution of the Judge's Decisions given the Pretrial
   Public Safety Assessment (PSA) among the Cases in the Treatment
   (Top Panel) and Control (Bottom Panel) Groups. There are three PSA
   scores, two of which are ordinal --- FTA and NCA --- while the
   other is dichotomous --- NVCA. The judge's decision is coded as a
   three-category ordinal variable based on the type and amount of
   bail: a signature bond, a small cash bond (less than \$1,000), and
   a large cash bond (greater than or equal to \$1,000). The DMF
   recommendation is presented as a binary variable: signature or cash
   bond. The width of each bar is proportional to the number of cases
   for each value of the corresponding PSA score. There exists a
   positive correlation between PSA scores and the severity of the
   judge's decisions in both treatment and control
   groups.} \label{fig:stackedbar}
\end{figure}

Figure~\ref{fig:stackedbar} presents the distribution of the judge's
decisions given each of the PSA scores among the cases in the
treatment (top panel) and control (bottom panel) groups. The overall
difference in the conditional distribution between the two groups is
small though there are some differences in some subgroups (see
Appendix~\ref{app:dist}). The PSA scores for FTA and NCA are ordinal,
ranging from 1 (safest) to 6 (riskiest), whereas the PSA score for
NVCA is binary, 0 (safe) and 1 (risky). We also plot the DMF
recommendation, which aggregates these three PSA scores as well as
other information such as types of charges. The DMF recommendation has
four categories (signature bond, modest cash bond, moderate cash bond,
and cash bond with maximum conditions), but we dichotomize it into
signature or cash bond given its skewed empirical distribution.

In general, we observe a positive association between the PSA scores
and judge's decisions, implying that a higher PSA score is associated
with a harsher decision. We also find that for FTA and NCA, the most
likely scores are in the medium range, while the vast majority of NVCA
cases were classified as no elevated risk. For NCA and FTA, the
judge's decisions varied little when the PSA score took a value in the
lower range. For the DMF recommendation, the judge is far more likely
to give a signature bond for the cases that are actually recommended
for a signature bond.

\begin{figure}[t]
 \spacingset{1}\centering  
 \includegraphics[width=0.485\linewidth]{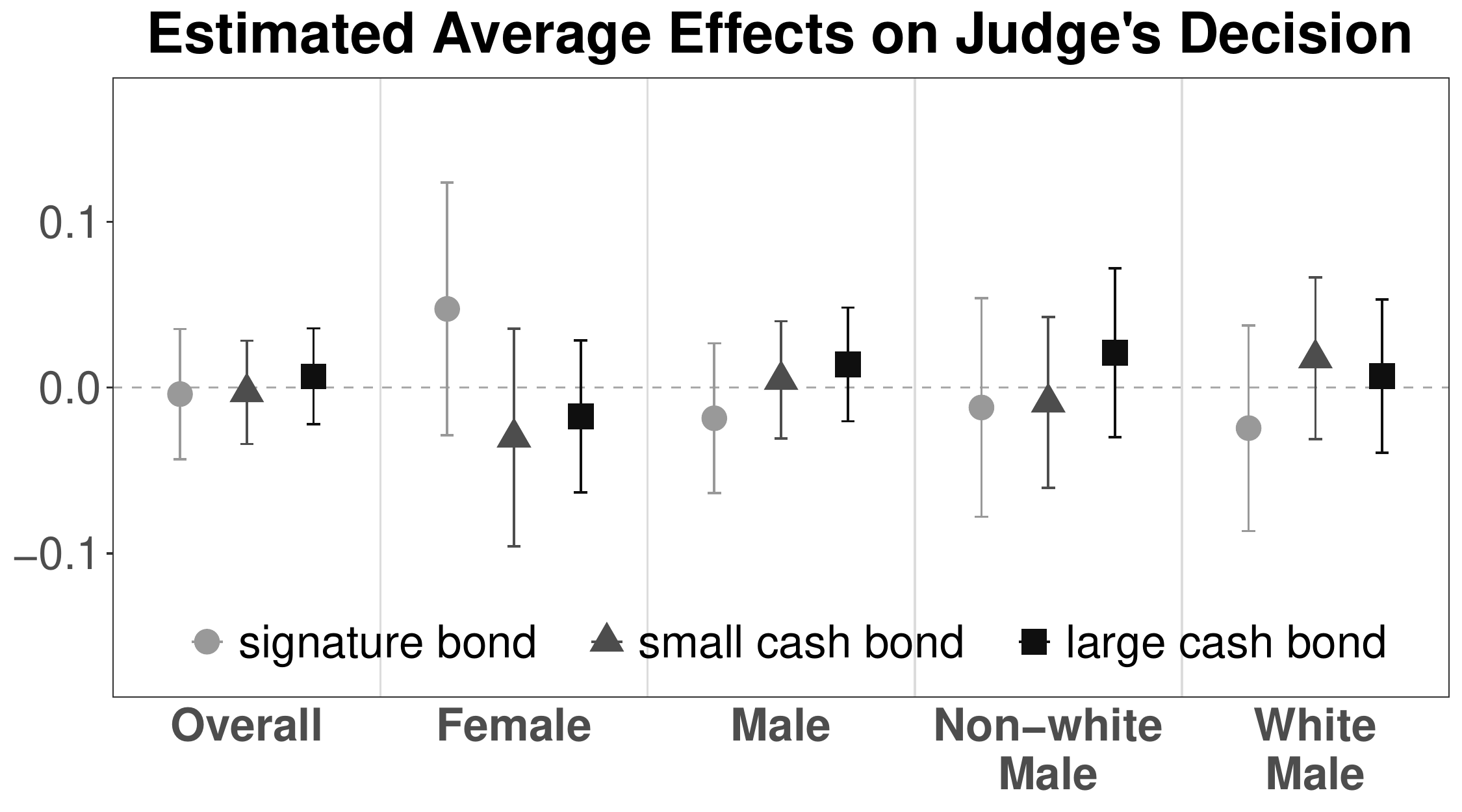}
 \includegraphics[width=0.485\linewidth]{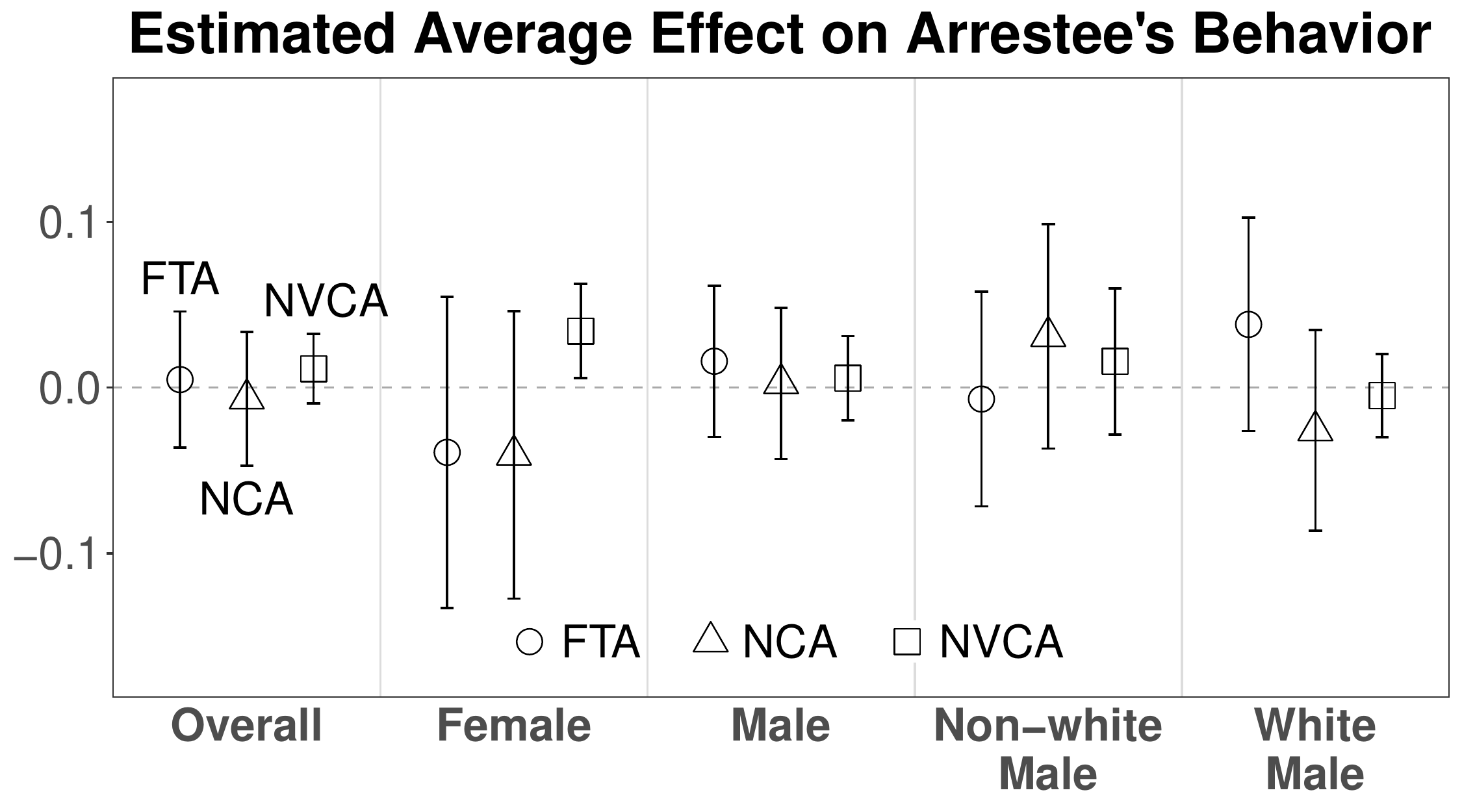}
 \caption{Estimated Average Causal Effects of PSA Provision on the
  Judge's Decisions and Outcome Variables. The results are based on
  the difference-in-means estimator. The vertical bars represent the
  95\% confidence intervals. In the left plot, we report the
  estimated effects of PSA provision on the judge's decision to
  charge a signature bond (solid circles), a small cash bail (\$1,000
  dollars or less; solid triangles), and a large cash bail (greater
  than \$1,000; solid squares). In the right plot, we report the
  estimated effects of PSA provision on the three different outcome
  variables: FTA (open circles), NCA (open triangles), and NVCA (open
  squares). PSA provision appears to have little overall effect on
  the judge's decision and arrestee's behavior, on average, though it
  may slightly increase NVCA among female arrestees.} \label{fig:ITT}
\end{figure}

Figure~\ref{fig:ITT} presents the estimated average causal effect of
PSA provision on the judge's decisions (left plot) and three outcomes
of interest (right plot). We use the difference-in-means estimator and
display the 95\% confidence intervals as well as the point
estimates. We do not compute separate estimates for white females and
non-white females because we have too few female arrestees (see
Table~\ref{tab:prop}). The results imply that PSA provision, on
average, has little effect on the judge's decisions. In addition, the
average effects of PSA provision on the three outcomes are also
largely ambiguous although there is suggestive evidence that it may
slightly increase NVCA among female arrestees. In
Appendix~\ref{subsec:age_summary}, we also explore the average causal
effects of PSA provision across different age groups. We find some
suggestive causal effects for the group of 29 -- 35 year old
arrestees.

Although these results show whether PSA provision leads to a harsher
or more lenient decision (and whether it increases or decreases the
proportions of negative outcomes), they are not informative about
whether it helps judges make better decisions. In the current context,
a primary goal of the judge is to make lenient decisions in low-risk
cases and less lenient decisions in high-risk cases. If the PSA is
helpful, therefore, its provision should encourage the judge to impose
small or no bail on safe cases and impose a greater amount of bail on
risky cases (we formally define ``safe'' and ``risky'' cases
below). This demands the study of an important causal heterogeneity by
distinguishing among cases with different risk levels. In addition, we
may also be interested in knowing how PSA provision affects the gender
and racial fairness of judges' decisions.  Thus, the goal of the
remainder of the paper is to develop statistical methods that directly
address these and other questions.

\section{The Proposed Evaluation Methodology}

In this section, we describe the proposed methodology for
experimentally evaluating the impacts of algorithmic recommendations
on human decision-making. Although we refer to our specific
application throughout, the proposed methodology can be applied or
extended to other settings, in which humans make decisions using
algorithmic recommendations as an input. We will begin by considering
a binary decision and then extend our methodology to an ordinal
decision in Section~\ref{sec:ordinal_decision}.

\subsection{The Setup}

Let $Z_i$ be a binary treatment variable indicating whether the PSA is
presented to the judge of case $i=1,2,\ldots,n$. We use $D_i$ to
denote the binary detention decision made by the judge to either
detain ($D_i = 1$) or release ($D_i = 0$) the arrestee prior to the
trial. In addition, let $Y_i$ represent the binary outcome: we code
all our outcomes --- NCA, NVCA, and FTA --- as binary variables. For
example, $Y_i = 1$ ($Y_i=0$) implies that the arrestee of case $i$
commits (does not commit) an NCA. Finally, we use $\bX_i$ to denote a
vector of observed pre-treatment covariates for case $i$. They
include age, gender, race, and prior criminal history.

We adopt the potential outcomes framework of causal inference and
assume the stable unit treatment value assumption (SUTVA)
\citep{rubi:90}. In particular, we assume no interference among cases,
implying that the treatment assignment for one case does not influence
the judge's decision and outcome variable in another case. This
assumption is reasonable in our analysis because we focus only on
first arrests and do not analyze cases with subsequent
arrests. Appendix~\ref{app:sutva} provides the empirical evidence in
support of this assumption.

Let $D_i(z)$ be the potential value of the pretrial detention
decision if case $i$ is assigned to the treatment condition
$z \in \{0, 1\}$. Furthermore, $Y_i(z,d)$ represents the potential
outcome under the scenario, in which case $i$ is assigned to the
treatment condition $z$ and the judge makes the decision
$d \in \{0, 1\}$. Then, the observed decision is given by
$D_i = D(Z_i)$ whereas the observed outcome is denoted by
$Y_i = Y_i(Z_i, D_i(Z_i))$.

Throughout this paper, we maintain the following three assumptions,
all of which we believe are reasonable in our application. First,
because the treatment assignment is essentially randomized, the
following independence assumption is automatically satisfied.
\begin{assumption}[Randomization of the Treatment Assignment]
\label{asm::rand} \spacingset{1}
$$\{D_i(z), Y_i(z,d),\bX_i\} \ \ind \ Z_i \quad {\rm for} \ z \in
\{0,1\} {\rm\ and\ all} \ d.$$
\end{assumption}

Second, we assume that the provision of the PSA influences the outcome
only through the judge's decision. Because an arrestee would not care
and, perhaps, would not even know whether the judge is presented with
the PSA at their first appearance, it is reasonable to assume that
their behavior, be it NCA, NVCA, or FTA, is not affected directly by
the treatment assignment.
\begin{assumption}[Exclusion Restriction]
\label{asm::ex} \spacingset{1}
$$Y_i(z,d) \ = \ Y_i(z',d) \quad {\rm for} \ z, z' \in \{0,1\} \ {\rm and\ all} \ i,d.$$
\end{assumption}
Under Assumption~\ref{asm::ex}, we can simplify our notation by
writing $Y_i(z,d)$ as $Y_i(d)$. A potential violation of this
assumption is that the PSA may directly influence the judge's decision
about release conditions, which can in turn affect the outcome. The
extension of the proposed methodology to multi-dimensional decisions
including the bail amount and monitoring conditions is left for future
research.

Finally, we assume that the judge's decision monotonically affects the
outcome. Thus, for NCA (NVCA), the assumption implies that each
arrestee is no less likely to commit a new (violent) crime if
released. If FTA is the outcome of interest, this assumption implies
that an arrestee is no more likely to appear in court if released.
The assumption is reasonable because being held in custody of a court
makes it difficult to engage in NCA, NVCA, and FTA.
\begin{assumption}[Monotonicity]
\label{asm::mon} \spacingset{1}
$$Y_i(1) \ \leq \ Y_i(0) \quad {\rm for\ all} \ i.$$
\end{assumption}

\subsection{Causal Quantities of Interest}

We define causal quantities of interest using principal strata that
are determined by the joint values of potential outcomes, i.e.,
$(Y_i(1), Y_i(0))=(y_1, y_0)$, where $y_1,y_0 \in \{0,1\}$
\citep{fran:rubi:02}. Since Assumption~\ref{asm::mon} eliminates one
principal stratum, $(Y_i(1), Y_i(0))=(1, 0)$, there are three
remaining principal strata. The stratum $(Y_i(1),Y_i(0))=(0,1)$
consists of those who would engage in NCA (NVCA or FTA) only if they
are released. We call members of this stratum as ``preventable
cases'' because keeping those arrestees in custody would prevent the
negative outcome (NCA, NVCA, or FTA). The stratum $(Y_i(1), Y_i(0))=(1,1)$ is called
``risky cases,'' and corresponds to those who always engage in NCA
(NVCA or FTA) regardless of the judge's decision. In contrast, the
stratum $(Y_i(1),Y_i(0))=(0, 0)$ represents ``safe cases,'' in which
the arrestees would never engage in NCA (NVCA or FTA) regardless of
the detention decision.

We are interested in examining how PSA provision influences the
judge's detention decisions across different types of cases. We define
the following three average principal causal effects (\ACE),
\begin{eqnarray}
 \ACEP & = &
 \E\{D_i(1)-D_i(0) \mid Y_i(1)=0,Y_i(0)=1\}, \\
 \ACER & = & \E\{D_i(1)-D_i(0) \mid Y_i(1)=1,Y_i(0)=1\},\\ 
 \ACES & = & \E\{D_i(1)-D_i(0) \mid Y_i(1)=0,Y_i(0)=0\}.
\end{eqnarray}
If the PSA is helpful, its provisions should make the judge more
likely to detain the arrestees of the preventable cases. That is, the
principal causal effect on the detention decision for the preventable
cases ($\ACEP$) should be positive. In addition, the PSA should
encourage the judge to release the arrestees of the safe cases,
implying that the principal causal effect for the safe cases ($\ACES$)
should be negative. The desirable direction of the principal causal
effect for risky cases ($\ACER$) depends on various factors including
the societal costs of holding the arrestees of this category in
custody.

\subsection{Nonparametric Identification}

We consider the nonparametric identification of the principal causal
effects defined above. The following theorem shows that under the
aforementioned assumptions, these effects can be identified up to the
marginal distributions of $Y_i(d)$ for $d=0,1$.
\begin{theorem}[Identification]
\label{thm::identification-mon} \spacingset{1}
Under Assumptions~\ref{asm::rand},~\ref{asm::ex},~and~\ref{asm::mon},
\begin{eqnarray*}
\ACEP &=&\frac{\Pr(Y_i=1 \mid Z_i=0)-\Pr(Y_i=1\mid Z_i=1)}{\Pr\{Y_i(0)=1\}- \Pr\{Y_i(1)=1\}},\\
 \ACER 
&=& \frac{\Pr(D_i=1,Y_i=1 \mid Z_i=1)-\Pr(D_i=1,Y_i=1 \mid Z_i=0)}{\Pr\{Y_i(1)=1\}}, \\
\ACES 
&=&\frac{ \Pr(D_i=0,Y_i=0 \mid Z_i=0) - \Pr(D_i=0,Y_i=0 \mid Z_i=1)}{1-\Pr\{Y_i(0)=1\}}. 
\end{eqnarray*}
\end{theorem}
Proof is given in Appendix~\ref{app:identification-mon}. Because
$\Pr\{Y_i(d)\}$ is not identifiable without additional assumptions, we
cannot estimate the causal effects based on
Theorem~\ref{thm::identification-mon}.  The denominators of the
expressions on the right-hand side of
Theorem~\ref{thm::identification-mon}, however, are positive under
Assumption~\ref{asm::mon}. As a result, the signs of the causal
effects are identified from Theorem~\ref{thm::identification-mon},
which allows us to draw qualitative conclusions.  In addition, the
theorem implies that the sign of $\ACEP$ is the opposite of the sign
of the average causal effect on the outcome. This is intuitive because
if the provision of the PSA increases the probability of NCA (NVCA or
FTA), then the judge must have released more arrestees for preventable
cases.

Furthermore, we can obtain the nonparametric bounds on these causal
quantities by bounding $\Pr\{Y_i(d) = y\}$ that appears in the
denominators. From Assumption~\ref{asm::rand} and the law of total probability,
\begin{eqnarray*}
\Pr\{Y_i(d)=1\} &=& \Pr\{Y_i(d)=1 \mid Z_i=z\}\\
 &=& \Pr(Y_i=1\mid D_i=d, Z_i = z) \Pr(D_i=d \mid Z_i =
  z)\\
  && \hspace{.2in}+ \Pr\{Y_i(d)=1\mid D_i=1-d, Z_i = z\} \Pr(D_i=1-d \mid Z_i = z) 
\end{eqnarray*}
for $z,d=0,1$. 
Under Assumption~\ref{asm::mon}, the bounds on the unidentifiable terms are
$\Pr\{Y_i = 1 \mid D_i = 1, Z_i = z\} \le \Pr\{Y_i(0) = 1 \mid D_i =
1, Z_i = z\} \le 1$ and
$0 \le \Pr\{Y_i(1) = 1 \mid D_i = 0, Z_i = z\} \le \Pr\{Y_i = 1 \mid
D_i = 0, Z_i = z\}$. This yields the following bounds on $\Pr\{Y_i(d)=1\} $,
\begin{eqnarray*}
\max_z \Pr(Y_i=1,D_i=1 \mid Z_i = z) \leq &\Pr\{Y_i(1)=1\} & \leq \min_z \Pr(Y_i=1\mid Z_i = z),\\
\max_z \Pr(Y_i=1\mid Z_i = z) \leq &\Pr\{Y_i(0)=1\}& \leq 1- \max_z \Pr(Y_i=0,D_i=0\mid Z_i = z).
\end{eqnarray*}


For point identification, we consider the following unconfoundedness
assumption, which states that conditional on a set of observed
pre-treatment covariates $\bX_i$ and PSA provision, the judge's
decision is independent of the potential outcomes.
\begin{assumption}[Unconfoundedness]
\label{asm::indep} \spacingset{1}
$$Y_i(d) \ \ind \
D_i \mid \bX_i = \bx, Z_i = z, $$ where we also assume
$0 < \Pr(D_i = d \mid \bX_i = \bx, Z_i = z) < 1$ for $z \in \{0,1\}$,
and all $\bx \in \mathcal{X} \ {\rm and} \ d$.

\end{assumption}
Assumption~\ref{asm::indep} holds if $\bX_i$ contains all the
information the judge has access to when making the detention decision
under each treatment condition.  As noted in
Section~\ref{subsec:experiment}, however, the judge may receive and
use additional information regarding whether the arrestee has a job or
a family in the jurisdiction, or perhaps regarding the length of time
the arrestee has lived in the jurisdiction. If these factors have an
impact on both the judge's decisions and arrestee's behaviors, then
the assumption is unlikely to be satisfied. Later, we address this
issue by developing a sensitivity analysis for the potential violation
of Assumption~\ref{asm::indep} (see Section~\ref{sec:sensitive}).

To derive the identification result, consider the following principal
scores \citep{ding:lu:17}, which represent in our application the
population proportion (conditional on $\bX_i)$ of preventable, risky,
and safe cases, respectively,
\begin{eqnarray*}
e_P(\bx)&=& \Pr\{Y_i(1)=0,Y_i(0)=1 \mid \bX_i=\bx\},\\
e_R(\bx)&=& \Pr\{Y_i(1)=1,Y_i(0)=1 \mid \bX_i=\bx\},\\
e_S(\bx)&=& \Pr\{Y_i(1)=0,Y_i(0)=0 \mid \bX_i=\bx\}.
\end{eqnarray*}
Under Assumptions~\ref{asm::ex},~\ref{asm::mon},~and~\ref{asm::indep},
we can identify the principal scores as,
\begin{eqnarray*}
 e_P(\bx) 
 & = & \Pr\{Y_i=1 \mid D_i=0, \bX_i=\bx\}- \Pr\{Y_i=1 \mid D_i=1, \bX_i=\bx\}, \\
 e_R(\bx) & = & 
  \Pr\{Y_i = 1 \mid D_i = 1, \bX_i=\bx\}, \\
 e_S(\bx) &=& 
  \Pr\{Y_i=0 \mid D_i=0, \bX_i=\bx\}.
\end{eqnarray*}

The next theorem shows that we can identify the \ACE\ as the
difference in the weighted average of judge's decisions between the
treatment and control groups.
\begin{theorem}[Identification under Unconfoundedness]
\label{thm::identification-mon-ipw} \spacingset{1}
Under Assumptions~\ref{asm::rand},~\ref{asm::ex},~\ref{asm::mon},~and~\ref{asm::indep}, $\ACEP$, $\ACER$ and $\ACES$ are identified as,
\begin{eqnarray*}
\ACEP &=& \E\{ w_P(\bX_i) D_i \mid Z_i=1\} - \E\{ w_P(\bX_i) D_i \mid Z_i=0\},\\
\ACER &=& \E\{ w_R(\bX_i) D_i \mid Z_i=1\} - \E\{ w_R(\bX_i) D_i \mid Z_i=0\},\\
\ACES &=& \E\{ w_S(\bX_i) D_i\mid Z_i=1\} - \E\{ w_S(\bX_i) D_i \mid Z_i=0\},
\end{eqnarray*}
where 
\begin{eqnarray*}
w_P(\bx) = \frac{e_P(\bx)}{\E\{e_P(\bX_i)\}}, \quad w_R(\bx) = \frac{e_R(\bx)}{\E\{e_R(\bX_i)\}}, \quad w_S(\bx) = \frac{e_S(\bx)}{\E\{e_S(\bX_i)\}}.
\end{eqnarray*}
\end{theorem}
Proof is given in Appendix~\ref{app:identification-mon}.  Although
\citet{ding:lu:17} also identify principal causal effects using
principal scores, they consider principal strata based on an
intermediate variable.  In contrast, we are interested in the causal
effects on the decision within each principal stratum defined by the
values of the potential outcomes.


In some situations, we might consider the following strong
monotonicity assumption instead of Assumption~\ref{asm::mon}.
\begin{assumption}[Strong Monotonicity]
\label{asm::strmon} \spacingset{1}
$$Y_i(1) \ = \ 0\quad {\rm for\ all} \ i.$$
\end{assumption}
The assumption implies that the detention decision prevents NCA, NVCA, 
or FTA. The assumption is plausible for FTA, but may not hold for
NCA/NVCA in some cases. In our data, for example, we find some NCA and
NVCA among the incarcerated arrestees.

Under Assumption~\ref{asm::strmon}, the risky cases do not exist and
hence the $\ACER$ is not defined. This leads to the following
identification result.
\begin{theorem}[Identification under Strong Monotonicity]
\label{thm::identification-strmon} \spacingset{1}
Under
Assumptions~\ref{asm::rand},~\ref{asm::ex},~and~\ref{asm::strmon},
 \begin{eqnarray*}
\ACEP &=& \frac{ \Pr(D_i=0,Y_i=1 \mid Z_i=0) - \Pr(D_i=0,Y_i=1 \mid Z_i=1) }{ \Pr\{Y_i(0)=1\}},\\
\ACES &=& \frac{ \Pr(D_i=0,Y_i=0 \mid Z_i=0) - \Pr(D_i=0,Y_i=0 \mid Z_i=1) }{ \Pr\{Y_i(0)=0\}}.
\end{eqnarray*}
\end{theorem}
Proof is given in Appendix~\ref{app:identification-strmon}. As in
Theorem~\ref{thm::identification-mon}, the $\ACEP$ and $\ACES$ depend
on the distribution of $Y_i(0)$, which is not identifiable. However,
as before, the sign of each effect is identifiable.

For point identification, we invoke the unconfoundedness assumption.
Note that under the strong monotonicity assumption,
Assumption~\ref{asm::indep} is equivalent to a weaker conditional
independence relation concerning only one of the two potential
outcomes,
\begin{equation*}
 Y_i(0) \ \ind \ D_i \mid \bX_i, Z_i = z
\end{equation*}
for $z=0,1$. We now present the identification result.
\begin{theorem}[Identification under Unconfoundedness and Strong
 Monotonicity] \spacingset{1} Under
 Assumptions~\ref{asm::rand},~\ref{asm::ex},~\ref{asm::indep}~and~\ref{asm::strmon},
\begin{eqnarray*}
\ACEP &=& \E\{ w_P(\bX_i) D_i \mid Z_i=1\} - \E\{ w_P(\bX_i) D_i \mid Z_i=0\},\\
\ACES &=& \E\{ w_S(\bX_i) D_i\mid Z_i=1\} - \E\{ w_S(\bX_i) D_i \mid Z_i=0\},
\end{eqnarray*}
where 
\begin{eqnarray*}
w_P(\bx) = \frac{e_P(\bx)}{\E\{e_P(\bX_i)\}}, \quad w_S(\bx) = \frac{e_S(\bx)}{\E\{e_S(\bX_i)\}}.
\end{eqnarray*}
\end{theorem}
Proof is straightforward and hence omitted. While the identification
formulas are identical to those in
Theorem~\ref{thm::identification-mon-ipw}, under
Assumption~\ref{asm::strmon}, we can simply compute the principal
score as $e_S(\bx) = \Pr(Y_i=0 \mid D_i=0, \bX_i=\bx)$ and set
$e_P(\bx)=1-e_S(\bx)$.

\subsection{Ordinal Decision}
\label{sec:ordinal_decision}

We generalize the above identification results to an ordinal decision.
In our application, this extension is important as the judge's release
decision often is based on different amounts of cash bail or varying
levels of supervision of an arrestee. We first generalize the
monotonicity assumption (Assumption~\ref{asm::mon}) by requiring that
a decision with a greater amount of bail is no less likely to make an
arrestee engage in NCA (NVCA or FTA). The assumption may be
reasonable, for example, because a greater amount of bail is expected
to imply a greater probability of being held in custody. The
assumption could be violated if arrestees experience financial strain
in an effort to post bail, causing them to commit NCA (NVCA or FTA).

Formally, let $D_i$ be an ordinal decision variable where $D_i=0$ is
the least amount of bail, and $D_i = 1, \ldots, k$ represents a bail
of increasing amount, i.e., $D_i = k$ is the largest bail
amount. Then, the monotonicity assumption for an ordinal decision is
given by,
\begin{assumption}[Monotonicity with Ordinal Decision]
\label{asm::mon-discrete} \spacingset{1}
$$Y_i(d_1) \ \leq \ Y_i(d_2)\quad {\rm for} \ d_1 \geq d_2.$$ 
\end{assumption}

To generalize the principal strata introduced in the binary decision
case, we define the decision with the least amount of bail that
prevents an arrestee from committing NCA (NVCA or FTA) as follows,
\begin{equation*}
 R_i \ = \ \begin{cases} \min \{d: Y_i(d)=0\} & {\rm if} \ Y_i(k) = 0,
 \\
 k+1 & {\rm if} \ Y_i(k) = 1. \end{cases}
\end{equation*}
We may view $R_i$ as an ordinal measure of risk with a greater value
indicating a higher degree of risk. When $D_i$ is binary, $R_i$ takes
one of the three values, $\{0, 1, 2\}$, representing safe,
preventable, and risky cases, respectively. Thus, $R_i$ generalizes
the principal strata to the ordinal case under the monotonicity
assumption.

Now, we define the principal causal effects in the ordinal decision
case. Specifically, for $r=1,\ldots,k$ (excluding the cases with $r=0$
and $r=k+1$), we define the average principal causal effect of the PSA
on the judge's decisions as a function of this ordinal risk measure,
\begin{equation}
 \ACEP(r) \ = \ \Pr\{D_i(1) \geq r \mid R_i=r\} -\Pr\{D_i(0) \geq r
 \mid R_i=r\}. \label{eq:principaleffect}
\end{equation}
Since the arrestees with $R_i = r$ would not commit NCA (NVCA or FTA)
under the decision with $D_i \geq r$, $\ACEP(r)$ represents a
reduction in the proportion of NCA (NVCA or FTA) that is attributable
to PSA provision among the cases with $R_i=r$. Thus, the expected
proportion of NCA (NVCA or FTA) that would be reduced by the PSA is
given by,
\begin{equation*}
 \sum_{r=1}^k\ACEP(r) \cdot \Pr (R_i=r). 
\end{equation*}
This quantity equals the overall Intention-to-Treat (ITT) effect of
PSA provision on NCA (NVCA or FTA).

Furthermore, the arrestees with $R_i = 0$ would never commit a new
crime regardless of the judge's decisions.  We may, therefore, be
interested in estimating the increase in the proportion of the most
lenient decision for these safest cases. This generalizes the \ACES{}
to the ordinal decision case,
\begin{eqnarray*}
\ACES \ = \ \Pr\{D_i(1) = 0 \mid R_i=0\} -\Pr\{D_i(0) = 0 \mid R_i=0\}.
\end{eqnarray*}
For the cases with $R_i = k+1$ that would always result in a new
criminal activity, a desirable decision may depend on a number of
factors. Note that if we assume the strong monotonicity, i.e.,
$Y_i(k) = 0$ for all $i$, then such cases do not exist.

Like the $\ACES$, the $\ACEP(r)$ can be expressed as a function of the
average principal causal effect ($\ACE$) for each decision
$d=0,1,2,\ldots,k$. This generalized \ACE{} is given by,
\begin{equation}
 \ACE(d,r) \ = \ \Pr\{D_i(1) = d \mid R_i=r\} -\Pr\{D_i(0) =d \mid
 R_i=r\}. \label{eq:ACE}
\end{equation}
In our empirical analysis, we estimate this causal quantity, which has
the same identification conditions.

The identification of these principal causal effects requires the
knowledge of the distribution of $R_i$. Fortunately, under the unconfoundedness
and monotonicity assumptions
(Assumptions~\ref{asm::indep}~and~\ref{asm::mon-discrete}), this
distribution is identifiable conditional on $\bX_i$,
\begin{eqnarray}
e_r(\bx) & = & \Pr(R_i = r \mid \bX_i = \bx) \nonumber\\
&=&\Pr(R_i \geq r \mid \bX_i =\bx)-\Pr(R_i \geq r+1 \mid \bX_i
 =\bx)\nonumber\\
 & = & \Pr\{Y_i(r-1)=1 \mid \bX_i = \bx\}- \Pr\{Y_i(r)=1 \mid
 \bX_i = \bx\} \nonumber\\
&=& \Pr\{Y_i=1 \mid D_i=r-1, \bX_i =\bx\}- \Pr\{Y_i=1 \mid 
 D_i=r, \bX_i = \bx\}, \text{ for } \ r =1,\ldots,k, \label{eq:principalstrata}\\
 e_{k+1}(\bx) &=& \Pr\{Y_i(k)=1 \mid \bX_i = \bx\}\ =\ \Pr\{Y_i=1 \mid
 D_i=k, \bX_i = \bx\}, \nonumber\\
 e_0(\bx) &=& \Pr\{Y_i(0)=0 \mid \bX_i = \bx\}\ =\ \Pr\{Y_i=0 \mid
   D_i=0, \bX_i = \bx\}. \nonumber
\end{eqnarray}
Since $e_r(\bx)$ cannot be negative for each $r$, this yields a set of
testable conditions for
Assumptions~\ref{asm::indep}~and~\ref{asm::mon-discrete}.  This
statement is also true in the binary decision case.



Finally, we formally present the identification result for the ordinal
decision case.
\begin{theorem}[Identification with Ordinal Decision] \spacingset{1}
\label{thm::identification-discrete-ipw}
Under
Assumptions~\ref{asm::rand},~\ref{asm::ex},~\ref{asm::indep}~and~\ref{asm::mon-discrete},
the \ACE{} is identified by
\begin{eqnarray*}
\ACEP(r) &=& \E\{ w_r(\bX_i) \bone(D_i \geq r) \mid Z_i=1\} - \E\{ w_r(\bX_i) \bone(D_i\geq r) \mid Z_i=0\},\\
\ACES &=& \E\{ w_0(\bX_i) \bone(D_i = 0) \mid Z_i=1\} - \E\{ w_0(\bX_i) \bone(D_i= 0) \mid Z_i=0\},
\end{eqnarray*}
where 
$w_r(\bx) = e_r( \bx)/\E\{e_r(\bX_i)\}$ and $\bone()$ is the indicator function.
\end{theorem}
Proof is given in Appendix~\ref{app:identification-discrete-ipw}.

\subsection{Sensitivity Analysis}
\label{sec:sensitive}

The unconfoundedness assumption, which enables the nonparametric
identification of causal effects, may be violated when researchers do
not observe some information that is used by the judge and is
predictive of arrestees' behavior. As noted in
Section~\ref{subsec:experiment}, the length of time the arrestee has
lived in the community may represent an example of such unobserved
confounders.  It is important, therefore, to develop a sensitivity
analysis for the potential violation of the unconfoundedness
assumption (Assumption~\ref{asm::indep}).

We propose a parametric sensitivity analysis (see
Appendix~\ref{sec:nonpara_sens} for a nonparametric sensitivity
analysis). We consider the following bivariate ordinal probit model
for the observed judge's decision $D$ and the latent risk measure
$R_i$,
\begin{eqnarray}
\label{eqn::bayes-Dz} D^\ast_i(z) &=& \beta_Z z+ \bX_i^\top \beta_X +
     z \bX_i^\top \beta_{ZX} + \epsilon_{i1},\\
\label{eqn::bayes-R} R^\ast_i&=& \bm{X}_i^\top \alpha_X+\epsilon_{i2},
\end{eqnarray}
where \begin{eqnarray*}
\begin{pmatrix}
\epsilon_{i1}\\
\epsilon_{i2}
\end{pmatrix} \sim N \left( 
\begin{pmatrix}
0\\
0
\end{pmatrix}, \begin{pmatrix}
1 & \rho\\
\rho & 1
\end{pmatrix}
\right),
\end{eqnarray*}
 and 
\begin{eqnarray*}
D_i(z)= \begin{cases}
0 & D^\ast(z) \leq \theta_{z1}\\
1& \theta_{z1} <D^\ast_i(z) \leq \theta_{z2}\\
\vdots & \vdots\\
k-1 & \theta_{z,k-1} <D^\ast_i(z) \leq \theta_{zk}\\
k & \theta_{zk} <D^\ast_i(z)
\end{cases}, \quad
R_i= \begin{cases}
0& R^\ast_i \leq \delta_0\\
1& \delta_0 <R^\ast_i \leq \delta_1\\
\vdots & \vdots\\
k & \delta_{k-1} <R^\ast_i \leq \delta_{k}\\
k+1 & \delta_{k} <R^\ast_i
\end{cases}.
\end{eqnarray*}
The error terms $ (\epsilon_{i1},\epsilon_{i2})$ are assumed to follow
a bivariate normal distribution. Under this model, $\rho$ represents a
sensitivity parameter since $\rho=0$ implies
Assumption~\ref{asm::indep}. If the value of $\rho$ is known, then the
other coefficients, i.e., $\beta_X$, $\alpha_X$ and $\beta_Z$, can be
estimated, which in turn enables the estimation of the \ACE. In the
literature, \citet{frangakis2002clustered},
\citet{barnard2003principal}, and \citet{forastiere2016identification}
also model the distribution of principal strata using the ordinal
probit model.

Because $R_i$ is a latent variable, the estimation of this model is
not straightforward. In our empirical application, we conduct a
Bayesian analysis to estimate the causal effects \citep[see e.g.,][for
other applications of Bayesian sensitivity
analysis]{hira:imbe:rubi:zhou:00,
 schwartz2011bayesian,mattei2013exploiting,jiang2016principal}.
Appendix~\ref{app:bayesdetails} presents the details of the Bayesian
estimation. We also perform a frequentist analysis, based on
Theorem~\ref{thm::identification-mon-ipw}, that does not require an
outcome model, assessing the robustness of the results to the outcome
model (though we assume $\rho=0$).
 

\subsection{Fairness}
\label{subsec:principalfairness}

Next, we discuss how the above causal effects relate to the fairness
of the judge's decision. In particular, \citet{imai:jian:20} introduce
the concept of ``principal fairness.'' The basic idea is that within
each principal stratum a fair decision should not depend on protected
attributes (race, gender, etc.). \citet{imai:jian:20} provide a
detailed discussion about how principal fairness is related to the
existing definitions of fairness \citep[see also][and references
therein]{corb:etal:17,chou:roth:20}. Although \citet{cost:etal:20}
consider the potential outcomes framework, they only focus on one
potential outcome $Y_i(0)$ rather than the joint potential outcomes
$(Y_i(0), Y_i(1))$.

Formally, let $A_i \in \mathcal{A}$ be a protected attribute such as
race and gender. We first consider a binary decision. We say that
decisions are fair on average with respect to $A_i$ if it does not
depend on the attribute within each principal stratum, i.e.,
\begin{equation}
\label{eqn::def-fairness}
 \Pr\{D_i = 1 \mid A_i, Y_i(1) = y_1, Y_i(0) = y_0\}\ = \ \Pr\{D_i =
 1 \mid Y_i(1) = y_1, Y_i(0) = y_0\}
\end{equation}
for all $y_1, y_0 \in \{0, 1\}$. We can generalize this definition to
the ordinal case as,
\begin{equation*}
 \Pr(D_i \ge d \mid A_i, R_i = r)\ = \ \Pr(D_i \ge d \mid R_i = r) 
\end{equation*}
for $1 \le d \le k$ and $0 \le r \le k+1$.

The degree of fairness for
principal stratum $R_i = r$ can be measured using the maximal
deviation among the distributions for different groups,
\begin{equation}
\Delta_r(z) \ = \ \max_{a, a^\prime,d} \left|\Pr\{D_i(z) \ge d
 \mid A_i = a, R_i = r\}\ - \ \Pr\{D_i(z) \ge d  
 \mid A_i = a^\prime, R_i = r\}\right| \label{eq:deltaz}
\end{equation}
for $z=0,1$. By estimating $\Delta_r(z)$, we can use the experimental
data to examine whether or not the provision of the PSA improves the
fairness of the judge's decisions. Specifically, if PSA provision
improves the fairness of judge's decisions for the principal stratum
$r$, we should have $\Delta_r(1) \le \Delta_r(0)$.

\subsection{Optimal Decision Rule}
\label{sec::optdecision}

The discussion so far has focused on estimating the impacts of
algorithmic recommendations on human decisions. We now show that the
experimental data can also be used to derive an optimal decision rule
given a certain objective. In addition, by comparing human decisions
and algorithmic recommendations with optimal decision rules, we can
evaluate their efficacy. In our application, one goal is to prevent as
many NCAs (NVCAs or FTAs) as possible while avoiding unnecessarily
harsh initial release decisions. To achieve this, we must carefully
weigh the cost of negative outcomes and that of unnecessarily harsh
decisions. Once these costs are specified as part of the utility
function, one can empirically assess this tradeoff using the
experimental data.

Formally, let $\delta$ be the judge's decision based on $\bX_i$, which
may include the PSA. We consider a deterministic decision rule, i.e.,
$\delta(\bx) = d$ if $\bx \in \cX_d$ where $\cX_d$ is a
non-overlapping partition of the covariate space $\cX$ with
$\cX=\bigcup_{r=0}^k \cX_r$ and
$\cX_r \cap \cX_{r^\prime} = \emptyset$. We consider the utility
function of the following form,
\begin{eqnarray*}
U_i(\delta) \ = \ \begin{cases}
-c_0 & \delta(\bX_i) < R_i\\
1 & \delta(\bX_i) = R_i\\
1-c_1 & \delta(\bX_i) > R_i
\end{cases},
\end{eqnarray*}
where $c_0$ and $c_1$ represent the cost of an NCA (NVCA or FTA) and
that of an unnecessarily harsh decision, respectively. Under this
setting, preventing an NCA (NVCA or FTA) with the most lenient
decision ($ \delta(\bX_i) = R_i$) yields the utility of one, while we
incur the cost $c_1$ for an unnecessarily harsh decision
($ \delta(\bX_i) > R_i$), leading to the net utility of $1-c_1$.

The relative magnitude of these two cost parameters, $c_0$ and $c_1$,
may depend on the consideration of various factors including the
potential harm to the public and arrestees caused by the negative
outcomes and unnecessarily harsh decisions, respectively. When
$c_0 =c_1=0$, for example, $U_i(\delta)$ reduces to
$\bone\{\delta(\bX_i) \geq R_i\}$, which is non-zero only if the
decision is sufficiently harsh so that it prevents the negative
outcome. The optimal decision under this utility is the most
stringent decision, i.e., $\delta(\bX_i)= k$, for all cases. If
$c_0=2$ and $c_1 =1$, the resulting utility function implies that the
cost of NCA (NVCA or FTA) is twice as large as that of an
unnecessarily harsh decision.

We derive the optimal decision rule $\delta^\ast$ that maximizes the
expected utility,
\begin{eqnarray*}
\delta^\ast & = & \argmax_\delta \E\{U_i(\delta)\}.
\end{eqnarray*}
For $r=0,\ldots,k+1$ and $d = 0,\ldots,k$, we can write,
\begin{eqnarray*}
\E[ \bone\{\delta(\bX_i)=d, R_i=r\}] & = & \E \{\bone(\bX_i \in \cX_d , R_i=r )\}
 \ = \ \E\left\{ \bone(\bX_i \in \cX_d) \cdot
      e_r(\bX_i)\right\}. 
\end{eqnarray*}
Thus, we can express the expected utility as,
\begin{eqnarray}
   && \E\{U_i(\delta)\} \\
  & = & \sum_{r=0}^{k+1} \left(\sum_{d> r}  (1-c_1)\E [\bone\{\delta(\bX_i)= d, R_i=r \} ] +\sum_{d= r}  \E [\bone\{\delta(\bX_i)= d, R_i=r \} ] -\sum_{d< r}  c_0\E [\bone\{\delta(\bX_i)= d, R_i=r \} ]\right) \nonumber\\
&=& \sum_{r=0}^{k+1} \left[ \sum_{d\geq r} \E\left\{ \bone(\bX_i \in \cX_d) \cdot
      e_r(\bX_i)\right\}-c_0 \sum_{d<r} \E\left\{ \bone(\bX_i \in \cX_d) \cdot
      e_r(\bX_i)\right\}-c_1 \sum_{d>r} \E\left\{ \bone(\bX_i \in \cX_d) \cdot
      e_r(\bX_i)\right\} \right]\nonumber\\
&=& \sum_{d=0}^k \E \left[ \bone(\bX_i \in \cX_d) \left\{
 \sum_{r\leq d} e_r(\bX_i) - c_0 \cdot \sum_{r>d} e_r(\bX_i) -c_1
 \cdot \sum_{r<d} e_r(\bX_i) \right\}
 \right]. \label{eq:expectedutility}
\end{eqnarray}
This yields the following optimal decision,
\begin{eqnarray}
\delta^\ast(\bx) \ = \ \argmax_{d \in \{0,\ldots,k\}} g_d(\bx)
\quad
{\rm where} \quad
g_d(\bx) \ = \ \sum_{r\leq d} e_r(\bx) - c_0 \cdot \sum_{r>d}
 e_r(\bx)- c_1 \cdot \sum_{r<d} e_r(\bx) . \label{eq:optimal}
\end{eqnarray}
We can, therefore, use the experimental estimate of $e_r(\bx)$ to
learn about the optimal decision.

Policy makers could derive the optimal decision rule by using the
above result and then adopt this rule as the recommendation for
judges. However, this may not be useful if the judge decides to follow
the algorithmic recommendation selectively for some cases or ignore it
altogether. Instead, we may wish to construct PSA scores that maximize
the optimality of the judge's decision. Unfortunately, the derivation
of such an optimal PSA score is challenging since the PSA scores were
not directly randomized in our experiment. We tackle this problem in a
separate paper \citep{benm:etal:21}. In
Appendix~\ref{app:optimalprovision}, we also consider the optimal
provision of the PSA given the same goal considered above (i.e.,
prevent as many NCAs (NVCAs or FTAs) as possible with the minimal
amount of bail).

\section{Empirical Analysis}
\label{sec:application}

In this section, we apply the proposed methodology to the data from
the field RCT described in Section~\ref{sec:RCT}.

\subsection{Preliminaries}

As explained in Section~\ref{subsec:data}, we use the ordinal decision
variable with three categories --- the signature bond ($D_i=0$), the
bail amount of \$1,000 or less ($D_i=1$), and the bail amount of
greater than \$1,000 ($D_i=2$). Given this ordinal decision, we label
the principal strata as safe ($R_i = 0$), easily preventable
($R_i = 1$), preventable ($R_i = 2$), and risky cases ($R_i = 3$).

We fit the Bayesian model defined in
Equations~\eqref{eqn::bayes-Dz}~and~\eqref{eqn::bayes-R} with a
diffuse prior distribution as specified in
Appendix~\ref{app:bayesdetails}, separately for each of three binary
outcome variables --- FTA, NCA, and NVCA. The model incorporates the
following pre-treatment covariates: gender (male or female), race
(white or non-white), the interaction between gender and race, age,
and several indicator variables regarding the current and past
charges. It also includes a binary variable for the presence of
pending charge (felony, misdemeanor, or both) at the time of offense,
four binary variables for current charges (non-violent misdemeanor,
violent misdemeanor, non-violent felony, and violent felony), a
four-level ordinal variable for the DMF recommendation, three
variables for prior conviction (binary variables for misdemeanor and
felony as well as a four-level factor variable for violent
conviction), a binary variable for prior sentence to incarceration,
and two variables for prior FTA (a three-level factor variable for
FTAs from past two years, and a binary variable for FTAs from over two
years ago).

We use the Gibbs sampling and run five Markov chains of 50,000
iterations each with random starting values independently drawn from
the prior distribution. Based on the Gelman-Rubin statistic for
convergence diagnostics, we retain the second half of each chain and
combine them to be used for our
analysis. Appendix~\ref{app:bayesdetails} presents the computational
details including the Gibbs sampling algorithm we use.

\begin{figure}[t!]
 \centering \spacingset{1}
 \includegraphics[width=0.32\textwidth]{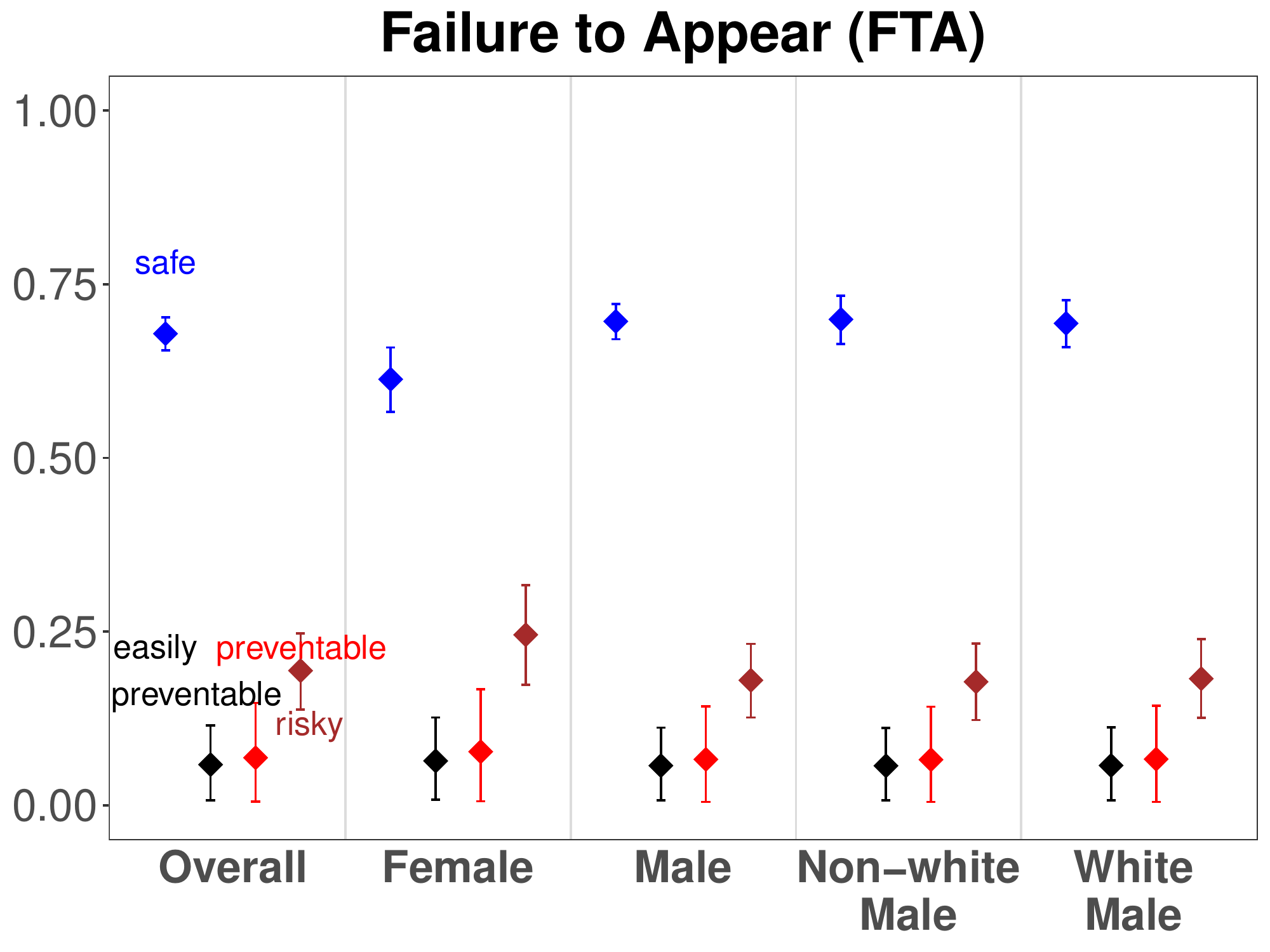}
 \includegraphics[width=0.32\textwidth]{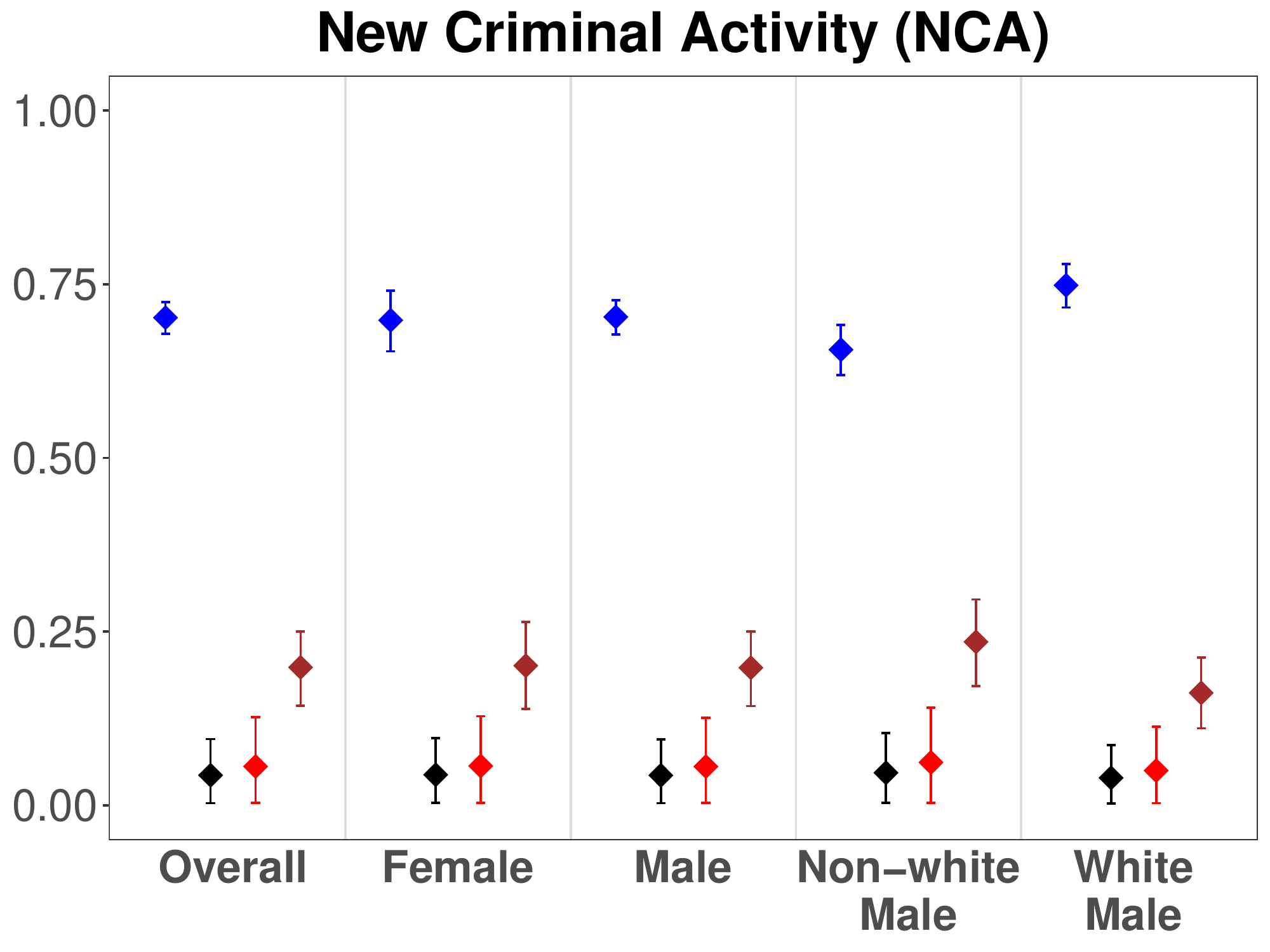}
 \includegraphics[width=0.32\textwidth]{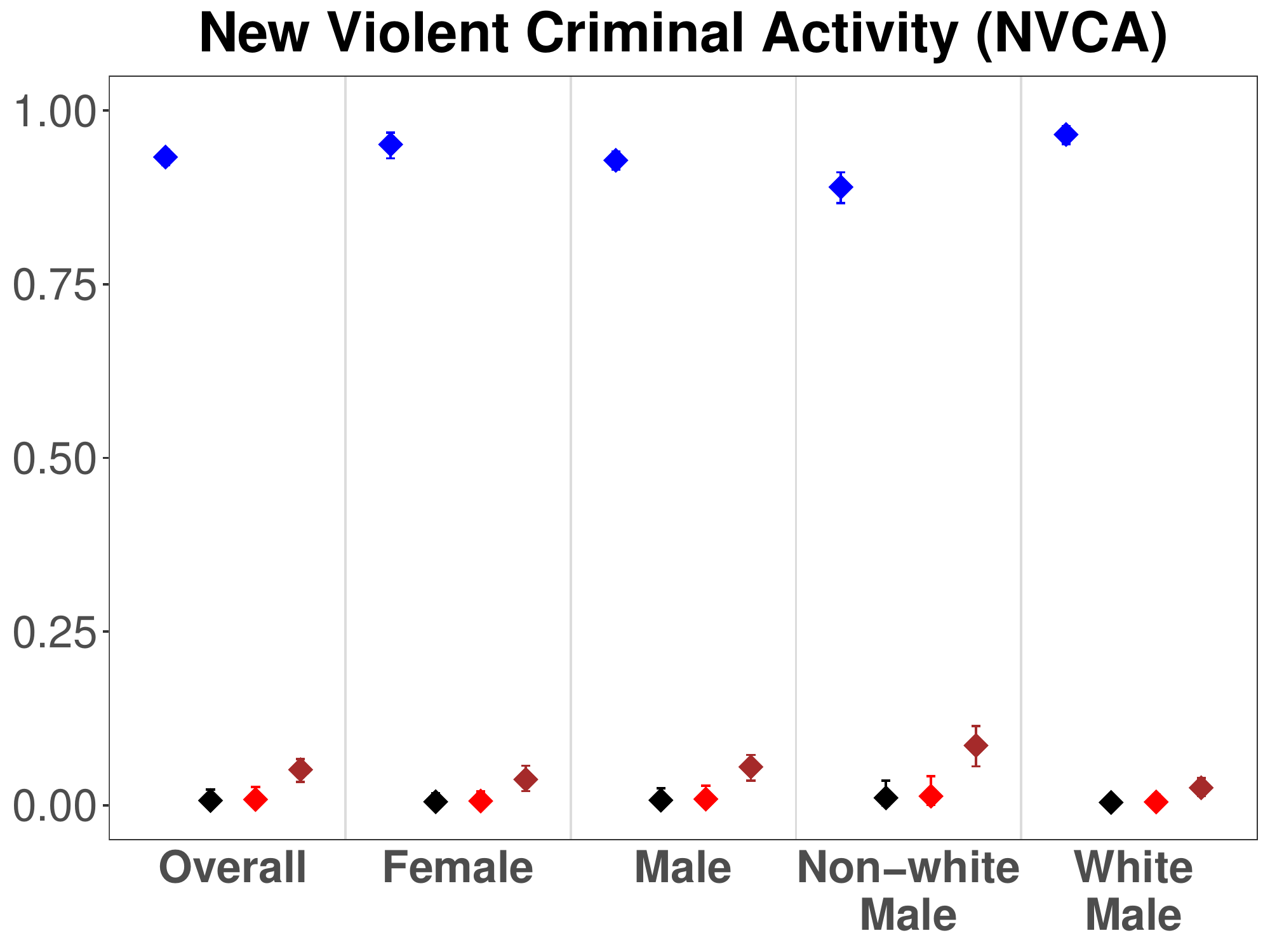}
 \caption{Estimated Proportion of Each Principal Stratum. Each plot
  represents the result using one of the three outcome variables
  (FTA, NCA, and NVCA), where the blue, black, red, and brown
  diamonds represent the estimates for safe, easily preventable,
  preventable, risky cases, respectively. The solid vertical lines
  represent the $95\%$ Bayesian credible intervals. The results show
  that a vast majority of cases are safe across subgroups and across
  different outcomes. The proportion of safe cases is estimated to be
  especially high for NVCA.} \label{fig:er}
\end{figure}
	
We begin by computing the estimated population proportion of each
principal stratum based on Equation~\eqref{eq:principalstrata}.
Figure~\ref{fig:er} presents the results. We find that for FTA, the
overall proportion of safe cases (blue) is estimated to be 67\%,
whereas those of easily preventable (black), preventable (red), and
risky (brown) cases are 6\%, 7\%, and 20\% respectively. A similar
pattern is observed for FTA and NCA across different racial and gender
groups, while the estimated overall proportion of safe cases is even
higher for NVCA, exceeding 90\%.

\subsection{Average Principal Causal Effects}
\label{subsec:estimatedeffects}

\begin{figure}[p]
 \centering \spacingset{1}
 \includegraphics[width=\textwidth]{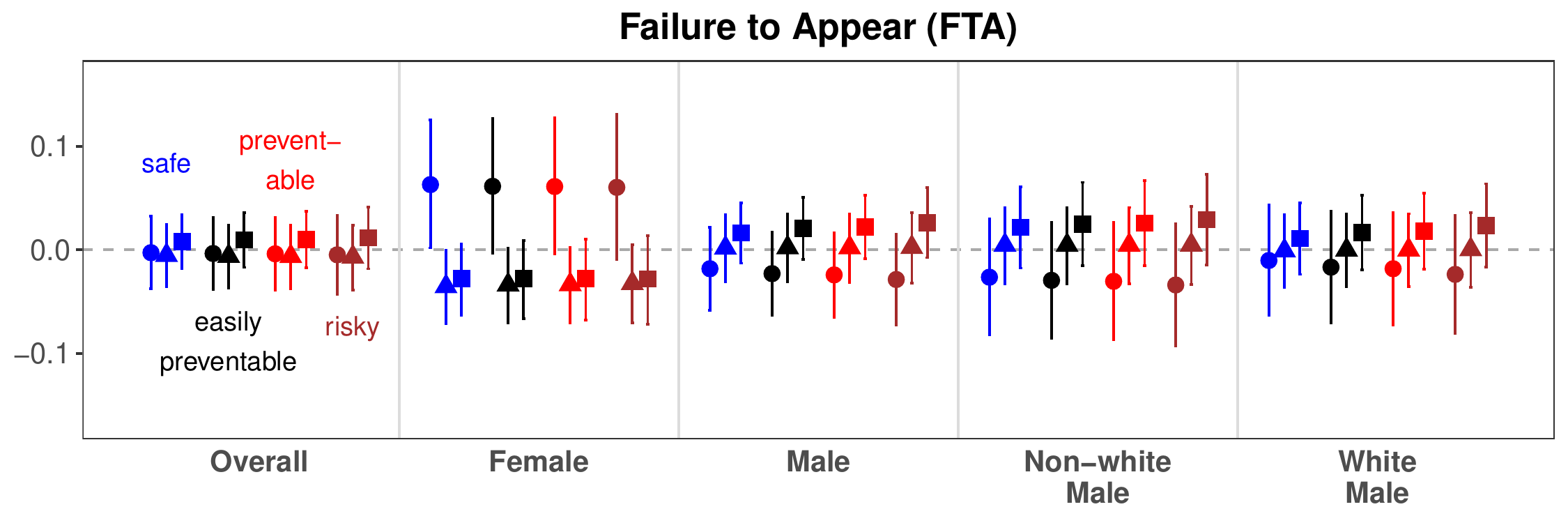}
 \includegraphics[width=\textwidth]{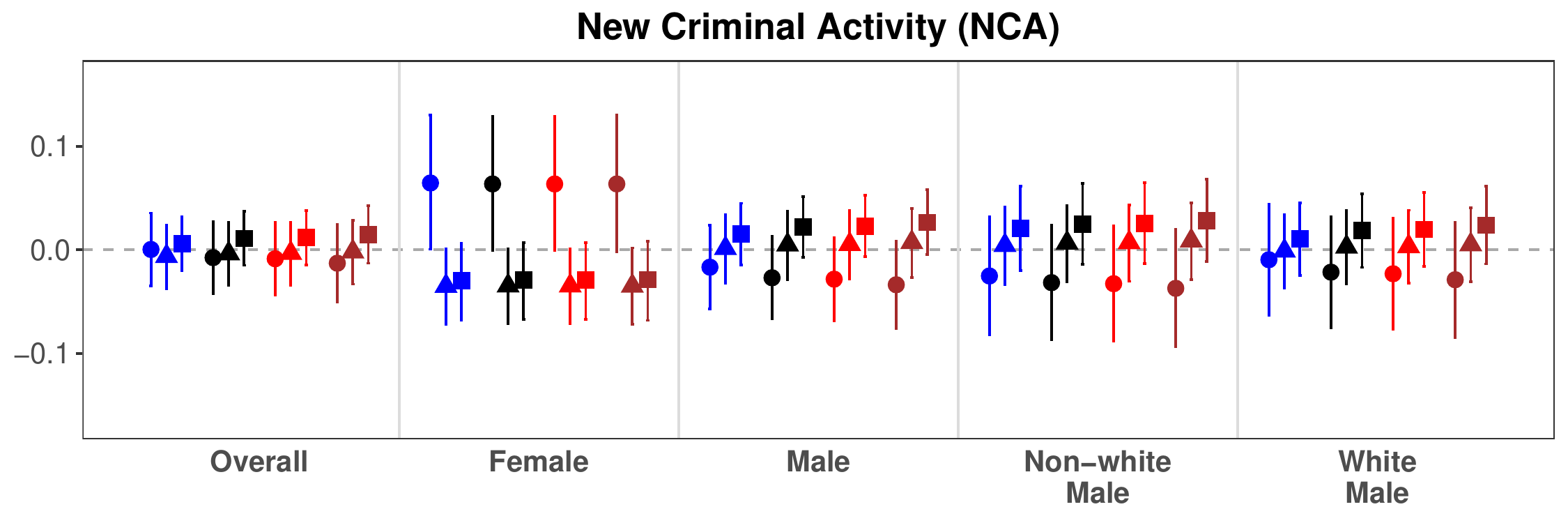}
 \includegraphics[width=\textwidth]{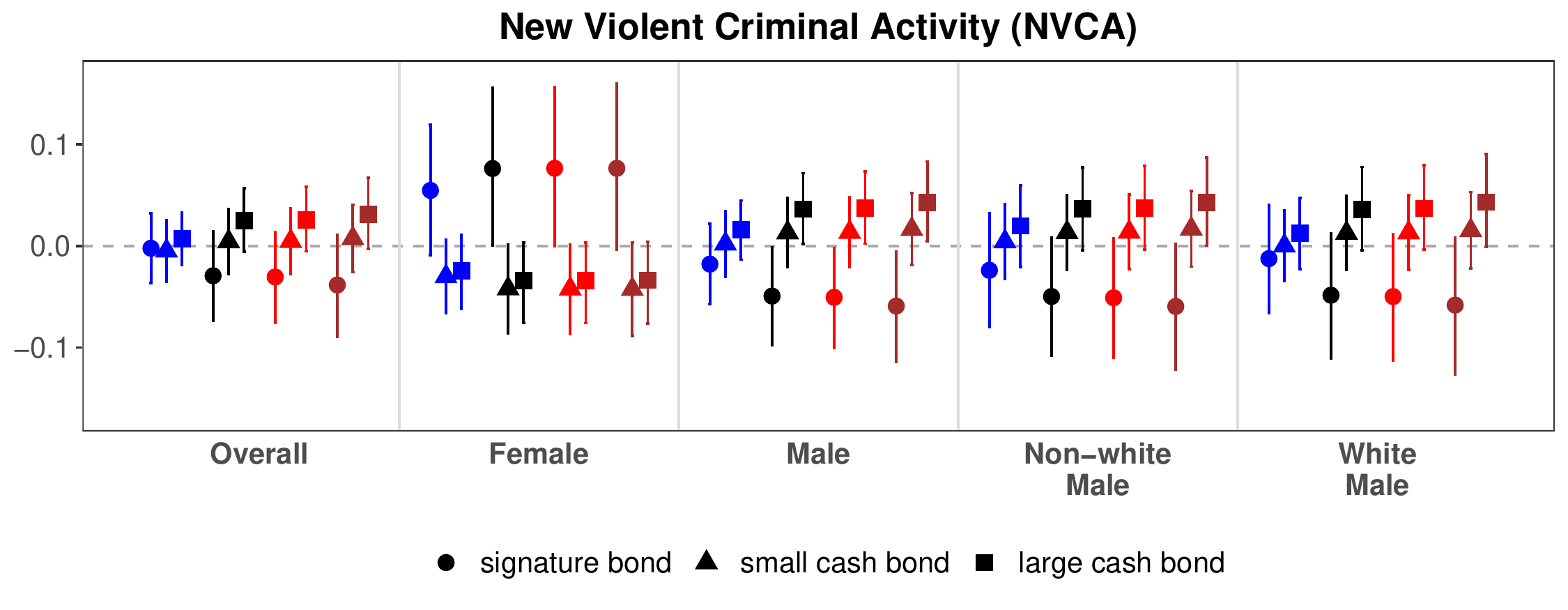}
 \caption{Estimated Average Principal Causal Effects (\ACE) of PSA
  Provision on the Judge's Decision. Each panel presents the overall
  and subgroup-specific results for a different outcome
  variable. Each column within a panel shows the estimated \ACE{} of
  PSA provision for safe (blue), easily preventable (black),
  preventable (red), and risky (brown) cases. For each of these
  principal strata, we report the estimated \ACE{} on the judge's
  decision to impose a signature bond (circles), a small cash bail
  amount of 1,000 dollars or less (triangles), and a large cash bail
  amount of greater than 1,000 (squares). The vertical line for each
  estimate represents the Bayesian $95\%$ credible interval. The
  results show that PSA provision may make the judge's decision more
  lenient for female arrestees regardless of their risk levels. PSA
  provision may also encourage the judge to make harsher decisions
  for male arrestees with a greater risk level though the effect
  sizes are relatively small.}
 \label{fig:ace}
\end{figure}

Figure~\ref{fig:ace} presents the estimated \ACE{} of PSA provision on
the three ordinal decision categories, separately for each of the
three outcomes and each principal stratum (see
Equation~\eqref{eq:ACE}). The overall and subgroup-specific results
are given for each of the four principal strata --- safe (blue),
easily preventable (black), preventable (red), and risky (brown)
cases. For a given principal stratum, we present the estimated \ACE{}
on each decision category --- signature bond (circle), small cash bond
(triangle), and large cash bond (square). The left column of each
panel shows that PSA provision has little overall impact on the
judge's decision across four principal strata for FTA and NCA. There
is a suggestive, but inconclusive, evidence that PSA provision leads
to an overall harsher decision for NVCA among easily preventable,
preventable, and risky cases.

We also present the estimated \ACE{} for different gender and racial
groups in the remaining columns of each panel. We find potentially
suggestive evidence that PSA provision may make it more likely for the
judge to impose signature bonds (circles) on female arrestees instead
of cash bonds (triangles and squares) across three
outcomes. Interestingly, for all outcomes, this pattern appears to
hold for any of the four principal strata, implying that PSA provision
might not help the judge distinguish different risk levels of female
arrestees. Our analysis also finds that for NVCA, PSA provision may
lead to a harsher decision for easily preventable, preventable, and
risky cases among male arrestees while it has little effect on the
safe cases. This suggests that PSA provision may help distinguish
different risk levels among male arrestees, resulting in improved
decisions at least in terms of the original goal of the PSA. There is
no discernible racial difference in these effects.

In Appendix~\ref{subsec:age_effects}, we explore the estimated \ACE{}
for different age groups. We find that PSA provision may lead to a
harsher decision for arrestees of the 29--35 years old group across
three outcomes. This pattern appears to generally hold across all
principal strata though for NVCA the effects are more pronounced for
easily preventable, preventable, and risky cases.  In addition, our
analysis yields suggestive evidence that across all outcomes, PSA
provision may make the judge's decision more lenient for the oldest
(46 years old or above) group. This appears to be true across all
three outcomes except that for NVCA the effect may exist only for safe
cases. We reiterate, however, that these results are based on
preliminary data and the effect sizes are relatively small.

We conduct two robustness analyses. First, we perform a frequentist
analysis that is based on Theorem~\ref{thm::identification-mon-ipw}
and does not assume an outcome model. The results are shown in
Appendix~\ref{app:frequentist}, and are largely consistent with those
shown here. As expected, the estimation uncertainty of the frequentist
analysis, which makes less stringent assumptions than Bayesian
analysis, is greater.  Second, we conduct parametric sensitivity
analyses using the methods described in Section~\ref{sec:sensitive}
(see Appendix~\ref{sec:sensitivity_results}). We set the value of
correlation parameter $\rho$ to 0.05, 0.1, and 0.3, and examine how
the estimated \ACE{} changes. The results (see
Figures~\ref{fig:sens1}--\ref{fig:sens3}) are largely consistent
across different values of $\rho$ although the effects for females
tend to exhibit a large degree of estimation uncertainty especially
when the correlation is high and particularly for NVCA. This is not
surprising. There are only a small number of female arrestees and only
a handful of NVCA events corresponding to them.

\subsection{Gender and Racial Fairness}

We now examine the impacts of PSA provision on gender and racial
fairness. Specifically, we evaluate the principal fairness of PSA
provision as discussed in Section~\ref{subsec:principalfairness}. We
use gender (female vs. male) and race (white male vs. non-white male)
separately as a protected attribute, and analyze whether or not the
provision of the PSA improves the fairness of the judge's decision in
terms of the protected attribute. While the gender analysis is based
on the entire sample, the racial analysis is based on the male sample
only due to the limited sample size for females.

\begin{figure}[t!]
 \spacingset{1}
 \begin{subfigure}[t]{\textwidth}
 \subcaption{Gender Fairness}
 \includegraphics[width=\textwidth]{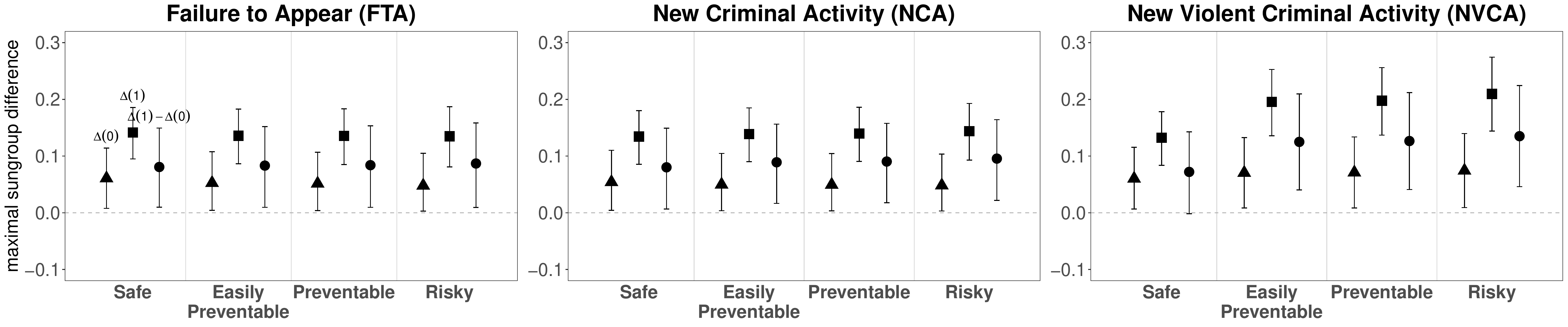} \\
 \end{subfigure} 
 \vspace{.2in}
 \begin{subfigure}[t]{\textwidth}
 \subcaption{Racial Fairness (Male Sample)}
 \includegraphics[width=\textwidth]{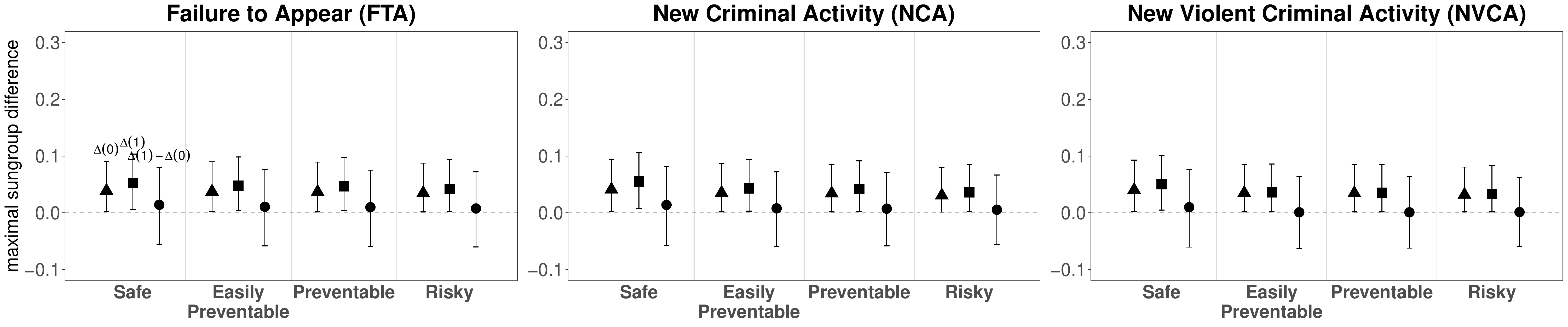}
 \end{subfigure}
 \vspace{-.2in}
 \caption{Gender and Racial Fairness of the Judge's Decisions. Within
  each plot, we show three estimates separately for each principal
  stratum --- the maximal subgroup difference in the judge's decision
  probability of imposing a cash bond with PSA provision (squares;
  $\Delta(1)$) and without it (triangles; $\Delta(0)$) as well as the
  difference between them (circles; $\Delta(1)-\Delta(0)$). The
  vertical solid lines represent the $95\%$ Bayesian credible
  intervals. A positive value of the difference would imply that the
  PSA reduces the fairness of the judge's decisions. For the gender
  analysis (top panel), even without the PSA, the judge seems to be
  more likely to impose a cash bond on male arrestees when compared
  to female arrestees with the same risk levels. PSA provision
  appears to increase this tendency. For the race analysis (bottom
  panel), PSA provision has little impact across all outcomes and
  risk levels. We reiterate, however, that these results are based
  on the preliminary data.}
 \label{fig:fair}
\end{figure}

Figure~\ref{fig:fair} presents the results for gender (top panel) and
racial (bottom panel) fairness across the principal strata and
separately for each of the three outcomes. Each column within a given
plot presents $\Delta_r(z)$ defined in equation~\eqref{eq:deltaz},
which represents the maximal subgroup difference in the judge's
decision probability distribution within the same principal stratum
$R_i = r$ under the provision of the PSA $z=1$ (no provision
$z=0$). In this application, the maximal difference always occurs at
$d=1$, allowing us to interpret $\Delta_r(z)$ as the difference in
probability of imposing a cash bond ($D\geq 1$) rather than a
signature bond. We also present the estimated difference caused by PSA
provision in the two maximal subgroup differences, i.e.,
$\Delta_r(1)-\Delta_r(0)$. If this difference is estimated to be
positive, then PSA provision reduces the fairness of judge's decisions
by increasing the maximal subgroup difference.

 
We find that PSA provision might worsen the gender fairness of the
judge's decisions. When the PSA is provided, the maximal gender
difference in the judge's decision probability is on average greater
than that when it is not provided. The effect is particularly large
and statistically significant for NVCA and for preventable, easily
preventable, and risky cases. This is consistent with our finding that
especially for NVCA, PSA provision might make the judge's decision
more lenient for female arrestees while it leads to a harsher decision
for male arrestees among preventable, easily preventable, and risky
cases. Thus, PSA provision appears to increase disparate
decision-making across gender.

PSA provision, however, does not have a statistically significant
impact on the racial fairness of the judges' decisions among male
arrestees. For instance, in the principal stratum of safe cases, we
find that PSA provision does not affect the maximal difference in the
judge's decision probability (between non-white males and white
males). This suggests that in terms of principal fairness, the PSA may
not alter any existing racial difference in the judge's decisions.

\subsection{Using Optimal Decision to Evaluate the DMF Recommendation}

Finally, we evaluate the DMF recommendation by comparing it with the
optimal decision under different values of the costs. For simplicity,
we consider a binary decision: signature or cash bond. As discussed in
Section~\ref{sec::optdecision}, given a specific pair of cost
parameters $(c_0, c_1)$ and the experimental estimate of $e_r(\bx)$
for $r=0,1,2$, we can compute the optimal decision for each case
according to Equation~\eqref{eq:optimal}.  We then obtain the
estimated proportion of cases, for which a cash bond is optimal. We
repeat this process for a grid of different values for the cost of a
negative outcome ($c_0$; FTA, NCA, and NVCA) and that of an
unnecessarily harsh decision ($c_1$).

\begin{figure}[t!]
 \vspace{-.2in}
 \centering \spacingset{1}
 \begin{subfigure}[t]{\textwidth}
 	\subcaption{The cases whose DMF recommendation is a signature bond}
 	\includegraphics[width = \textwidth]{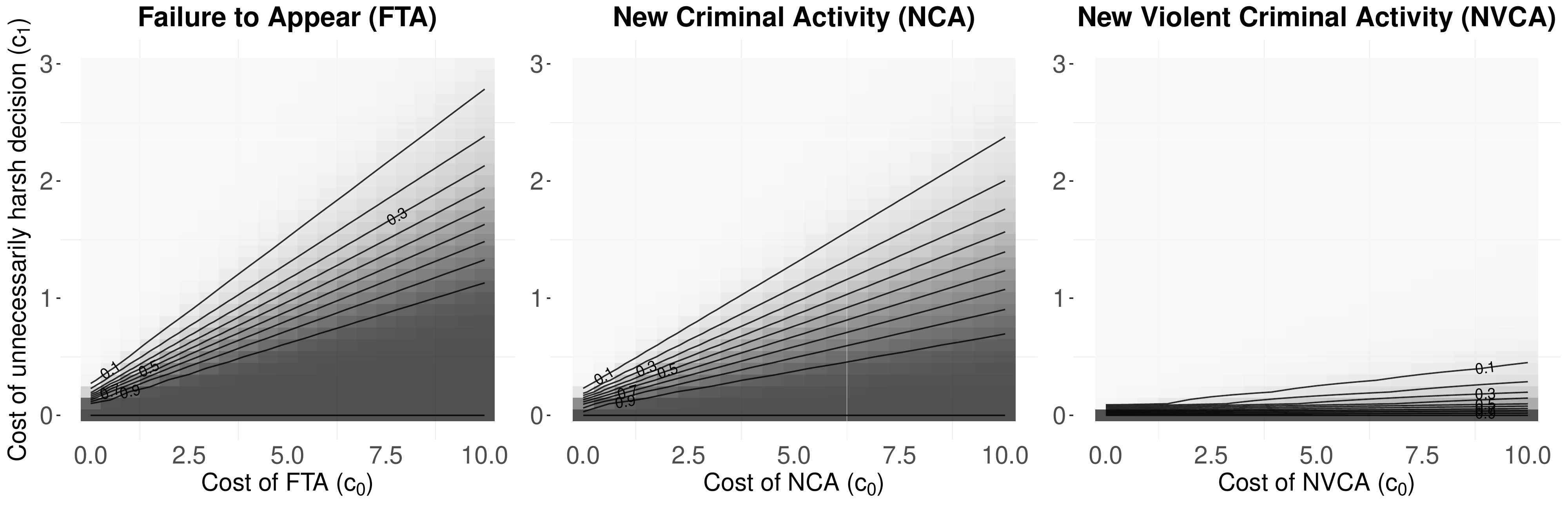}
 \end{subfigure} \\
 \vspace{.25in}
 \begin{subfigure}[t]{\textwidth}
 	\subcaption{The cases whose DMF recommendation is a cash bond}
 	\includegraphics[width = \textwidth]{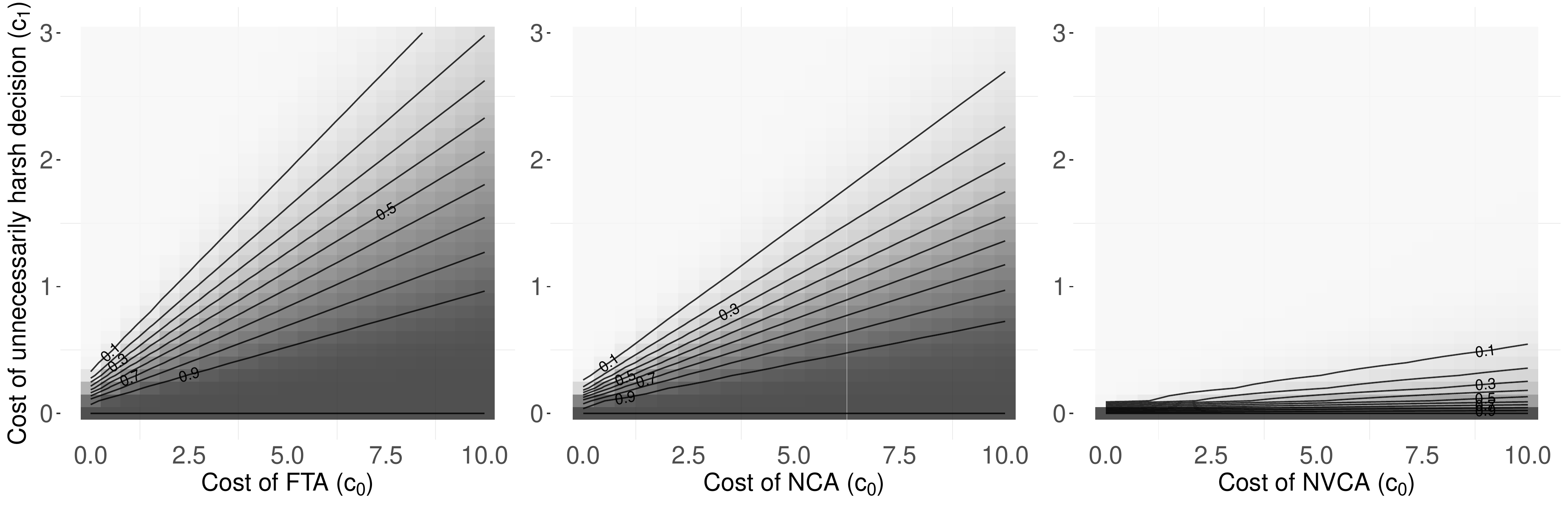}
 \end{subfigure}
 \caption{Estimated Proportion of Cases for Which Cash Bond is
  Optimal. Each column represents the results based on one of the
  three outcomes (FTA, NCA, and NVCA). The top (bottom) panel shows
  the results for the cases whose DMF recommendation is a signature
  (cash) bond. In each plot, the contour lines represent the
  estimated proportions of cases for which a cash bond is optimal,
  given the cost of an unnecessarily harsh decision ($c_1$; $y$-axis)
  and that of a negative outcome ($c_0$; $x$-axis). A dark grey area
  represents a greater proportion of such cases. The results show
  that regardless of DMF recommendation, a signature bond is optimal
  unless the cost of a negative outcome is much greater than the cost
  of an unnecessarily harsh decision.}
 \label{fig:opt_comb}
\end{figure}

The top panel of Figure~\ref{fig:opt_comb} presents the results for
the cases whose DMF recommendation is a signature bond. In contrast,
the bottom panel of the figure shows the results for the other cases
(i.e., the DMF recommendation is a cash bond). In each plot, a darker
grey region represents a greater proportion of cases, for which a cash
bond is optimal. The results suggest that unless the cost of a
negative outcome is much higher than the cost of an unnecessarily
harsh decision, imposing a signature bond is the optimal decision for
a vast majority of cases.

We also find that for all three outcomes, a cash bond is optimal for a
greater proportion of cases when the DMF recommendation is indeed a
cash bond. However, this difference is small, suggesting that the DMF
recommendation is only mildly informative. Similar results are found
even if we separately examine three PSA scores (see
Figure~\ref{fig:opt_sep} in Appendix~\ref{app:optimalresults}).

\subsection{Comparison between the Judge's Decisions and DMF Recommendations}

\begin{figure}[t!]
 \vspace{-.2in}
 \centering \spacingset{1}
 \begin{subfigure}[t]{\textwidth}
	\subcaption{Treatment Group}
	\includegraphics[width = \textwidth]{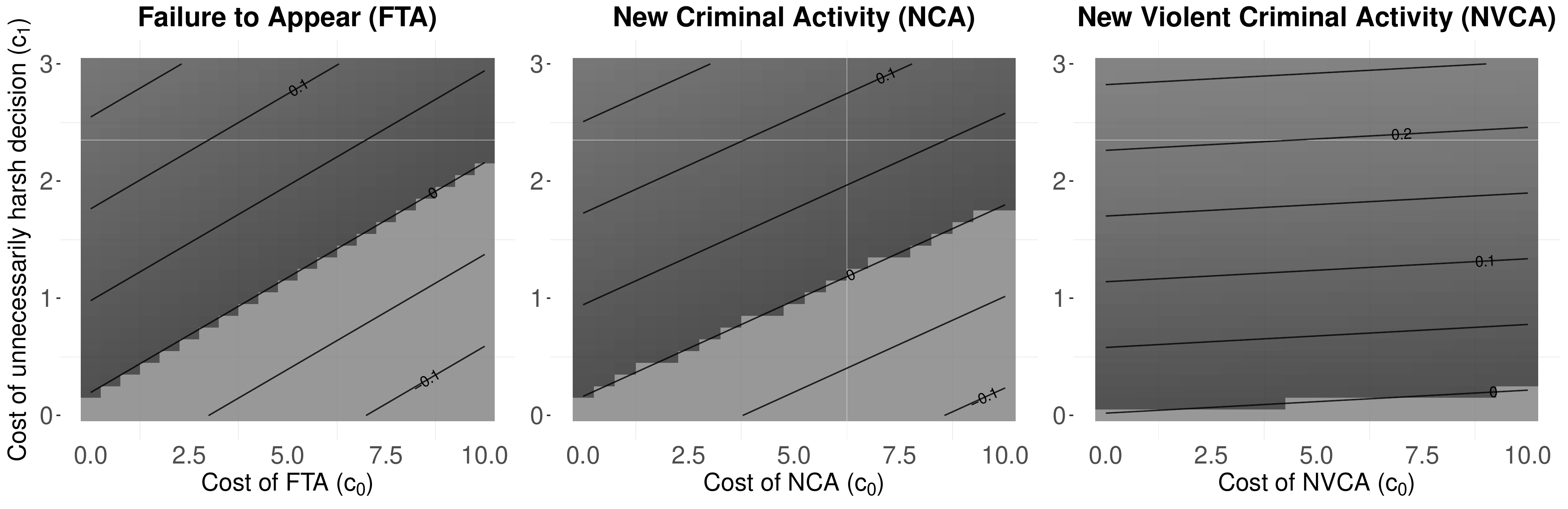}
\end{subfigure} \\
\vspace{.25in}
\begin{subfigure}[t]{\textwidth}
	\subcaption{Control Group}
	\includegraphics[width = \textwidth]{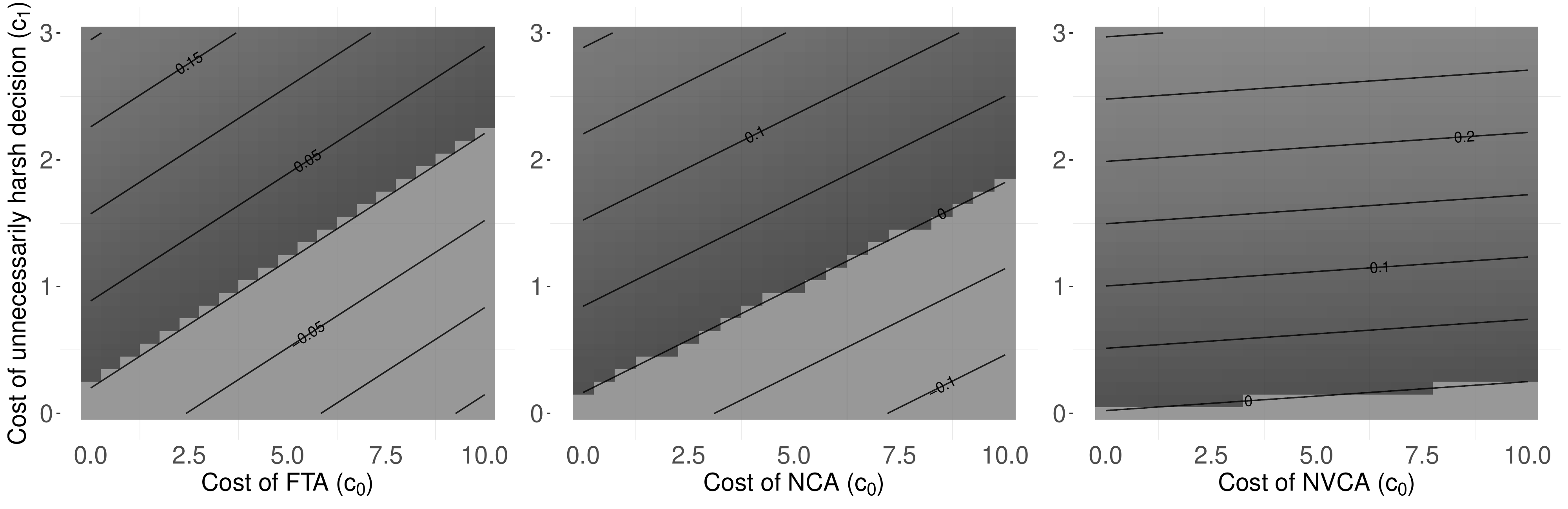}
\end{subfigure}
\caption{Estimated Difference in the Expected Utility between Judge's
  Decisions and DMF Recommendations for the Treatment (top panel) and
  Control (bottom panel) Group. Each column represents the results
  based on one of the three outcomes with a darker region indicating
  the values of the costs (the cost of a negative outcome and the cost
  of an unnecessarily harsh decision) for which the Judge's decision
  yields a higher expected utility than the corresponding DMF
  recommendation. The results show that the judge's decision yields a
  higher expected utility than the DMF recommendation unless the cost
  of a negative outcome is much higher than that of an unnecessarily
  harsh decision. This pattern holds for all outcomes and is unchanged
  by the provision of the PSA.}
 \label{fig:opt_utility}
\end{figure}

Lastly, we compare the judge's actual decision with the DMF
recommendation in terms of the expected utility given in
Equation~\eqref{eq:expectedutility}.
The top panel of Figure~\ref{fig:opt_utility} represents the results
for the treatment group (i.e., judge's decisions with the PSA),
whereas the bottom panel represents those for the control group (i.e.,
judge's decisions without the PSA). A darker grey area indicates that
the expected utility for the judge's decision is estimated to be
greater than the DMF recommendation. Most of these estimates are
statistically significant (see Figure~\ref{fig:utility_ci} for more
details). Therefore, unless the cost of a negative outcome is much
greater than the cost of an unnecessarily harsh decision, the judge's
decision (with or without the PSA scores) yields a greater expected
utility than the DMF recommendation. This is especially true for
NVCA. Altogether, our analysis implies that the DMF recommendations
may be unnecessarily harsher than the judge's decisions.

\section{Concluding Remarks}
\label{sec:conclusion}

In today's data-rich society, many human decisions are guided by
algorithmic recommendations. While some of these algorithmic-assisted
human decisions may be trivial and routine (e.g., online shopping and
movie suggestions), others that are much more consequential include
judicial and medical decision-making. As algorithmic recommendation
systems play increasingly important roles in our lives, we believe
that a policy-relevant question is how such systems influence human
decisions and how the biases of algorithmic recommendations interact
with those of human decisions. These questions necessitate the
empirical evaluation of the impacts of algorithmic recommendations on
human decisions.

In this paper, we present a set of general statistical methods that
can be used for the experimental evaluation of algorithm-assisted
human decision-making. We applied these methods to the preliminary
data from the first-ever randomized controlled trial for assessing the
impacts of PSA provision on judges' pretrial decisions. There are
several findings that emerge from our initial analysis. First, we find
that PSA provision has little overall impact on the judge's
decisions. Second, we find potentially suggestive evidence PSA
provision may encourage the judge to make more lenient decisions for
female arrestees regardless of their risk levels while leading to more
stringent decisions for males who are classified as risky. Third, PSA
provision appears to widen the existing gender difference of the
judge's decisions against male arrestees whereas it does not seem to
alter decision-making across race among male arrestees. We caution,
however, that these findings could be explained by other factors that
are correlated with gender and race. Finally, we find that for a vast
majority of cases, the optimal decision is to impose a signature bond
rather than a cash bond unless the cost of a negative outcome is much
higher than that of a decision that may result in unnecessary
incarceration. This suggests that the PSA's recommendations may be
harsher than necessary.

\newpage

\pdfbookmark[1]{References}{References}
\spacingset{1.5}
\bibliography{psa-ref,my,imai}

\newpage

\appendix
\setcounter{page}{1}

\setcounter{equation}{0}
\setcounter{figure}{0}
\setcounter{theorem}{0}
\setcounter{lemma}{0}
\setcounter{section}{0}
\renewcommand {\theequation} {S\arabic{equation}}
\renewcommand {\thefigure} {S\arabic{figure}}
\renewcommand {\thetheorem} {S\arabic{theorem}}
\renewcommand {\theproposition} {S\arabic{proposition}}
\renewcommand {\thelemma} {S\arabic{lemma}}
\renewcommand {\thesection} {S\arabic{section}}

\spacingset{1}

\begin{center}
  \LARGE {\bf Supplementary Appendix for\\ Imai, K., Z. Jiang,
    D. J. Greiner, R. Halen, and S. Shin. ``Experimental Evaluation of
    Algorithm-Assisted Human Decision-Making: Application to Pretrial
    Public Safety Assessment.''}
\end{center}

\section{Distribution of Judge's Decisions Given the PSA for Subgroups}
\label{app:dist}

\subsection{Female Arrestees}

\begin{figure}[h!]
 \centering \spacingset{1}
 \begin{subfigure}[t]{\textwidth}
 \subcaption{Treatment Group}
 \includegraphics[width = \textwidth]{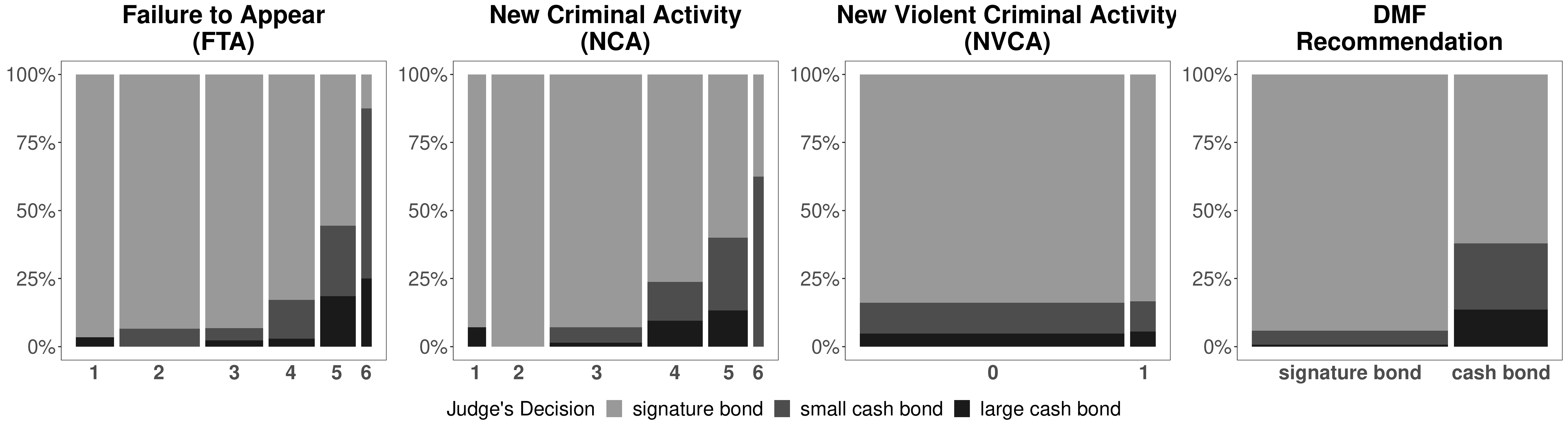}
 \end{subfigure} \\
 \vspace{.25in}
 \begin{subfigure}[t]{\textwidth}
 \subcaption{Control Group}
 \includegraphics[width = \textwidth, trim = 0 0 0 57, clip]{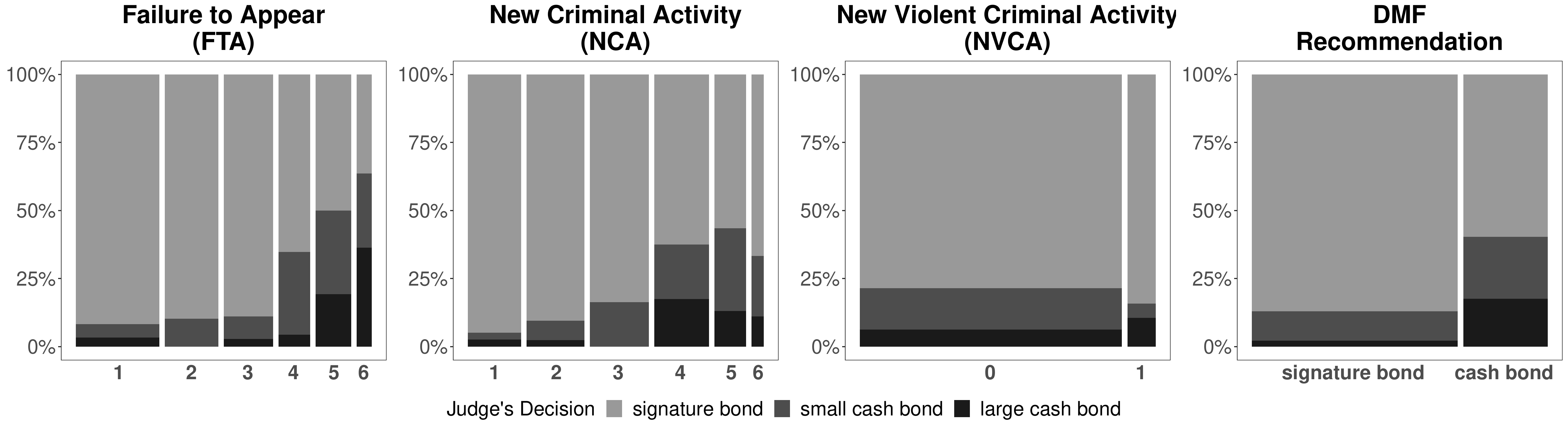}
 \end{subfigure}
 \caption{The Distribution of Judge's Decisions Given the Pretrial
 Public Safety Assessment (PSA) among the Cases in the Treatment
 (Top Panel) and Control (Bottom Panel) Groups Among Female Arrestees.}
\end{figure}

\clearpage
\subsection{Non-white Male Arrestees}

\begin{figure}[h!]
 \centering \spacingset{1}
 \begin{subfigure}[t]{\textwidth}
 \subcaption{Treatment Group}
 \includegraphics[width = \textwidth]{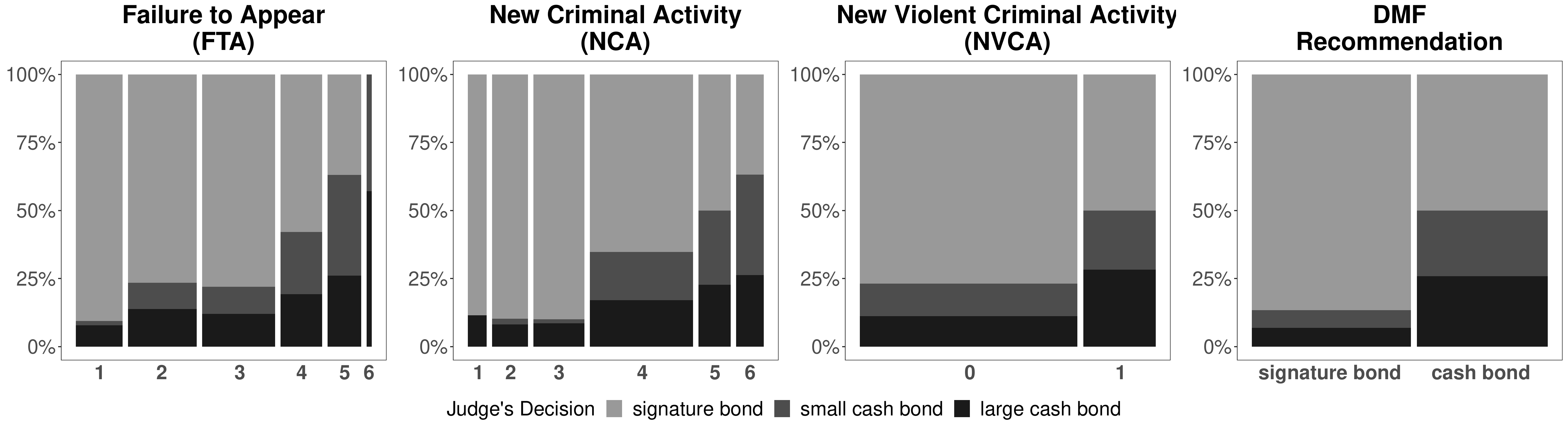}
 \end{subfigure} \\
 \vspace{.25in}
 \begin{subfigure}[t]{\textwidth}
 \subcaption{Control Group}
 \includegraphics[width = \textwidth, trim = 0 0 0 57, clip]{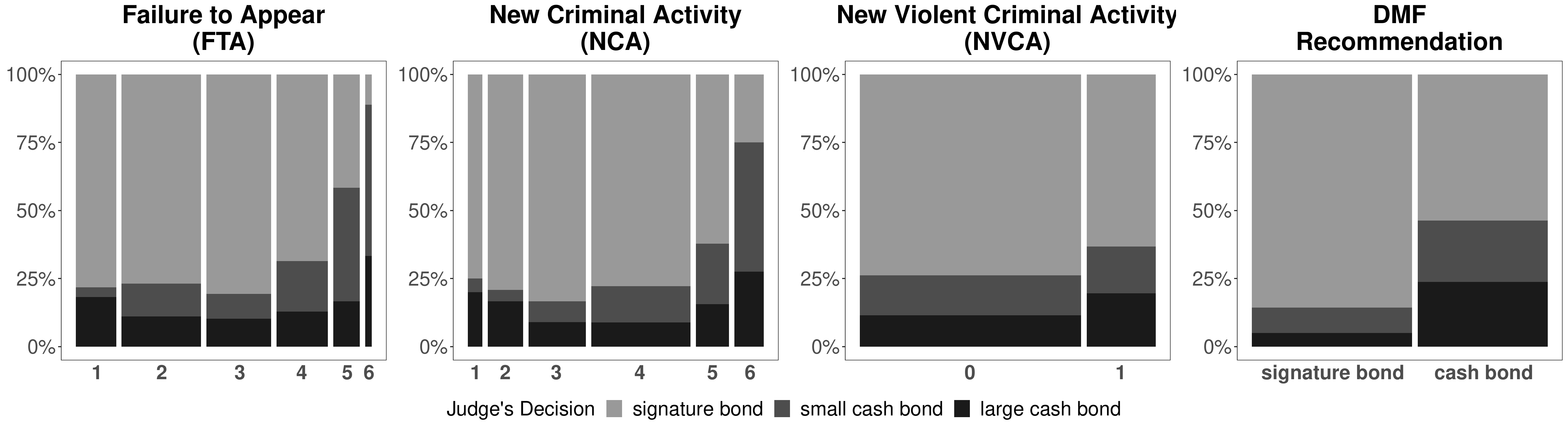}
 \end{subfigure}
 \caption{The Distribution of Judge's Decisions given the Pretrial
 Public Safety Assessment (PSA) among the Cases in the Treatment
 (Top Panel) and Control (Bottom Panel)
 Groups Among Non-white  Male Arrestees.}
\end{figure}

\clearpage
\subsection{White Male Arrestees}

\begin{figure}[h!]
 \centering \spacingset{1}
 \begin{subfigure}[t]{\textwidth}
 \subcaption{Treatment Group}
 \includegraphics[width = \textwidth]{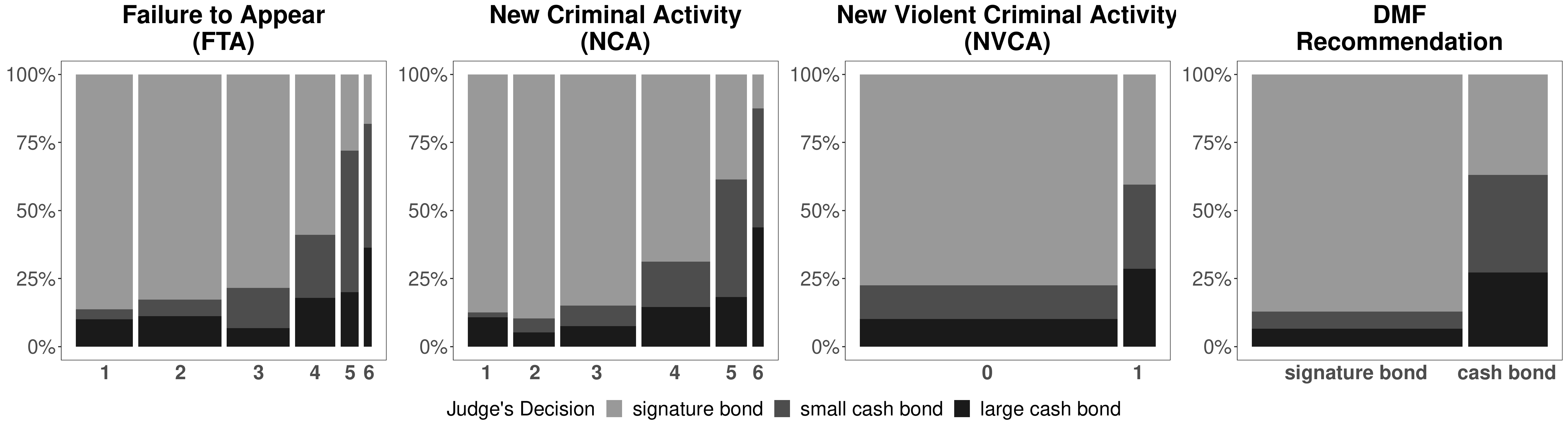}
 \end{subfigure} \\
 \vspace{.25in}
 \begin{subfigure}[t]{\textwidth}
 \subcaption{Control Group}
 \includegraphics[width = \textwidth, trim = 0 0 0 57, clip]{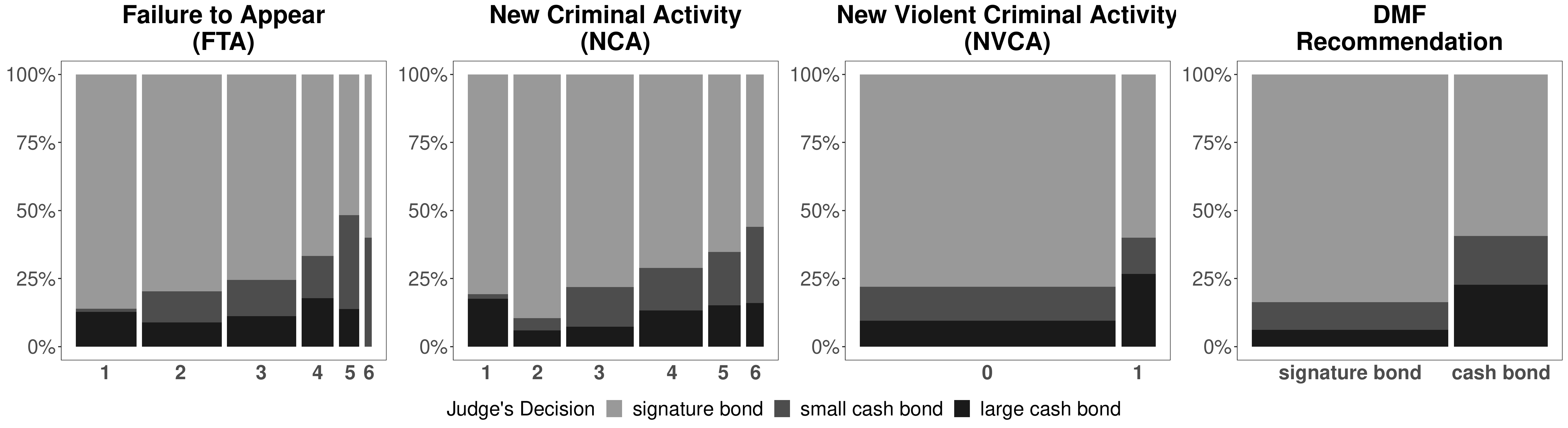}
 \end{subfigure}
 \caption{The Distribution of Judge's Decisions Given the Pretrial
 Public Safety Assessment (PSA) among the Cases in the Treatment
 (Top Panel) and Control (Bottom Panel)
 Groups Among White Male Arrestees.}
\end{figure}

\newpage
\section{Subgroup Analysis for Age Groups}
\label{app:age}

In this appendix, we conduct the subgroup analysis for different age
groups.

\subsection{Age Distribution, Descriptive Statistics, and
 Average Causal Effects}
\label{subsec:age_summary}

\begin{figure}[h!]
 \centering \spacingset{1}
 \begin{subfigure}[t]{0.45\textwidth}
 \subcaption{Treatment Group}
 \includegraphics[width = \textwidth]{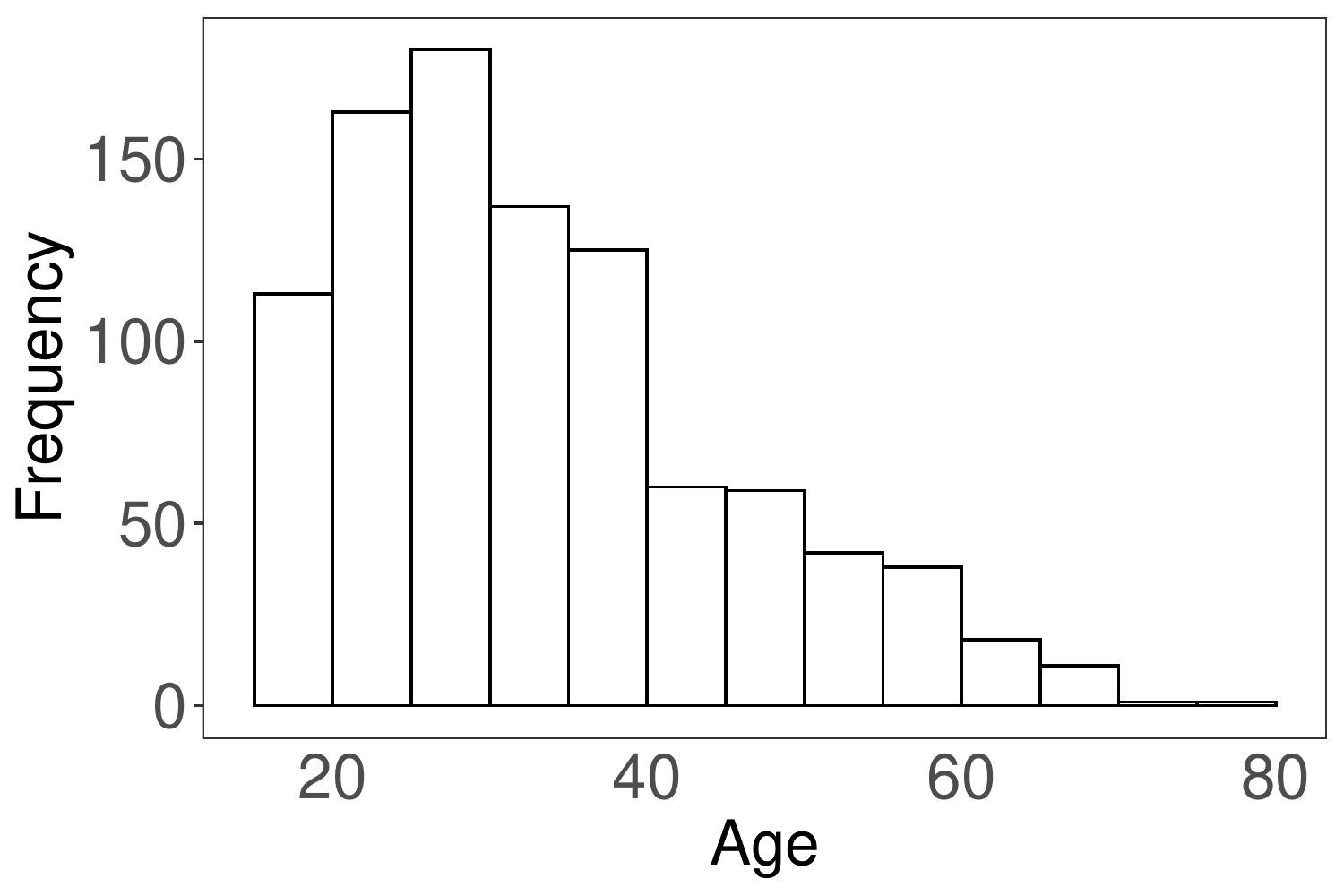}
 \end{subfigure} 
 \begin{subfigure}[t]{0.45\textwidth}
 \subcaption{Control Group}
 \includegraphics[width = \textwidth]{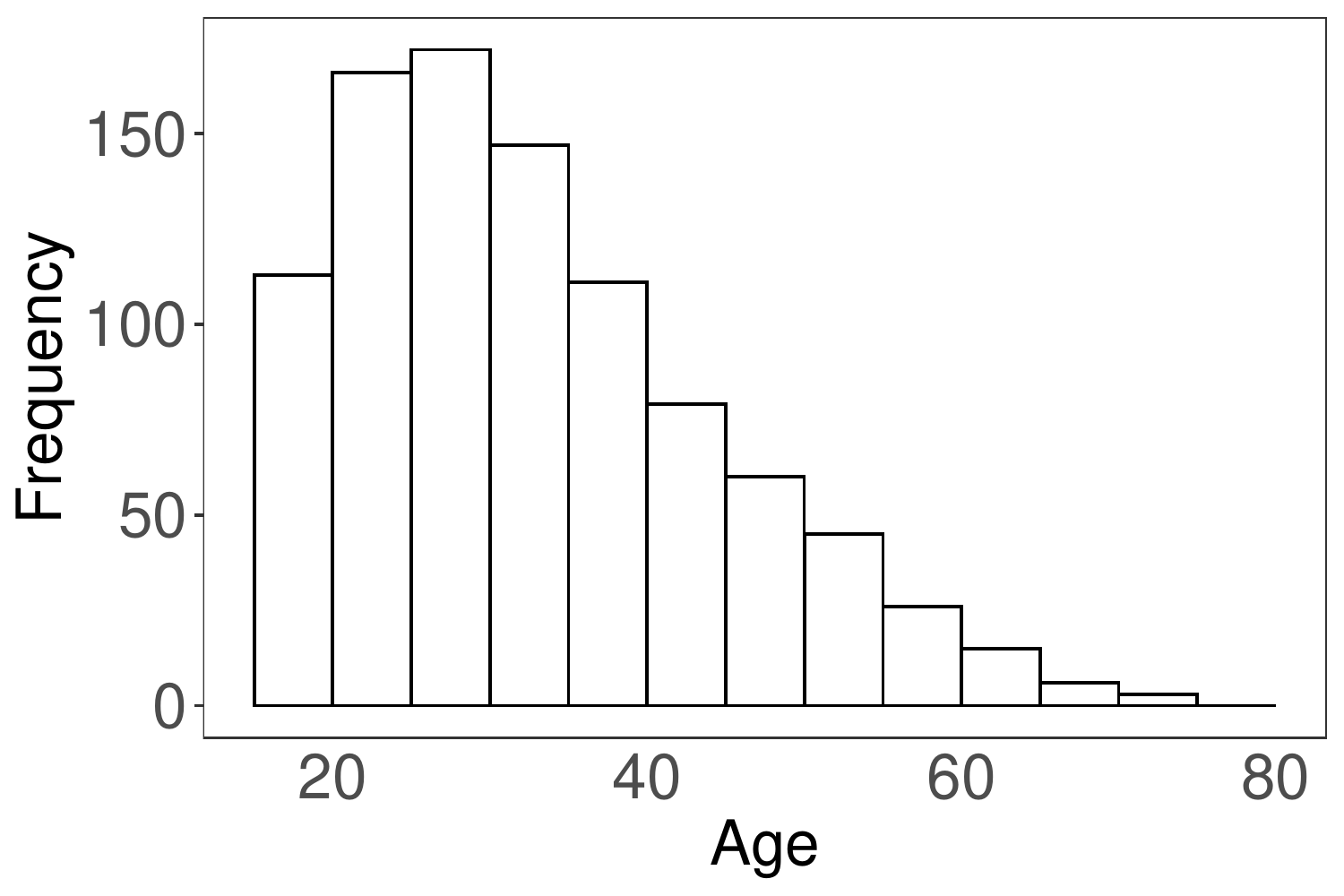}
 \end{subfigure}
 \caption{The Distribution of Age in the Treatment (Left Panel) and
 Control (Right Panel) Groups Among
 Arrestees.} \label{fig:age_hist}
\end{figure}

\begin{table}[h!] \centering \spacingset{1} \setlength{\tabcolsep}{3.5pt}
 \begin{tabular}{@{\extracolsep{5pt}} rrrrrrrr}
 \\[-1.8ex]\hline
 \hline \\[-1.8ex]
 & \multicolumn{3}{c}{\textit{no} PSA} & \multicolumn{3}{c}{PSA}
 \\\cmidrule(lr){2-4}\cmidrule(lr){5-7}
 & \multicolumn{1}{c}{Signature} & \multicolumn{2}{c}{Cash bond} & \multicolumn{1}{c}{Signature} & \multicolumn{2}{c}{Cash bond} \\
 & \multicolumn{1}{c}{bond} & $\leq$\$1000 & $>$\$1000 & \multicolumn{1}{c}{bond} & $\leq$\$1000 & $>$\$1000 & Total (\%) \\
 \hline \\[-1.8ex]
 22 or below& 135 & 24 & 22 & 136 & 24 & 16 & 357 \\
 & (7.1) & (1.3) & (1.2) & (7.2) & (1.3) & (0.8) & (18.9) \\
 23 -- 28& 158 & 25 & 23 & 148 & 29 & 28 & 411 \\
 & (8.4) & (1.3) & (1.2) & (7.8) & (1.5) & (1.5) & (21.7) \\
 29 -- 35 & 157 & 40 & 14 & 151 & 33 & 28 & 423 \\
 & (8.3) & (2.1) & (0.7) & (8.0) & (1.7) & (1.5) & (22.3) \\
 36 -- 45 & 142 & 22 & 26 & 133 & 30 & 22 & 375 \\
 & (7.5) & (1.2) & (1.4) & (7.0) & (1.6) & (1.2) & (19.9) \\
 46 or above & 113 & 21 & 21 & 137 & 14 & 19 & 325 \\
 & (6.0) & (1.1) & (1.1) & (7.2) & (0.7) & (1.0) & (17.1) \\ \hline \\[-1.8ex]
 \end{tabular}
 \caption{The Joint Distribution of Treatment Assignment, Decisions,
 and Age. The table shows the number of cases in each category
 with the corresponding percentage in parentheses.} \label{tab:age}
\end{table}

Figure~\ref{fig:age_hist} presents the distribution of age for the
treatment and control groups. As expected, the two distributions are
similar. We observe that the age distribution is right skewed with
many more young arrestees. Table~\ref{tab:age} presents the
descriptive statistics for different age groups examined here. We
divide the arrestees into five subgroups with different ranges of age
(aged 22 or below, between 23 to 28, between 29 to 35, between 36 to 45, 46 or above).
Within each age group, the signature bond appears to be the dominant
decision.

\clearpage

\begin{figure}[h!]
	\spacingset{1}\centering  
	\includegraphics[width=0.485\linewidth]{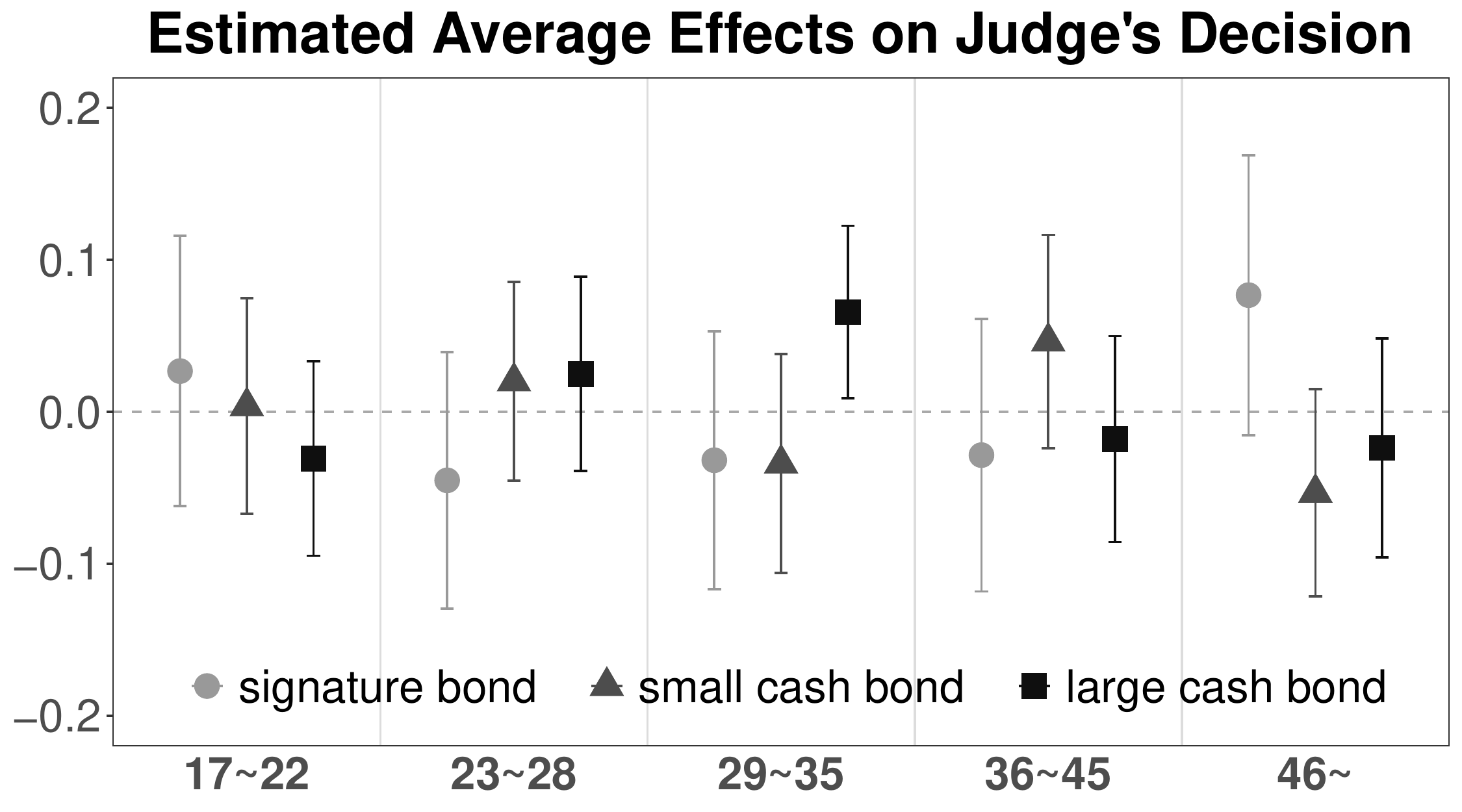}
	\includegraphics[width=0.485\linewidth]{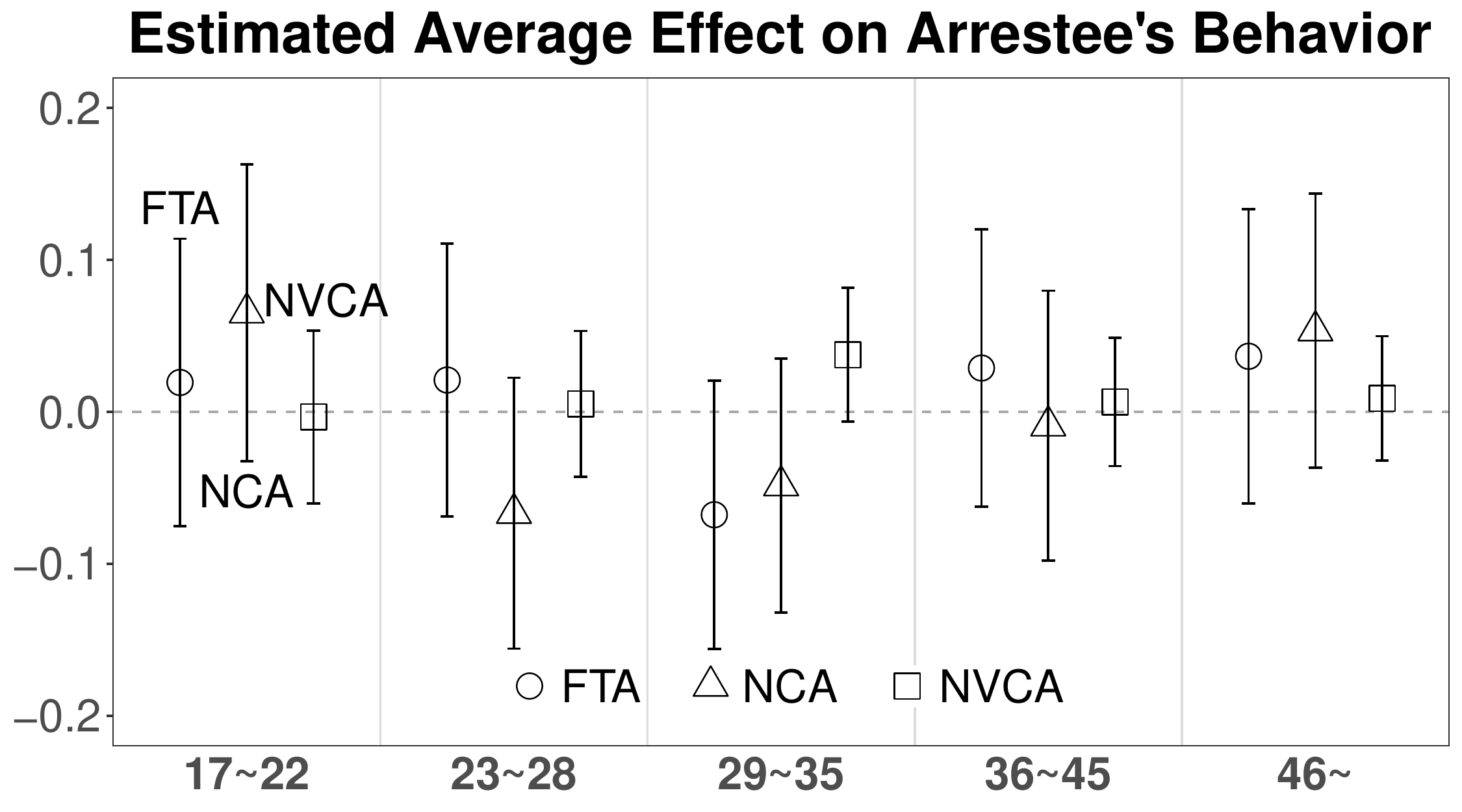}
	\caption{Estimated Average Causal Effects of PSA Provision on
          Judge's Decisions and Outcome Variables for First Arrest
          Cases (FTA, NCA, and NVCA). The results are based on the
          difference-in-means estimator. The vertical bars represent
          the 95\% confidence intervals. In the left figure, we report
          the estimated average causal effect of PSA provision on the
          decision to charge a signature bond (circles), a small cash
          bail (\$1,000 dollars or less; triangles), and a large cash
          bail (greater than \$1,000; squares). In the right figure,
          we report the estimated average causal effect of PSA
          provision on the three different outcome variables: FTA
          (open circles), NCA (open triangles), and NVCA (open
          squares).} \label{fig:ITTage}
\end{figure}

Figure~\ref{fig:ITTage} presents the estimated Intention-to-Treat
(ITT) effects of PSA provision on judge's decisions (left panel) and
arrestee's behaviors (right panel). We find that PSA provision has
little effect on the judge's decisions with the exception of the 29 --
35 years old group and the oldest group. For the 29 -- 35 years old
group, the PSA appears to lead to a harsher decision while for the 46
or older group the effect is the opposite. As for the effects on
arrestee's behavior, our analysis suggests that PSA provision may
increase NVCA among the 29 -- 35 years old group though the estimate
is only marginally significant.

\subsection{Principal Stratum Proportion and Average Principal Strata
 Effects}
\label{subsec:age_effects}

\begin{figure}[h!]
	\centering \spacingset{1}
	\includegraphics[width=0.32\textwidth]{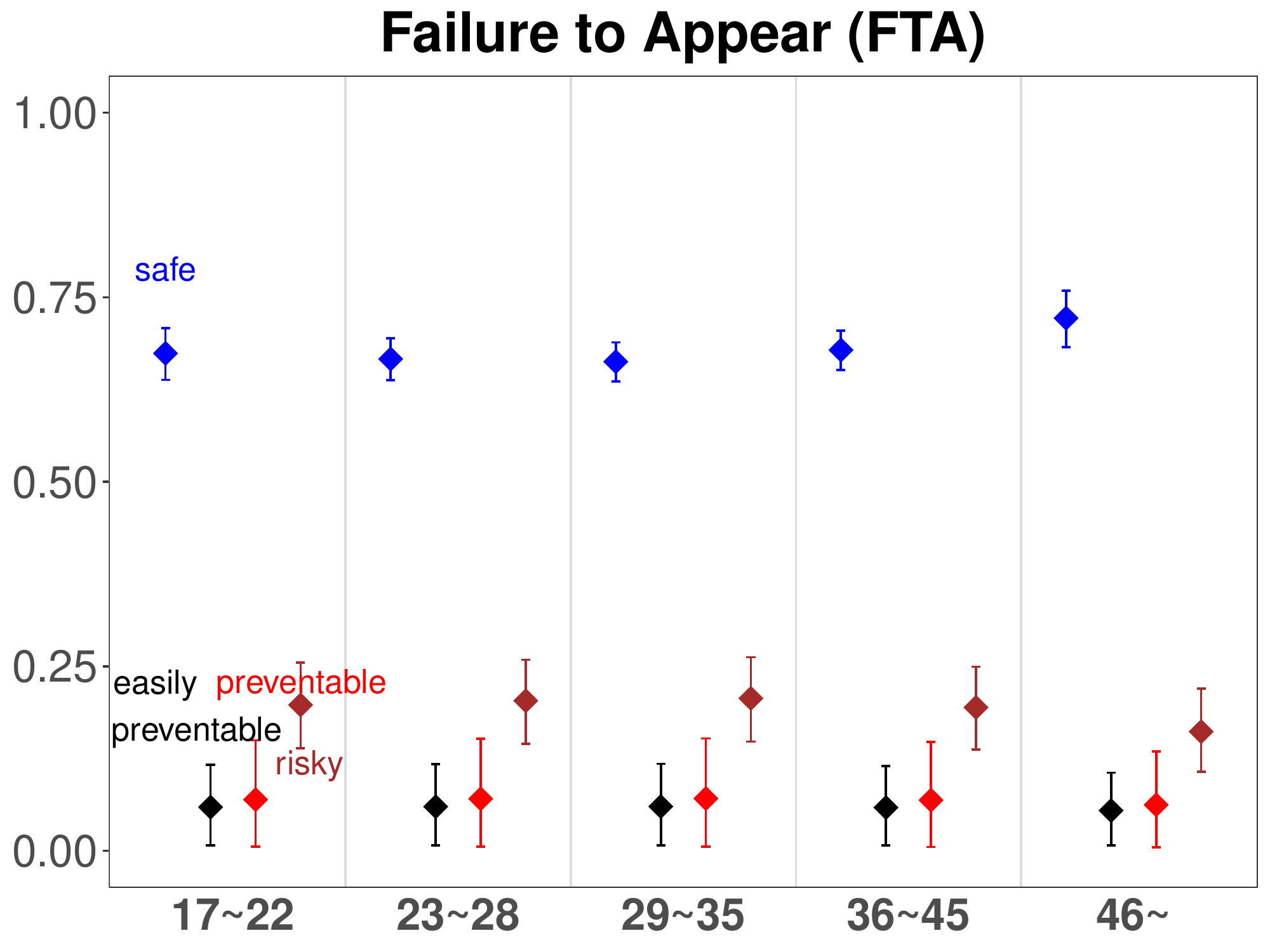}
	\includegraphics[width=0.32\textwidth]{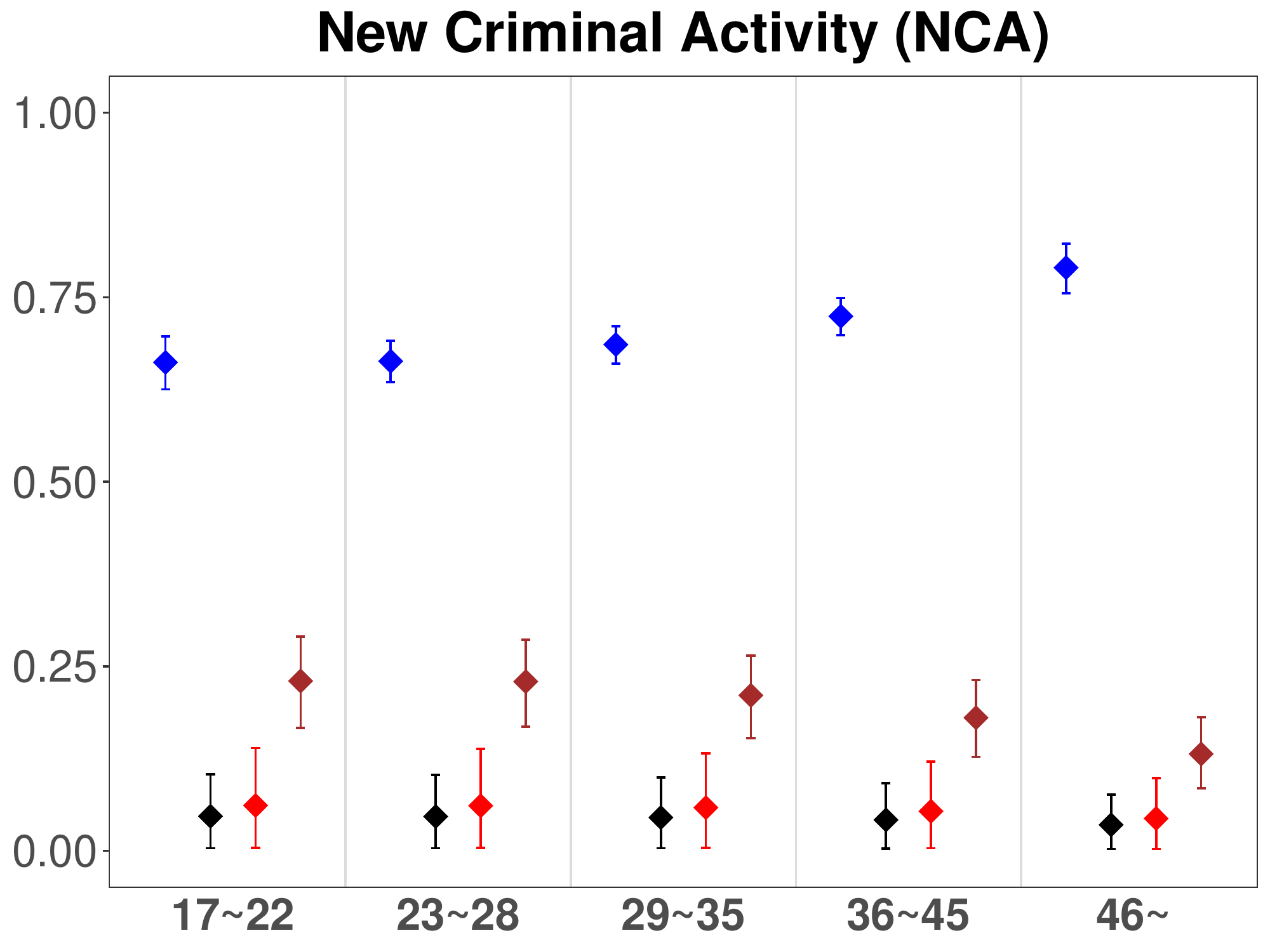}
	\includegraphics[width=0.32\textwidth]{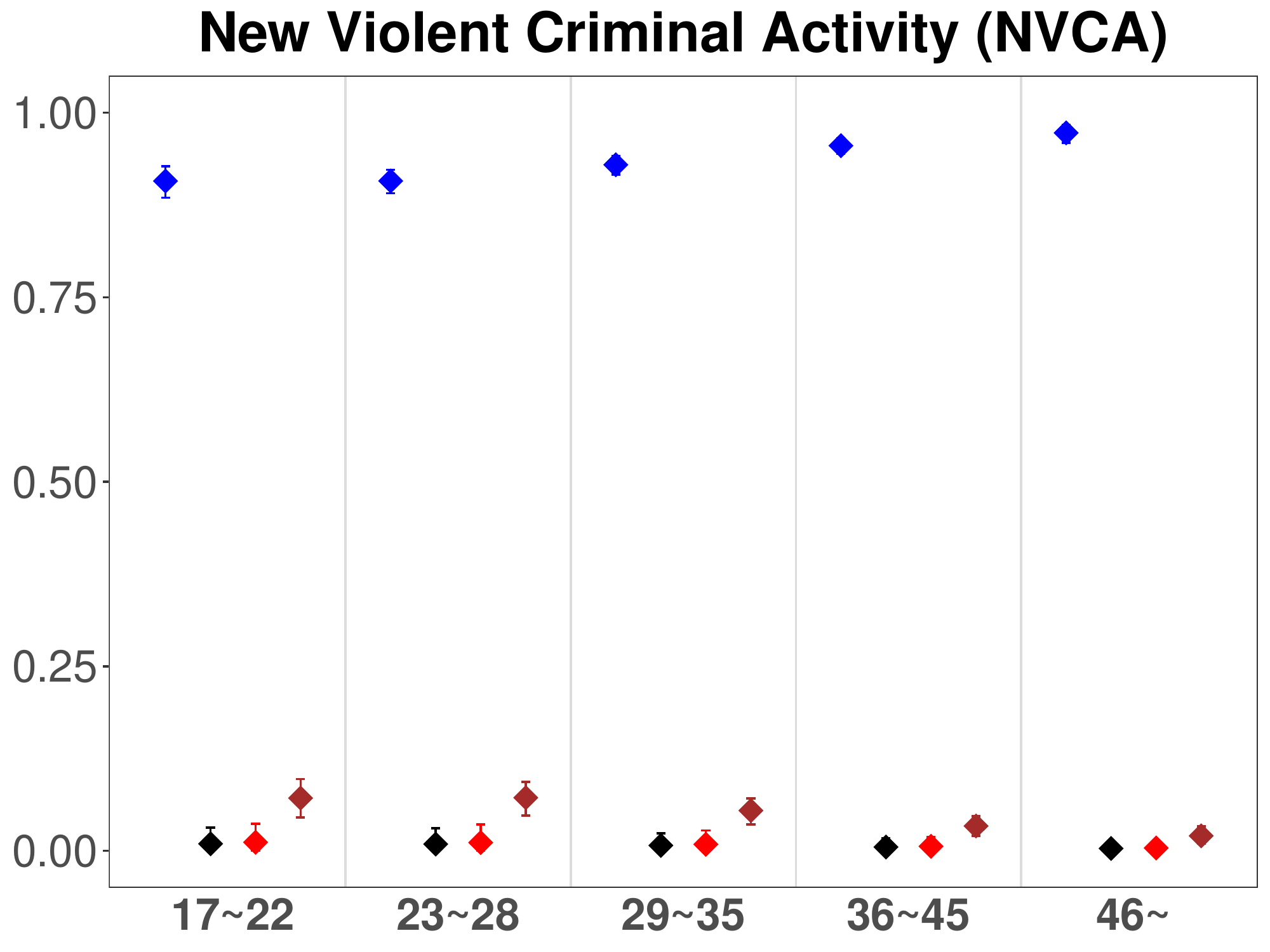}
	\caption{Estimated Proportion of Each Principal Stratum.  Each
          plot represents the result using one of the three outcome
          variables (FTA, NCA, and NVCA), where the blue, black, red,
          and brown diamonds represent the estimates for safe, easily
          preventable, preventable, risky cases, respectively. The
          solid vertical lines represent the $95\%$ Bayesian credible
          intervals.} \label{fig:er_age}
\end{figure}

\begin{figure}[p]
	\centering \spacingset{1}
	\includegraphics[width=\textwidth]{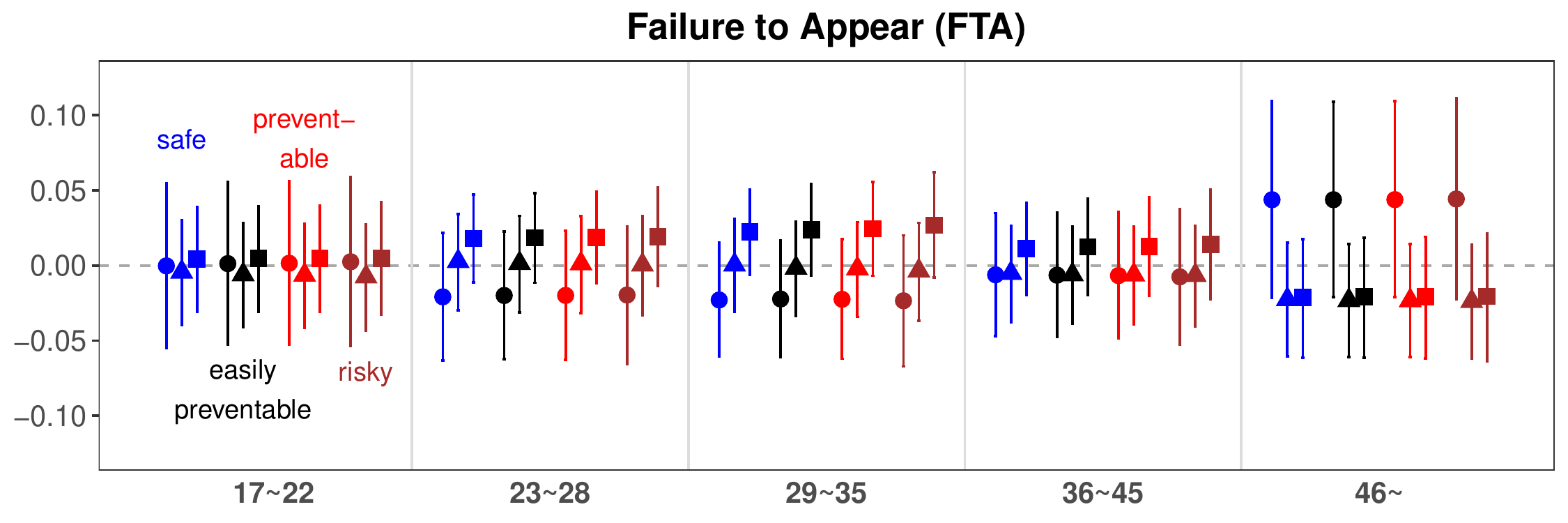}
	\includegraphics[width=\textwidth]{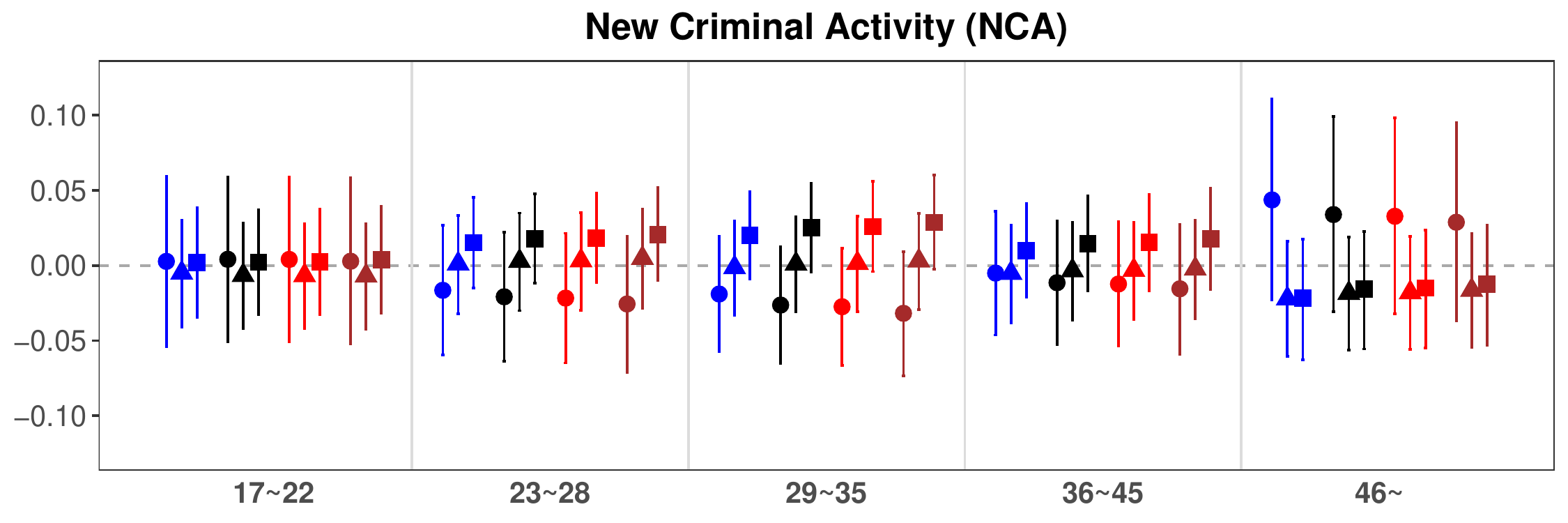}
	\includegraphics[width=\textwidth]{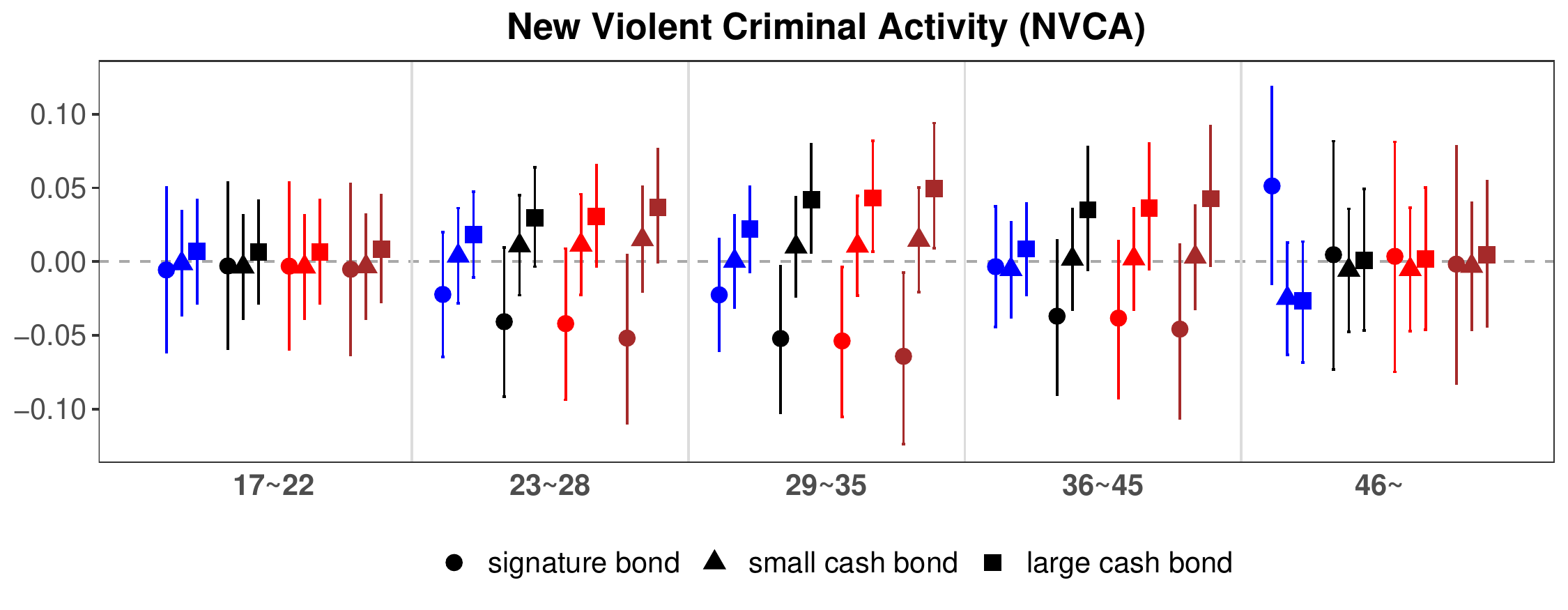}
	\caption{Estimated Average Principal Causal Effects (\ACE) of PSA
		Provision on the Judge's Decision. Each panel presents the age group-specific results for a different outcome
		variable. Each column within a panel shows the estimated \ACE{} of
		PSA provision for  safe (blue), easily preventable (black),
			preventable (red), and risky (brown) cases. For each of these principal strata, we
		report the estimated \ACE{} on the judge's decision to impose a
		signature bond (circles), a small cash bail amount of 1,000 dollars
		or less (triangles), and a large cash bail amount of greater than
		1,000 (squares). The vertical line for each estimate represents the
		Bayesian $95\%$ credible interval.}
	\label{fig:ace_age}
\end{figure}

Figure~\ref{fig:er_age} presents the estimated proportion of each
principal stratum for different age groups. We observe that the
principal stratum size is similar across age groups with the safe
cases being the most dominant. The proportion of safe cases appears to
be greater for older age groups though the rate of increase is
modest. The interpretation of Figure~\ref{fig:ace_age} is given in
the last paragraph of Section~\ref{subsec:estimatedeffects}.

\newpage
\section{Testing the Potential Existence of Spillover Effects}
\label{app:sutva}

\subsection{Conditional Randomization Test}
\label{subsec:CRT}

We examine the possible existence of spillover effects. In particular,
we use a conditional randomization test to examine whether or not PSA
provision of prior cases affects the judge's decision in later cases
\citep[e.g.,][]{aron:12,athe:eckl:imbe:18,candes2018panning}. The
basic idea is to test whether the decision, $D_i$, is conditionally
independent of the treatment assignment of the other cases whose court
hearing date is prior to that of case $i$, given its own treatment
assignment status $Z_i$. The judge made decision for $1,891$ cases on
$274$ different dates. Unfortunately, we do not have information about
the ordering of decisions within each hearing date. Let
$O_i \in \{1,2,\ldots,274\}$ denote the order of the hearing date of
case $i$. Let
$\widetilde{Z}_{i} = |\{i^\prime \in\{1,2,\ldots,n\}: O_{i^\prime} =
O_i - 1\}|$ denote the proportion of treated cases whose hearing date order is immediately
before that of case $i$. Then, the null hypothesis is given by
$H_0: \widetilde{Z}_{i} \perp \!\!\! \perp D_i
\mid Z_i$. We conduct a conditional randomization test as follows:
\begin{enumerate}
	\item Create a new treatment assignment $Z_i^\prime$ as follows:
	\begin{enumerate}
		\item For each $i$, if $O_i$ is even then $Z_i^\prime = Z_i$
		\item For each $i$, if $O_i$ is odd then randomly sample $Z_i^\prime \sim \text{Bernoulli}(1/2)$
	\end{enumerate}
	Then compute $\widetilde{Z}_{i}^\prime$ based on $Z_i^\prime$, i.e.,
    $\widetilde{Z}_{i}^\prime = |\{i^\prime \in \{1,2,\ldots,n\}: O_{i^\prime} = O_i -
    1\}|$.
   \item Regress $D_i$ on
    $(1, Z_i, \widetilde{Z}_{i}^\prime)$ only using the subset of
    observations whose $O_i$ is even. Let our test statistic $T$ be the squared term of estimate of coefficient of $\widetilde{Z}_{i}^\prime$.
   \item Repeat the above $S$ times and compute (one-sided)
    p-value:
    $\frac{1}{S}\sum_{s=1}^S \bone\{T^{(s)} \geq T_{\text{obs}}\}$
    where $T^{(s)}$ is the test statistic for the $s$th iteration
    and $T_{\text{obs}}$ is the observed test statistic.
\end{enumerate}

\begin{figure}[h!]
	\centering \spacingset{1}
	\includegraphics[width = 0.5\textwidth]{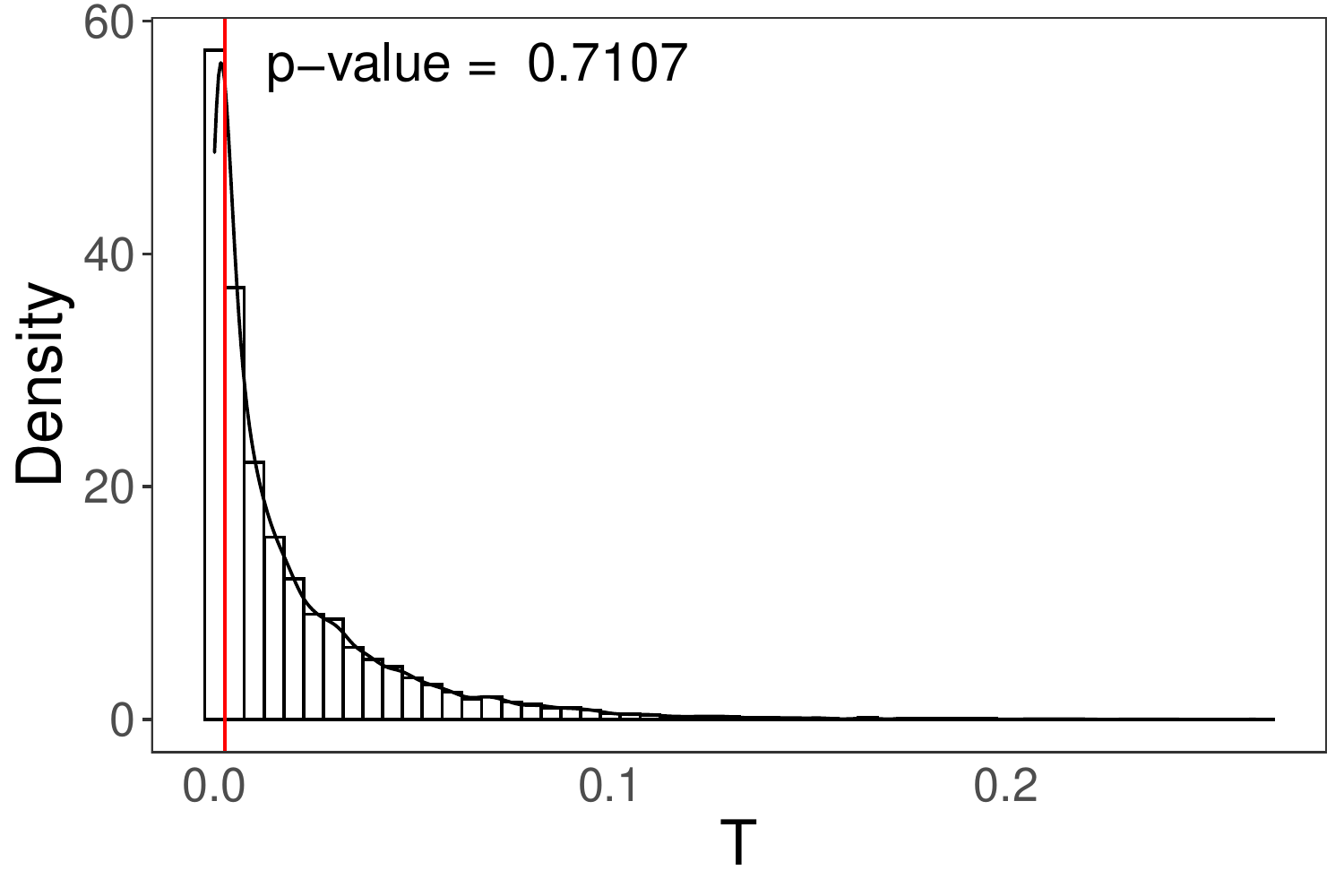}
	\caption{The Distributions of Test Statistics. The red vertical lines indicate the observed test statistics.} \label{fig:crt}
\end{figure}

Figures~\ref{fig:crt} presents the resulting distribution of our test
statistics. The $p$-value is $0.71$ for the test statistics $T$, and
thus we fail to reject the null hypothesis. That is, we find no
statistically significant evidence that the judge's decision is
influenced by PSA provision of the prior cases. This is consistent
with the assumption of no inference among the cases, which is made
throughout our analysis.

\subsection{Power: A Simulation Analysis}

We examine the power of the statistical test used above via a
simulation study.  Our simulation procedure is as follows:
\begin{enumerate}
\item Regress $D_i$ on $(1, Z_i, \widetilde{Z}_{i})$ using the ordinal
  logistic regression model based on the observed data. Let $\omega$
  denote the coefficient for $\widetilde{Z}_{i}$.
\item Choose a value of $\omega$, and set the other model parameters
  to their estimated values.  Using this mode, generate $D_i$ with the
  same sample size and observed treatment variable.
\item Conduct the conditional randomization test as described
  in Section~\ref{subsec:CRT}. Repeat this for $1,000$ times
  and calculate the proportion of rejecting the null hypothesis at the
  $0.05$ level.
\item Repeat the above procedure for each value of $\omega \in \{-1.5, -1, -0.5, 0, 0.5, 1, 1.5\}$.
\end{enumerate}

\begin{figure}[h!]
	\centering \spacingset{1}
	\includegraphics[width = 0.8\textwidth]{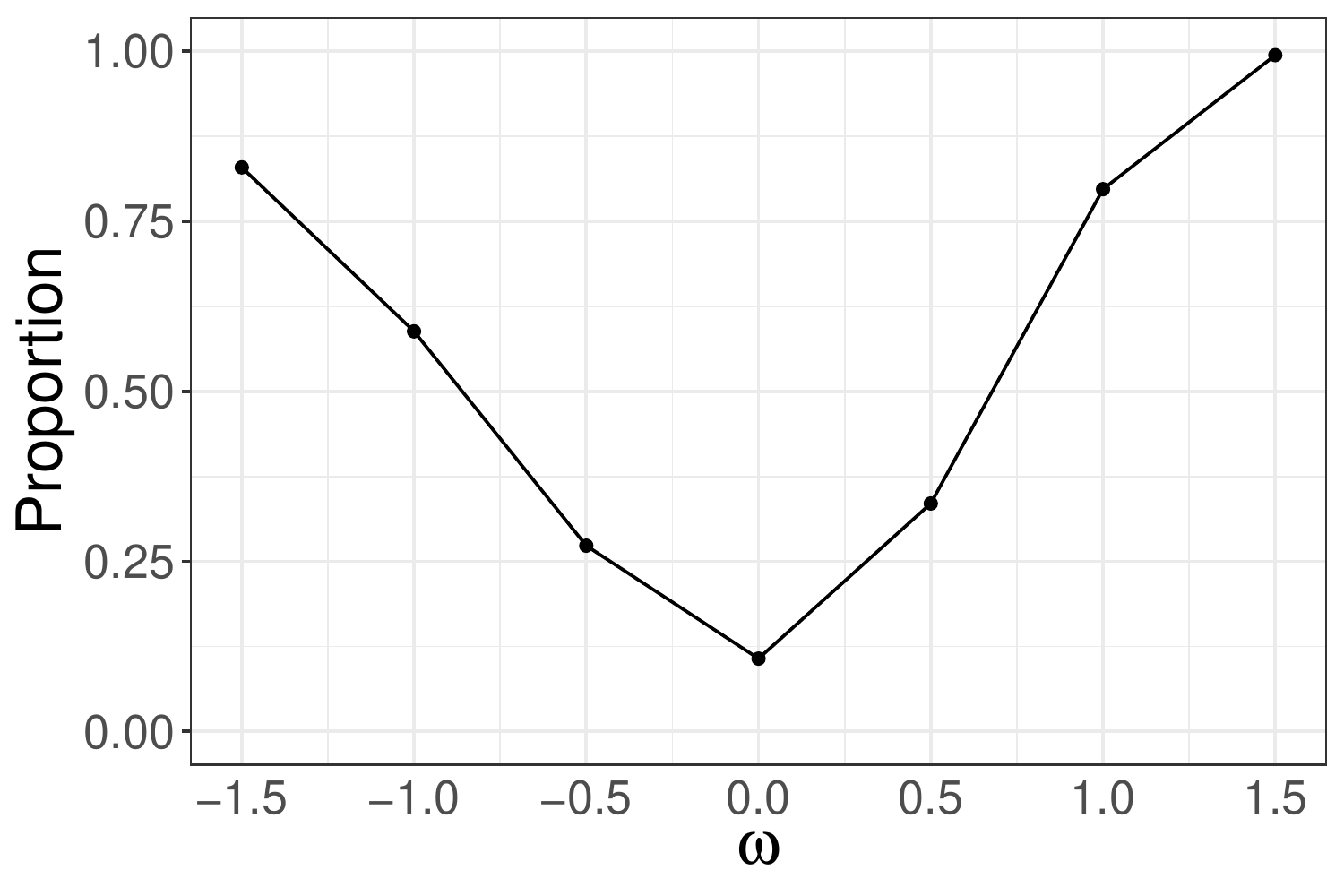}
	\caption{The Proportion of Rejecting the Null Hypothesis at the $0.05$ Level.} \label{fig:crtpower}
\end{figure}

Figures~\ref{fig:crtpower} presents the results of our simulation
study for calculating the power of the test. Here, if the proportion
of treated cases whose hearing date order is immediately before is
$1$, the odds of judges making harsher decision is $\exp \omega$ times
that of the arrestees whose proportion of treated cases whose hearing
date order is immediately before is $0$.  According to the simulation,
the power of the test reaches about 0.8 when $\omega = 1$ or
$\exp \omega = 2.72$.  Thus, it is possible that with the given sample
size, only the relatively large effect can be detected.  This suggests
that we must interpret the result of this test presented in
Section~\ref{subsec:CRT} with caution. 

\clearpage
\section{Proofs of the Theorems}

\subsection{Lemmas}
To prove the theorems, we need some lemmas.
\begin{lemma}
\label{lem::1} 
Consider two random variables $X$ and $Y$. Suppose that they have finite moments and the support of $Y$ contains that of $X$. Let $f_1(x)$ and $f_2(y)$ be their density functions. Then, any function $g(\cdot)$,
\begin{eqnarray*}
\E \{g(X)\} = \E\left\{ \frac{f_1(Y)}{f_2(Y)} g(Y)\right\}.
\end{eqnarray*}
\end{lemma}
Proof is straightforward and hence omitted.

\begin{lemma}
\label{lem::2} 
For a binary decision, Assumption~\ref{asm::indep} implies $\{Y_i(1),Y_i(0)\} \ind D_i \mid \bX_i, Z_i=z$
under Assumption~\ref{asm::mon}. For an ordinal decision, Assumption~\ref{asm::indep} implies $R_i \ind D_i \mid \bX_i, Z_i=z$
under Assumption~\ref{asm::mon-discrete}. 
\end{lemma}
\noindent {\it Proof of Lemma~\ref{lem::2}}. For a binary decision, we have 
\begin{eqnarray*}
\Pr\{Y_i(1)=1, Y_i(0)=1\mid D_i, \bX_i, Z_i=z\} &=& \Pr\{Y_i(1)=1\mid D_i, \bX_i, Z_i=z\}\\
 &=&\Pr\{ Y_i(1)=1\mid \bX_i, Z_i=z\} \\
 &=&\Pr\{ Y_i(1)=1, Y_i(0)=1\mid \bX_i, Z_i=z\},
\end{eqnarray*}
where the first and third equality follow from Assumption~\ref{asm::mon} and the second equality follows from Assumption~\ref{asm::indep}.
Similarly, we have 
\begin{eqnarray*}
\Pr\{Y_i(1)=0, Y_i(0)=0\mid D_i, \bX_i, Z_i=z\} &=& \Pr\{Y_i(0)=0\mid D_i, \bX_i, Z_i=z\}\\
 &=&\Pr\{ Y_i(0)=0\mid \bX_i, Z_i=z\} \\
 &=&\Pr\{ Y_i(1)=0, Y_i(0)=0\mid \bX_i, Z_i=z\},
\end{eqnarray*}
where the first and third equality follow from Assumption~\ref{asm::mon} and the second equality follows from Assumption~\ref{asm::indep}.
As a result, $\{Y_i(1),Y_i(0)\} \ind D_i \mid \bX_i, Z_i=z$ because $\{Y_i(1),Y_i(0)\}$ takes only three values.

For a discrete decision $D_i $ taking values in $\{0,\ldots,k\}$, we have 
\begin{eqnarray*}
\Pr(R_i=r \mid D_i, \bX_i,Z_i=z) &=& \Pr(R_i \geq r \mid D_i, \bX_i,Z_i=z) - \Pr(R_i \geq r+1 \mid D_i, \bX_i,Z_i=z) \\
&=& \Pr( Y_i(r-1)=1 \mid D_i, \bX_i,Z_i=z) - \Pr(Y_i(r)=1 \mid D_i, \bX_i,Z_i=z) \\
&=& \Pr( Y_i(r-1)=1 \mid \bX_i,Z_i=z) - \Pr(Y_i(r)=1 \mid \bX_i,Z_i=z) \\
&=& \Pr(R_i \geq r \mid \bX_i,Z_i=z) - \Pr(R_i \geq r+1 \mid \bX_i,Z_i=z)\\
 &=& \Pr(R_i = r \mid D_i, \bX_i,Z_i=z)
\end{eqnarray*}
for $r=1,\ldots,k$,
where the second and the fourth equality follow from the definition of $R_i$ and the third equality follows from Assumption~\ref{asm::indep}. Similarly, we have
\begin{eqnarray*}
\Pr(R_i=0 \mid D_i, \bX_i,Z_i=z) &=& \Pr( Y_i(0)=0 \mid D_i, \bX_i,Z_i=z) \\
&=& \Pr( Y_i(0)=0 \mid D_i, Z_i=z) \\
&=&\Pr(R_i=0 \mid D_i,Z_i=z), \\
\Pr(R_i=k+1 \mid D_i, \bX_i,Z_i=z) &=& \Pr( Y_i(k)=1 \mid D_i, \bX_i,Z_i=z) \\
&=& \Pr( Y_i(k)=1 \mid D_i, Z_i=z) \\
&=&\Pr(R_i=k+1 \mid D_i,Z_i=z).
\end{eqnarray*}
As a result, $R_i \ind D_i \mid \bX_i, Z_i=z$. \QEDB

\subsection{Proof of Theorem~\ref{thm::identification-mon}}
\label{app:identification-mon}

First, Assumption~\ref{asm::mon} implies,
\begin{eqnarray*}
\Pr\{Y_i(0)=0,Y_i(1)=0\} &=& \Pr\{Y_i(0)=0\}, \quad \Pr\{Y_i(0)=1,Y_i(1)=1\}\ =\ \Pr\{Y_i(1)=1\},\\
\Pr\{Y_i(0)=1,Y_i(1)=0\} &=&1- \Pr\{Y_i(0)=0\}- \Pr\{Y_i(1)=1\}.
\end{eqnarray*}
Second, we have
\begin{eqnarray*}
&&\Pr\{D_i(z)=1,Y_i(0)=0,Y_i(1)=0\}\\
& =& \Pr\{Y_i(0)=0,Y_i(1)=0\}- \Pr\{D_i(z)=0,Y_i(0)=0,Y_i(1)=0\}\\
& =& \Pr\{Y_i(0)=0\}- \Pr\{D_i(z)=0,Y_i(0)=0\} \\
&=& \Pr\{Y_i(0)=0\}- \Pr\{D_i(z)=0,Y_i(D_i(z))=0 \mid Z_i=z\} \\
&=& \Pr\{Y_i(0)=0\}- \Pr(D_i=0,Y_i=0 \mid Z_i=z),
\end{eqnarray*}
where the second equality follows from Assumption~\ref{asm::mon} and the third equality follows from Assumption~\ref{asm::rand}.
Similarly, we can obtain 
\begin{eqnarray*}
\Pr\{D_i(z)=1,Y_i(0)=1,Y_i(1)=1\}& =& \Pr\{D_i(z)=1,Y_i(1)=1\} \\
&=&\Pr\{D_i(z)=1,Y_i(D_i(z))=1 \mid Z_i=z\} \\
&=& \Pr(D_i=1,Y_i=1 \mid Z_i=z).
\end{eqnarray*}
Therefore,
\begin{eqnarray*}
&&\Pr\{D_i(z)=1,Y_i(0)=1,Y_i(1)=0\}\\
& =& \pr\{D_i(z)=1\}- \Pr\{D_i(z)=1,Y_i(0)=0,Y_i(1)=0\} -\Pr\{D_i(z)=1,Y_i(0)=1,Y_i(1)=1\} \\
& =& \pr\{D_i=1\mid Z_i=z\}- \Pr\{Y_i(0)=0\}+ \Pr(D_i=0,Y_i=0 \mid Z_i=z)- \Pr(D_i=1,Y_i=1 \mid Z_i=z) \\
&=& \Pr(Y_i=0 \mid Z_i=z)- \Pr\{Y_i(0)=0\}.
\end{eqnarray*}
Finally, we have,
\begin{eqnarray*}
\ACEP&=& \frac{\Pr\{D_i(1)=1,Y_i(0)=1,Y_i(1)=0\}-\Pr\{D_i(0)=1,Y_i(0)=1,Y_i(1)=0\}}{\Pr\{Y_i(0)=1,Y_i(1)=0\} }\\
&=& \frac{\Pr(Y_i=1 \mid Z_i=0)-\Pr(Y_i=1 \mid Z_i=1)}{\Pr\{Y_i(0)=1\}- \Pr\{Y_i(1)=1\}},
\end{eqnarray*}
\begin{eqnarray*}
\ACER&=& \frac{\Pr\{D_i(1)=1,Y_i(0)=1,Y_i(1)=1\}-\Pr\{D_i(0)=1,Y_i(0)=1,Y_i(1)=1\}}{\Pr\{Y_i(0)=1,Y_i(1)=1\} }\\
&=& \frac{\Pr(D_i=1,Y_i=1 \mid Z_i=1)-\Pr(D_i=1,Y_i=1 \mid Z_i=0)}{\Pr\{Y_i(1)=1\}},
\end{eqnarray*}
and 
 \begin{eqnarray*}
\ACES&=& \frac{\Pr\{D_i(1)=1,Y_i(0)=0,Y_i(1)=0\}-\Pr\{D_i(0)=1,Y_i(0)=0,Y_i(1)=0\}}{\Pr\{Y_i(0)=0\} }\\
&=&\frac{ \Pr(D_i=0,Y_i=0 \mid Z_i=0) - \Pr(D_i=0,Y_i=0 \mid Z_i=1)}{\Pr\{Y_i(0)=0\}}.
\end{eqnarray*}
\QEDB

%

\subsection{ Proof of Theorem~\ref{thm::identification-mon-ipw}}
\label{app:identification-mon-ipw}

Assumption~\ref{asm::indep} and Lemma~\ref{lem::2} imply,
\begin{eqnarray*}
\E\{D_i(z) \mid Y_i(1)=y_1,Y_i(0)=y_0\} &=& \E \left[\E\{D_i(z) \mid \bX_i,Y_i(1)=y_1,Y_i(0)=y_0\} \mid Y_i(1)=y_1,Y_i(0)=y_0 \right]\\
&=& \E \left[\E\{D_i(z) \mid \bX_i\} \mid Y_i(1)=y_1,Y_i(0)=y_0 \right].
\end{eqnarray*}
Based on Lemma~\ref{lem::1},
\begin{eqnarray}
\nonumber&&\E \left[\E\{D_i(z) \mid \bX_i\} \mid Y_i(1)=y_1,Y_i(0)=y_0 \right] \\
\nonumber&=& \E \left[ \frac{\Pr\{\bX_i \mid Y_i(1)=y_1,Y_i(0)=y_0 \}}{\Pr(\bX_i)} \E\{D_i(z) \mid \bX_i\} \right]\\
\nonumber&=&\E\left( \E \left[ \frac{\Pr\{\bX_i \mid Y_i(1)=y_1,Y_i(0)=y_0 \}}{\Pr(\bX_i)} D_i(z) \Bigg | \bX_i \right]\right)\\
\nonumber&=&\E\left( \E \left[ \frac{\Pr\{Y_i(1)=y_1,Y_i(0)=y_0 \mid \bX_i\}}{\Pr\{Y_i(1)=y_1,Y_i(0)=y_0 \}} D_i(z) \Bigg | \bX_i \right]\right)\\
\nonumber&=& \E \left[  \frac{\Pr\{Y_i(1)=y_1,Y_i(0)=y_0 \mid \bX_i\}}{\Pr\{Y_i(1)=y_1,Y_i(0)=y_0 \}} D_i(z) \right]\\
\label{eqn::mon-ipw}&=& \E \left[ \frac{\Pr\{Y_i(1)=y_1,Y_i(0)=y_0 \mid \bX_i\}}{\Pr\{Y_i(1)=y_1,Y_i(0)=y_0 \}} D_i \Bigg | Z_i=z \right],
\end{eqnarray}
where the last equality follows from Assumption~\ref{asm::rand}. We can then obtain the expressions for $\ACEP$, $\ACER$, and $\ACES$ by choosing different values of $y_1$ and $y_0$ in~\eqref{eqn::mon-ipw}. \QEDB
 
\subsection{Proof of Theorem~\ref{thm::identification-strmon}}
\label{app:identification-strmon}

Assumption~\ref{asm::rand} implies,
\begin{eqnarray*}
\Pr\{D_i(z)=d,Y_i(d)=y\}\ =\ \Pr\{D_i(z)=d,Y_i(D_i(z))=y \mid Z_i=z\}\ =\ \Pr(D_i=d,Y_i=y \mid Z_i=z). 
\end{eqnarray*}
 Therefore,
 \begin{eqnarray*}
\Pr\{D_i(z)=1\mid Y_i(0)=y\}&=&\frac{\Pr\{D_i(z)=1, Y_i(0)=y\}}{ \Pr\{Y_i(0)=y\}}\\
&=&\frac{\Pr\{ Y_i(0)=y\}-\Pr\{D_i(z)=0, Y_i(0)=y\}}{ \Pr\{Y_i(0)=y\}}\\
&=&\frac{\Pr\{ Y_i(0)=y\}- \Pr(D_i=0,Y_i=y \mid Z_i=z) }{ \Pr\{Y_i(0)=y\}}.
\end{eqnarray*}
As a result, we have 
\begin{eqnarray*}
\ACEP &=& \frac{ \Pr(D_i=0,Y_i=1 \mid Z_i=0) - \Pr(D_i=0,Y_i=1 \mid Z_i=1) }{ \Pr\{Y_i(0)=1\}},\\
\ACES &=& \frac{ \Pr(D_i=0,Y_i=0 \mid Z_i=0) - \Pr(D_i=0,Y_i=0 \mid Z_i=1) }{ \Pr\{Y_i(0)=0\}}.
\end{eqnarray*}
\QEDB

\subsection{Proof of Theorem~\ref{thm::identification-discrete-ipw}}
\label{app:identification-discrete-ipw}

Using the law of total expectation, we have
\begin{eqnarray*}
\E[\bone\{D_i(z)\geq r \}\mid R_i=r] &=&\E ( \E[\bone\{D_i(z)\geq r \}\mid \bX_i, R_i=r] \mid R_i=r)\\
&=& \E ( \E[\bone\{D_i(z)\geq r \}\mid \bX_i ] \mid R_i=r)\\
&=& \E \left( \frac{\Pr(\bX_i \mid R_i=r)}{\Pr(\bX_i)} \E[\bone\{D_i(z)\geq r \}\mid \bX_i ] \right)\\
&=& \E \left( \frac{\Pr(R_i=r \mid \bX_i )}{\Pr(R_i=r)} \E[\bone\{D_i(z)\geq r \}\mid \bX_i ] \right)\\
&=& \E \left[ \frac{\Pr(R_i=r \mid \bX_i )}{\Pr(R_i=r)} \bone\{D_i(z)\geq r \} \right]\\
&=& \E \left[ \frac{\Pr(R_i=r \mid \bX_i )}{\Pr(R_i=r)} \bone\{D_i\geq r \} \mid Z_i=z \right],
\end{eqnarray*}
where the second equality follows from Assumption~\ref{asm::indep} and Lemma~\ref{lem::2}, and the last equality follows from Assumption~\ref{asm::rand}. Thus, 
\begin{eqnarray*}
\ACEP(r) &=& \E\{ w_r(\bX_i) \bone(D_i \geq r) \mid Z_i=1\} - \E\{ w_r(\bX_i) \bone(D_i\geq r) \mid Z_i=0\}.
\end{eqnarray*}
We can prove the expression for \ACES{} similarly. \QEDB

\section{Details of the Bayesian Estimation}
\label{app:bayesdetails}

We only consider the algorithm for sensitivity analysis with ordinal decision since the computation of the original analysis is straightforward by setting the sensitivity parameters to zero.
Consider the model given in
Equations~\eqref{eqn::bayes-Dz}~and~\eqref{eqn::bayes-R}.
We can write Equation~\eqref{eqn::bayes-Dz} in terms of the observed data as,
\begin{eqnarray}
\label{eqn::bayes-Dstar}
D^\ast_i &=& \beta_Z Z_i+ \bX_i^\top \beta_X+Z_i\bX_i^\top \beta_{ZX} + \epsilon_{i1},
\end{eqnarray}
 where 
 \begin{eqnarray*}
D_i= \begin{cases}
0 & D^\ast \leq \theta_{Z_i,1}\\
1& \theta_{Z_i,1} <D^\ast_i \leq \theta_{Z_i,2}\\
\vdots & \vdots\\
k-1 & \theta_{Z_i,k-1} <D^\ast_i \leq \theta_{Z_i,k}\\
k & \theta_{Z_i,k} <D^\ast_i
\end{cases}.
\end{eqnarray*}
We then consider Equation~\eqref{eqn::bayes-R}.
For $r=0,\ldots,k$,
because $R_i \geq r+1$ is equivalent to $Y_i(r)=1$,
we have
\begin{eqnarray*}
\Pr\{Y(r) =1\} &=&\Pr(R^\ast_i > \delta_{r}) =\Pr( \bm{X}_i^\top \alpha_X+\epsilon_{i2}> \delta_{r})=\Pr( - \delta_{r}+\bm{X}_i ^\top\alpha_X+\epsilon_{i2} >0).
\end{eqnarray*}
Therefore,
we can introduce a latent variable $Y^\ast(r)$, and write
\begin{eqnarray}
\label{eqn::bayes-Yr}
Y_i^\ast(r) =- \delta_{r} + \bm{X}_i^\top \alpha_X+\epsilon_{i2},
\end{eqnarray}
where $Y_i(r)=1$ if $Y_i^\ast(r) >0$ and $Y_i(r)=0$ if
$Y_i^\ast(r) \leq 0$.  We can further write
Equation~\eqref{eqn::bayes-Yr} in terms of the observed data as
\begin{eqnarray}
\label{eqn::bayes-Ystar}
Y_i^\ast = -\sum_{r=0}^k \delta_{r} \bone(D_i=r) + \bm{X}_i^\top \alpha_X+\epsilon_{i2},
\end{eqnarray}
where $Y_i=1$ if $Y_i^\ast>0$ and $Y_i=0$ if $Y_i^\ast \leq 0$.

Combining
Equations~\eqref{eqn::bayes-Dstar}~and~\eqref{eqn::bayes-Ystar}, we
have
\begin{eqnarray}
\label{eqn::bayes-Dstar2}
D^\ast_i &=& \beta_Z Z_i+ \bX_i^\top \beta_X +Z_i\bX_i^\top \beta_{ZX}+ \epsilon_{i1},\\
\label{eqn::bayes-Ystar2}
Y^\ast_i&=& - \sum_{d=0}^k \delta_{d} \bone(D_i=d) +\bX_i^\top \alpha_X+\epsilon_{i2},
\end{eqnarray}
where \begin{eqnarray*}
\begin{pmatrix}
\epsilon_{i1}\\
\epsilon_{i2}
\end{pmatrix} \sim N \left( 
\begin{pmatrix}
0\\
0
\end{pmatrix}, \begin{pmatrix}
1 & \rho\\
\rho & 1
\end{pmatrix}
\right),
\end{eqnarray*}
 and 
\begin{eqnarray*}
D_i=\begin{cases}
0 & D^\ast \leq \theta_{Z_i,1}\\
1& \theta_{Z_i,1} <D^\ast_i \leq \theta_{Z_i,2}\\
\vdots & \vdots\\
k-1 & \theta_{Z_i,k-1} <D^\ast_i \leq \theta_{Z_i,k}\\
k & \theta_{Z_i,k} <D^\ast_i
\end{cases}, \quad
Y_i= \begin{cases}
0& Y^\ast_i \leq 0\\
1 & Y^\ast_i>0
\end{cases}
\end{eqnarray*}
with $\delta_{d} \leq \delta_{d'}$ for $d\leq d'$.

We choose multivariate normal priors for the regression coefficients,
$(\beta_Z,\beta^\top_X, \beta^\top_{ZX})\sim
\bm{N}_{2p+1}(\bm{0},\bm{\Sigma}_D)$ and
$\alpha_X\sim \bm{N}_p(\bm{0},\bm{\Sigma}_R)$. We choose the priors
for $\theta$ and $\delta$ in the following manner. We first choose a
normal prior for $\theta_{z1}$ and $\delta_0$,
$\theta_{z1} \sim N(0,\sigma_0^2)$ and $\delta_0 \sim N(0,\sigma_0^2)$
for $z=0,1$. We then choose truncated normal priors for other
parameters,
$\theta_{zj} \sim N(0,\sigma_0^2) \bone(\theta_{zj} \geq
\theta_{z,j-1})$ for $j=2,\ldots, k$ and
$\delta_l \sim N(0,\sigma_0^2) \bone(\delta_l \geq \delta_{l-1})$ for
$l=1,\ldots, k$. In this way, we guarantee that $\theta$'s and
$\delta$'s are increasing. In our empirical analysis, we choose $\bm{\Sigma}_D^{-1} = 0.01\cdot\bI_{2p+1}$, $\bm{\Sigma}_R^{-1} = 0.01\cdot\bI_{p}$, and $\sigma_0=10$.

Treating $Y^\ast_i$ and $D^\ast_i$ as missing data, we can write the complete-data likelihood as
\begin{eqnarray*}
 & & L(\theta,\beta,\delta,\alpha)\\
 &=& \prod_{i=1}^nL_i(\theta,\beta,\delta,\alpha) \\
&\propto& \prod_{i=1}^n \exp \left( -\frac{1}{2(1-\rho^2)}\left[(D^\ast-\bX_i^\top \beta_X-\beta_Z Z_i-Z_i \bX_i^\top \beta_{ZX})^2 + \left\{Y^\ast_i+\sum_{d=0}^k \delta_{d} \bone(D_i=d) -\bX_i^\top \alpha_X \right\}^2\right. \right.\\
&& \hspace{1.4cm}\left.\left. - 2\rho(D^\ast-\bX_i^\top \beta_X-\beta_Z Z_i) \left\{Y^\ast_i+ \sum_{d=0}^k \delta_{d} \bone(D_i=d) -\bX_i^\top \alpha_X\right\} \right] \right).
\end{eqnarray*}

\paragraph{Imputation Step.}
We first impute the missing data given the observed data and
parameters. Using R package {\it tmvtnorm}, we can jointly sample
$Y^\ast_i$ and $D^\ast_i$. Given
$(D_i,Y_i,Z_i,\bX_i^\top,\theta,\beta,\alpha,\delta)$,
$(D^\ast_i,Y_i^\ast)$ follows a truncated bivariate normal
distribution whose means are given by
$\bX_i^\top \beta_X+\beta_Z Z_i+Z_i \bX_i^\top \beta_{ZX}$ and
$-\sum_{d=0}^k \delta_{d} \bone(D_i=d) +\bX_i^\top \alpha_X$, and
whose covariance matrix has unit variances and correlation $\rho$
where $D^\ast$ is truncated within interval $[\theta_{zd},\theta_{z,d+1}]$ if $Z_i=z$ and $D_i=d$ (we define $\theta_0=-\infty$ and $\theta_{k+1}=\infty$) and $Y^*_i$ is truncated within $(-\infty,0)$ if $Y_i=0$ and $[1,\infty)$ if $Y_i=1$.

\paragraph{Posterior Sampling Step.}
The posterior distribution is proportional to
\begin{eqnarray*}
& & \prod_{i=1}^n \exp \left( -\frac{1}{2(1-\rho^2)}\left[(D^\ast-\bX_i^\top \beta_X-\beta_Z Z_i-Z_i \bX_i^\top \beta_{ZX})^2 + \left\{Y^\ast_i+\sum_{d=0}^k \delta_{d} \bone(D_i=d) -\bX_i^\top \alpha_X \right\}^2\right. \right.\\
&& \hspace{1.4cm}\left.\left. - 2\rho(D^\ast-\bX_i^\top \beta_X-\beta_Z Z_i-Z_i \bX_i^\top \beta_{ZX}) \left\{Y^\ast_i+\sum_{d=0}^k \delta_{d} \bone(D_i=d) -\bX_i^\top \alpha_X\right\} \right] \right)\\
&&\cdot \exp\left\{-\frac{(\beta_Z,\beta^\top_X,\beta^\top_{ZX})\bm{\Sigma}_D^{-1}(\beta_Z,\beta^\top_X,\beta^\top_{ZX}) ^\top}{2}\right\} \cdot \exp\left(-\frac{\alpha_X ^\top\bm{\Sigma}_R^{-1}\alpha_X}{2}\right)\\
&&\cdot \exp \left(-\frac{\theta_{11}^2}{2\sigma_0^2}\right) \exp \left(-\frac{\delta_0^2}{2\sigma_0^2}\right) \prod_{j=2}^k \left\{ \exp \left(-\frac{\theta_{1j}^2}{2\sigma_0^2}\right)\bone(\theta_{1j} \geq \theta_{1,j-1})\right\} \prod_{l=1}^k \left\{ \exp \left(-\frac{\delta_l^2}{2\sigma_0^2}\right)\bone(\delta_l \geq \delta_{l-1})\right\}\\
&&\cdot \exp \left(-\frac{\theta_{01}^2}{2\sigma_0^2}\right) \prod_{j=2}^k \left\{ \exp \left(-\frac{\theta_{0j}^2}{2\sigma_0^2}\right)\bone(\theta_{0j} \geq \theta_{0,j-1})\right\}.
\end{eqnarray*}
We first sample $(\beta_Z,\beta^\top_X,\beta^\top_{ZX})$. From the posterior distribution, we have 
{\footnotesize
\begin{eqnarray*}
&&f(\beta_Z,\beta^\top_X,\beta^\top_{ZX} \mid \cdot )\\
 &\propto& \prod_{i=1}^n \exp \left( -\frac{1}{2(1-\rho^2)}\left[(D_i^\ast-\bX_i^\top \beta_X-\beta_Z Z_i-Z_i \bX_i^\top \beta_{ZX})^2 \right.\right.\\
 && \left.\left. - 2\rho(D_i^\ast-\bX_i^\top \beta_X-\beta_Z Z_i-Z_i \bX_i^\top \beta_{ZX}) \left\{Y^\ast_i+\sum_{d=0}^k \delta_{d} \bone(D_i=d) -\bX_i^\top \alpha_X\right\} \right] \right)\cdot \exp\left\{-\frac{(\beta_Z,\beta^\top_X,\beta^\top_{ZX}) ^\top\bm{\Sigma}_D^{-1}(\beta_Z,\beta^\top_X,\beta^\top_{ZX})}{2}\right\}\\
 &\propto& \prod_{i=1}^n \exp \left( -\frac{1}{2(1-\rho^2)}\left[ (\beta_Z,\beta_X^\top,\beta^\top_{ZX}) (Z_i,\bX^\top_i,Z_i \bX_i^\top )^\top(Z_i,\bX_i^\top,Z_i \bX_i^\top) (\beta_Z,\beta_X^\top,\beta^\top_{ZX})^\top - 2 D^\ast_i (Z_i,\bX_i^\top,Z_i \bX_i^\top ) (\beta_Z,\beta_X^\top,\beta^\top_{ZX})^\top \right.\right.
 \\ 
&&\left.\left. \hspace{0.4cm} + 2\rho \left\{Y^\ast_i+\sum_{d=0}^k \delta_{d} \bone(D_i=d) -\bX_i^\top \alpha_X\right\} (Z_i,\bX_i^\top,Z_i \bX_i^\top ) (\beta_Z,\beta_X^\top, \beta_{ZX}^\top )^\top \right] \right) \cdot \exp\left\{-\frac{(\beta_Z,\beta^\top_X,\beta^\top_{ZX} ) ^\top\bm{\Sigma}_D^{-1}(\beta_Z,\beta^\top_X,\beta^\top_{ZX})}{2}\right\}.
\end{eqnarray*}
}
Therefore, we can sample
\begin{eqnarray*}
(\beta_Z,\beta^\top_X,\beta^\top_{ZX})^\top\mid \cdot \sim \bm{N}_{p+1}(\widehat{\mu}_D, \widehat{\bm{\Sigma}}_D),
\end{eqnarray*}
where 
\begin{eqnarray*}
\widehat{\bm{\Sigma}}_D&=& \left\{ \frac{1}{1-\rho^2}\sum_{i=1}^n(Z_i,\bm{X}^\top_i,Z_i \bX_i^\top )^\top(Z_i,\bm{X}^\top_i,Z_i \bX_i^\top )+\bm{\Sigma}_D^{-1}\right\}^{-1},\\
\widehat{\mu}_D&=&\widehat{\bm{\Sigma}}_D \left( \frac{1}{1-\rho^2} \sum_{i=1}^n(Z_i,\bm{X}^\top_i,Z_i \bX_i^\top)^\top \left[D_i^\ast-\rho \left\{Y^\ast_i+\sum_{d=0}^k \delta_{d} \bone(D_i=d) -\bX_i^\top \alpha_X\right\} \right]\right).
\end{eqnarray*}

We then consider sampling $\alpha_X$. We have 
{\footnotesize
\begin{eqnarray*}
&&f(\alpha_X \mid \cdot )\\
 &\propto& \prod_{i=1}^n \exp \left( -\frac{1}{2(1-\rho^2)}\left[ \left\{Y^\ast_i+\sum_{d=0}^k \delta_{d} \bone(D_i=d) -\bX_i^\top \alpha_X \right\}^2 \right.\right. \\
 && \hspace{0.5cm} \left.\left.- 2\rho(D_i^\ast-\bX_i^\top \beta_X-\beta_Z Z_i-Z_i \bX_i^\top \beta_{ZX}) \left\{Y^\ast_i+\sum_{d=0}^k \delta_{d} \bone(D_i=d) -\bX_i^\top \alpha_X\right\} \right] \right)\cdot \exp\left(-\frac{\alpha_X ^\top\bm{\Sigma}_R^{-1}\alpha_X}{2}\right)\\
 &\propto& \prod_{i=1}^n \exp \left( -\frac{1}{2(1-\rho^2)}\left[ \alpha_X^\top \bX_i^\top\bX_i \alpha_X-2 \left\{Y^\ast_i+ \sum_{d=0}^k \delta_{d} \bone(D_i=d) \right\} \bX_i \alpha_X + 2\rho(D_i^\ast-\bX_i^\top \beta_X-\beta_Z Z_i-Z_i \bX_i^\top \beta_{ZX}) \bX_i \alpha_X \right] \right)\\
&& \hspace{0.4cm} \cdot \exp\left(-\frac{\alpha_X ^\top\bm{\Sigma}_R^{-1}\alpha_X}{2}\right).
\end{eqnarray*}
}
Therefore, we can sample
\begin{eqnarray*}
\alpha_X\mid \cdot \sim \bm{N}_{p}(\widehat{\mu}_R, \widehat{\bm{\Sigma}}_R),
\end{eqnarray*}
where 
\begin{eqnarray*}
 \widehat{\bm{\Sigma}}_R&=& \left\{ \frac{1}{1-\rho^2}\sum_{i=1}^n\bm{X}_i^\top \bm{X}_i+\bm{\Sigma}_R^{-1}\right\}^{-1},\\
\widehat{\mu}_R&=&\widehat{\bm{\Sigma}}_R \left( \frac{1}{1-\rho^2}\sum_{i=1}^n\bm{X}_i \left[\left\{Y^\ast_i+ \sum_{d=0}^k \delta_{d} \bone(D_i=d) \right\} -\rho(D_i^\ast-\bX_i^\top \beta_X-\beta_Z Z_i-Z_i \bX_i^\top \beta_{ZX}) \right] \right).
\end{eqnarray*}

To sample $\delta$'s, we write $ \sum_{d=0}^k \delta_{d} \bone(D_i=d)=\delta_0+ \sum_{d=1}^k (\delta_d - \delta_{d-1}) \bone(D_i\geq d)$ and denote $\bW_i = (1, \bone(D_i \geq 1),\ldots, \bone(D_i \geq k))$ and $\delta= (\delta_0,\delta_1-\delta_0,\ldots,\delta_k-\delta_{k-1})$. Thus, we have 
{\footnotesize
\begin{eqnarray*}
&&f(\delta \mid \cdot )\\
 &\propto& \prod_{i=1}^n \exp \left( -\frac{1}{2(1-\rho^2)}\left[ \left\{Y^\ast_i+\bW_i\delta -\bX_i^\top \alpha_X \right\}^2 - 2\rho(D_i^\ast-\bX_i^\top \beta_X-\beta_Z Z_i-Z_i \bX_i^\top \beta_{ZX}) \left\{Y^\ast_i+ \bW_i\delta -\bX_i^\top \alpha_X\right\} \right] \right)\\
&& \hspace{0.5cm}\cdot \exp \left(-\frac{\delta_0^2}{2\sigma_0^2}\right) \prod_{l=1}^k \left\{ \exp \left(-\frac{\delta_l^        2}{2\sigma_0^2}\right)\bone(\delta_{l}-\delta_{l-1} \geq 0)\right\}\\
 &\propto& \prod_{i=1}^n \exp \left( -\frac{1}{2(1-\rho^2)}\left[ \delta^\top \bW_i^\top \bW_i \delta+2 \left(Y^\ast_i- \bX^\top_i\alpha_X \right) \bW_i \delta -2\rho(D_i^\ast-\bX_i^\top \beta_X-\beta_Z Z_i-Z_i \bX_i^\top \beta_{ZX}) \bW_i \delta \right] \right)\\
&& \hspace{0.5cm}\cdot \exp \left(-\frac{\delta_0^2}{2\sigma_0^2}\right) \prod_{l=1}^k \left\{ \exp \left(-\frac{\delta_l^2}{2\sigma_0^2}\right)\bone(\delta_{l}-\delta_{l-1} \geq 0)\right\}\\
&\propto& \prod_{i=1}^n \exp \left( -\frac{1}{2(1-\rho^2)}\left[ \delta^\top \bW_i^\top \bW_i \delta+2 \left(Y^\ast_i- \bX^\top_i\alpha_X \right) \bW_i \delta -2\rho(D_i^\ast-\bX_i^\top \beta_X-\beta_Z Z_i-Z_i \bX_i^\top \beta_{ZX}) \bW_i \delta \right] \right)\\
&& \hspace{0.5cm}\cdot \exp \left(-\frac{\delta^\top C^\top C \delta  }{2\sigma_0^2}\right) \prod_{l=1}^k \bone(\delta_{l}-\delta_{l-1} \geq 0),
\end{eqnarray*} }
where $C$ is a $(k+1)\times (k+1)$ lower triangular matrix with all non-zero entries equal to 1.
Therefore, we can draw from a truncated normal distribution with mean and covariance matrix
\begin{eqnarray*}
 \widehat{\bm{\Sigma}}_\delta&=& \left\{ \frac{1}{1-\rho^2}\sum_{i=1}^n\bm{W}_i^\top \bm{W}_i+ \frac{ C^\top C}{\sigma_0^{2}}\right\}^{-1},\\
\widehat{\mu}_\delta&=&\widehat{\bm{\Sigma}}_\delta \left[ \frac{1}{1-\rho^2} \sum_{i=1}^n\bW_i^\top \left\{\rho(D_i^\ast-\bX_i^\top \beta_X-\beta_Z Z_i-Z_i \bX_i^\top \beta_{ZX}) -\left(Y^\ast_i- \bX^\top_i\alpha_X\right) \right\} \right],
\end{eqnarray*}
where the $2$-th to ($k+1$)-th element is truncated within interval $[0,\infty)$.
We can then transform $\delta$ to obtain $(\delta_0,\delta_1,\ldots,\delta_k)$.

Finally, we sample
\begin{eqnarray*}
\theta_{z1} \mid \cdot \sim TN(0,\sigma_0^2; \max_{i:Z_i=z,D_i=0} D_i^\ast, \min_{i:Z_i=z,D_i=1}(D_i^\ast,\theta_2) ).
\end{eqnarray*}
We then sample
\begin{eqnarray*}
\theta_{zj} \mid \cdot \sim TN(0,\sigma_0^2; \max_{i:Z_i=z,D_i=j-1} (D_i^\ast,\theta_{j-1}), \min_{i:Z_i=z,D_i=j}(D_i^\ast,\theta_{j+1}) )
\end{eqnarray*}
for $j=2,\ldots,k-1$, and 
\begin{eqnarray*}
\theta_{zk} \mid \cdot \sim TN(0,\sigma_0^2; \max_{i:Z_i=z,D_i=k-1} (D_i^\ast,\theta_{k-1}), \min_{i:Z_i=z,D_i=k}D_i^\ast ).
\end{eqnarray*}

The MCMC gives the posterior distributions of the parameters and therefore we can obtain the posterior distributions of $\Pr(D_i \mid R_i, \bX_i=\bx, Z_i=z)$ and $\Pr(R_i \mid \bX_i=\bx)$. As a result, for $r=1,\ldots,k$, we have 
\begin{eqnarray*}
\ACEP(r)&=& \ \Pr\{D_i(1) \geq r \mid R_i=r\} -\Pr\{D_i(0) \geq r \mid R_i=r\}\\
&=& \frac{\E \left\{\Pr (D_i(1) \geq r ,R_i=r \mid \bX_i)\right\} }{\E \{\Pr(R_i=r\mid \bX_i)\}}-\frac{\E \left\{\Pr (D_i(0) \geq r ,R_i=r \mid \bX_i)\right\} }{\E \{\Pr(R_i=r\mid \bX_i)\}},\\
\ACES&=& \ \Pr\{D_i(1) = 0 \mid R_i=0\} -\Pr\{D_i(0) = 0 \mid R_i=0\}\\
&=& \frac{\E \left\{\Pr (D_i(1) =0 ,R_i=0 \mid \bX_i)\right\} }{\E \{\Pr(R_i=0\mid \bX_i)\}}-\frac{\E \left\{\Pr (D_i(0) =0 ,R_i=0 \mid \bX_i)\right\} }{\E \{\Pr(R_i=0\mid \bX_i)\}}.
\end{eqnarray*}
We can calculate the conditional probabilities
$\Pr\{D_i(z),R_i \mid \bX_i\}$ and $\Pr(R_i \mid \bX_i)$ based on the
posterior sample of the coefficients, and then replace the expectation
with the empirical average to obtain the estimates.

\section{Optimal PSA Provision}
\label{app:optimalprovision}

In this appendix, we consider the optimal PSA provision rule and
conduct an empirical analysis. Let $\xi$ be a PSA provision rule,
i.e., $\xi(\bx)= 1$ (the PSA is provided) if $\bx \in \mathcal{B}_1$
and $\xi(\bx)= 0$ (the PSA is not provided) if
$\bx \in \mathcal{B}_0$, where
$\cX = \mathcal{B}_0 \bigcup \mathcal{B}_1$ and
$\mathcal{B}_0 \cap \mathcal{B}_1=\emptyset$. The judges will make
their decisions based on the PSA and other available information
included in $\bX_i = \bx$. To consider the influence of the PSA on
judges' decision, we define $\delta_{i1}$ the potential decision rule
of case $i$ if the judge received the PSA and $\delta_{i0}$ if
not. Thus, $\delta_{iz}(\bx) = d$ if $\bx \in \cX_{i,zd}$ where
$\cX_{i,zd}$ is a partition of the covariate space with
$\cX=\bigcup_{d=0}^k \cX_{i,zd}$ and
$\cX_{i,zd} \cap \cX_{i,zd^\prime} = \emptyset$ for $z=0,1$. Although
we allow the judge to make a different decision even if the observed
case characteristics $\bX_i$ are identical, we assume that the judges'
decisions are identically distributed given the observed case
characteristics and PSA provision. That is, we assume
$\Pr\{\delta_{iz}(\bx) =d\} = \Pr\{\delta_{i^\prime z}(\bx) =d\} $ for
fixed $\bx$, $z$ and $i \ne i^\prime$, where the probability is taken
with respect to the super population of all cases.

Given this setup, we derive the optimal PSA provision rule. We consider the 0--1 utility
$U_i(\xi) = \bone\{\delta_{i,\xi(\bX_i)}(\bX_i) = R_i\}$. This utility equals
one, if the judge makes the most lenient decision to prevent an
arrestee from engaging in NCA (NVCA or FTA), and equals zero
otherwise. As before, we begin by rewriting the expected utility in the
following manner,
\begin{eqnarray*}
\E\{U_i(\xi)\}&=& \E \left[\bone\{R_i=\delta_{i,\xi(\bX_i)}(\bX_i)\}\right]\\
&=& \sum_{r=0}^k\E\left[ \bone\{R_i=r, \delta_{i,\xi(\bX_i)}(\bX_i)=r\}\right]\\
&=& \sum_{r=0}^k \sum_{z=0}^1\E[ \bone\{R_i=r, \delta_{iz}(\bX_i)=r, \bX_i \in \mathcal{B}_z\}].
\end{eqnarray*}
Under the unconfoundedness assumption, we can write,
\begin{eqnarray*}
\E[ \bone\{R_i=r, \delta_{iz}(\bX_i)=r, \bX_i \in \mathcal{B}_z\}] &=& \E[ \Pr(R_i=r\mid \bX_i) \cdot \Pr\{\delta_{iz}(\bX_i)=r \mid \bX_i \} \cdot \bone\{ \bX_i \in \mathcal{B}_z\}]\\
&=& \E[ e_r(\bX_i) \cdot \Pr\{\delta_{iz}(\bX_i)=r \} \cdot \bone\{ \bX_i \in \mathcal{B}_z\}].
\end{eqnarray*}

Because in the experiment, the provision of the PSA is randomized, we
can estimate
$\Pr\{\delta_{iz}(\bX_i)=r \} = \Pr(D_i=r\mid Z_i=z,\bX_i )$ from the
data. Therefore, we obtain
\begin{eqnarray*}
\E\{U_i(\xi)\}&=& \sum_{z=0,1} \E \left( \left[ \sum_{r=0}^k e_r(\bX_i) \cdot \Pr(D_i=r\mid Z_i=z,\bX_i ) \right]\cdot \bone\{ \bX_i \in \mathcal{B}_z\}\right).
\end{eqnarray*}
Then, the
optimal PSA provision rule is, 
\begin{eqnarray}
\xi(\bx) \ = \ \argmax_{z=0,1} h_z(\bx)
\quad
{\rm where} \quad
h_z(\bx) \ = \ \sum_{r=0}^k e_r(\bx) \cdot \Pr(D_i=r\mid Z_i=z,\bX_i
 ). \label{eq:optimalprovision}
\end{eqnarray}
Thus, we can use the experimental data to derive the optimal PSA
provision rule.

\section{Frequentist Analysis}
\label{app:frequentist}

In this appendix, we implement frequentist analysis and present the
results. We fit the model defined in Equation~\eqref{eqn::bayes-Ystar}
with probit regression. Recall that for $r=0,\ldots,k$, $R_i \geq r+1$
is equivalent to $Y_i(r)=1$. Hence, we can estimate the conditional
probabilities $e_r(\bX_i)$ for each $r=0,\ldots,k+1$ based on the
estimates of the regression coefficients,
\begin{eqnarray*}
	\hat e_r(\bx) & = & \Phi(-\hat \delta_{r-1}+\bx^\top\hat\alpha_X) - \Phi(-\hat \delta_r+\bx^\top\hat\alpha_X), \text{ for } \ r =1,\ldots,k,\\
	\hat e_{k+1}(\bx) &=& \Phi(-\hat \delta_k+\bx^\top\hat\alpha_X), \\
	\hat e_0(\bx) &=& 1 - \Phi(-\hat \delta_0+\bx ^\top\hat\alpha_X),
\end{eqnarray*}
where $\Phi(\cdot)$ denotes the cumulative distribution function of the standard normal distribution. We estimate $\ACEP(r)$ and
$\ACES$ using Hajek estimator as follows,
\begin{eqnarray*}
	\widehat\ACEP(r) &=& \frac{\sum_i \hat w_r(\bX_i) \bone(D_i\geq r) \bone(Z_i=1)}{\sum_i \hat w_r(\bX_i) \bone(Z_i=1)}
	 - \frac{\sum_i \hat w_r(\bX_i) \bone(D_i\geq r) \bone(Z_i=0)}{\sum_i \hat w_r(\bX_i) \bone(Z_i=0)},\\
	\widehat\ACES &=& \frac{\sum_i \hat w_0(\bX_i) \bone(D_i=0) \bone(Z_i=1)}{\sum_i \hat w_0(\bX_i) \bone(Z_i=1)} - \frac{\sum_i \hat w_0(\bX_i) \bone(D_i=0) \bone(Z_i=0)}{\sum_i \hat w_0(\bX_i) \bone(Z_i=0)},
\end{eqnarray*}
where 
$\hat w_r(\bx) = \hat e_r( \bx)/\{\frac{1}{n}\sum_i \hat
e_r(\bX_i)\}$. We use bootstrap to compute the 95\% confidence
interval. 

\begin{figure}[b!]
	\centering \spacingset{1}
	\includegraphics[width=\textwidth]{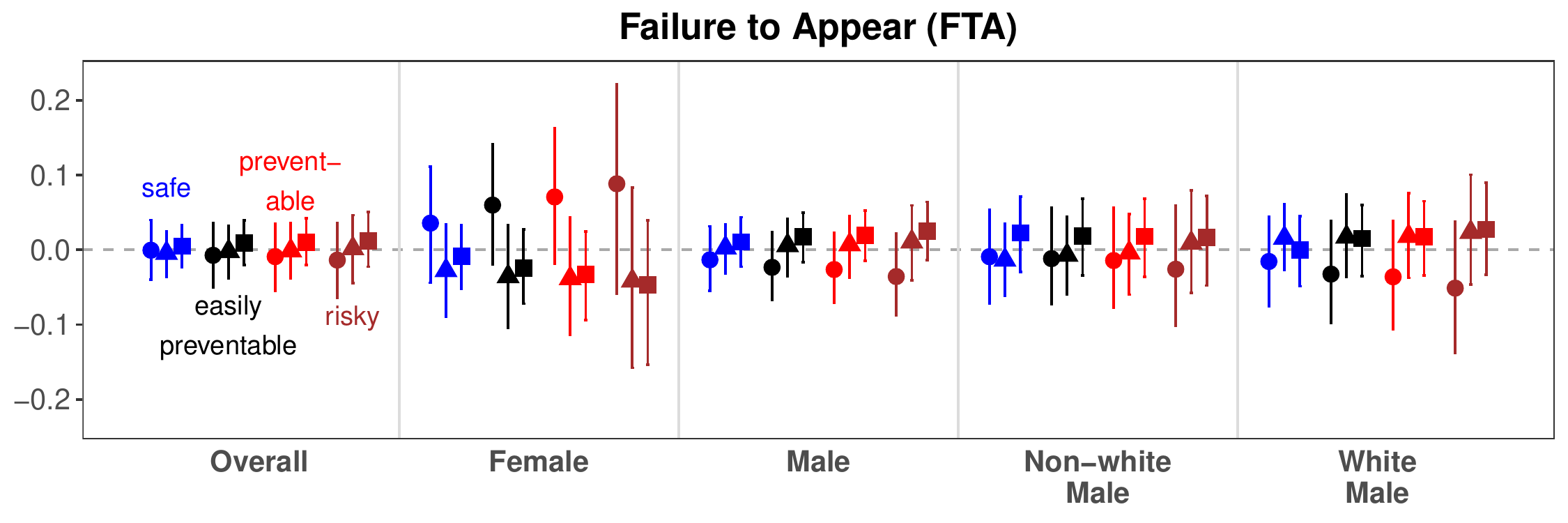}
	\includegraphics[width=\textwidth]{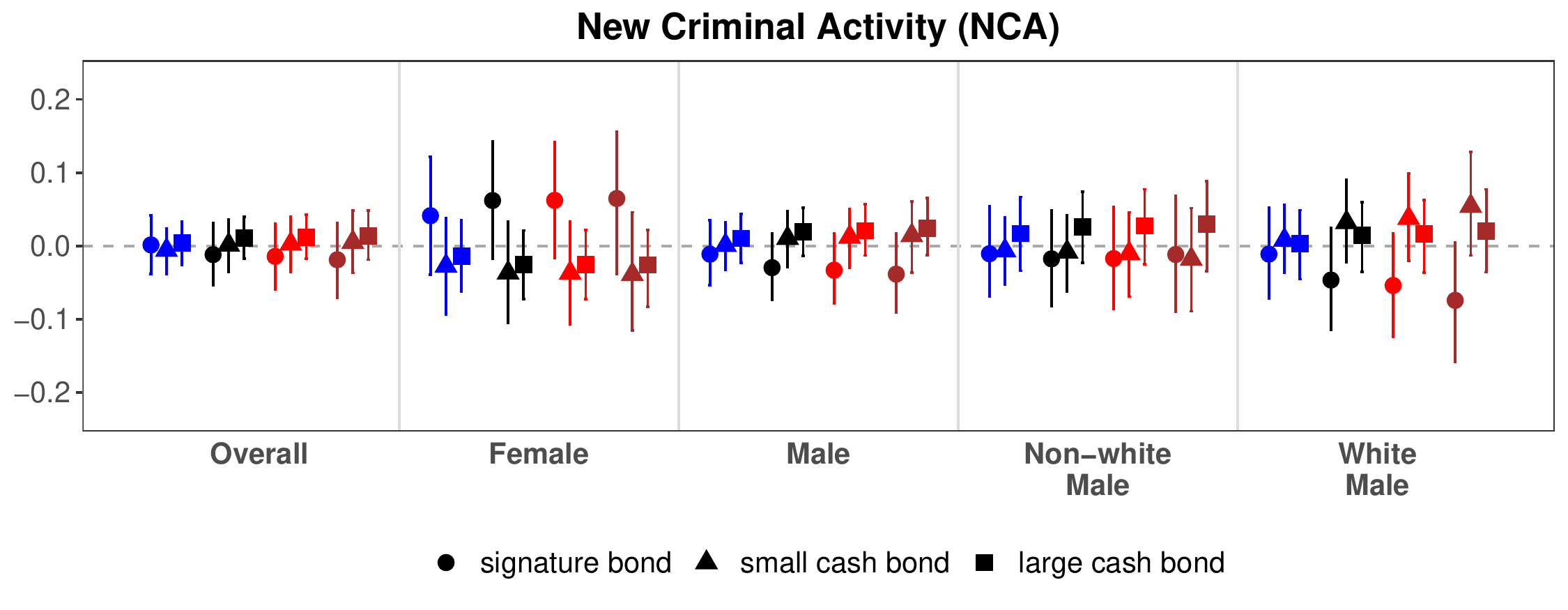}
 	\caption{Estimated Average Principal Causal Effects (\ACE) of
          PSA Provision on the Judge's Decision based on Frequentist
          Analysis. Each panel presents the overall and
          subgroup-specific results for a different outcome
          variable. Each column within a panel shows the estimated
          \ACE{} of PSA provision for safe (blue), easily preventable
          (black), preventable (red), and risky (brown) cases. For
          each of these principal strata, we report the estimated
          \ACE{} on the judge's decision to impose a signature bond
          (circles), a small cash bail amount of 1,000 dollars or less
          (triangles), and a large cash bail amount of greater than
          1,000 (squares). The vertical line for each estimate
          represents the $95\%$ credible interval.}
	\label{fig:freq}
\end{figure}

\clearpage

\begin{figure}[h!]
	\centering \spacingset{1}
	\includegraphics[width=\textwidth]{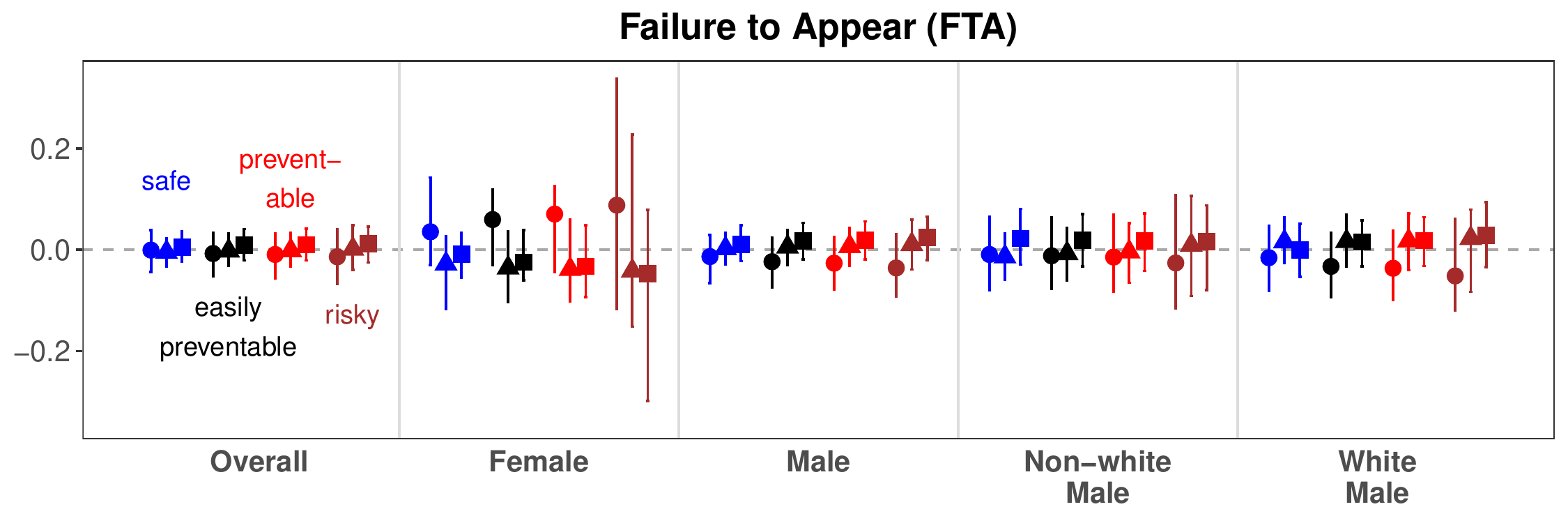}
	\includegraphics[width=\textwidth]{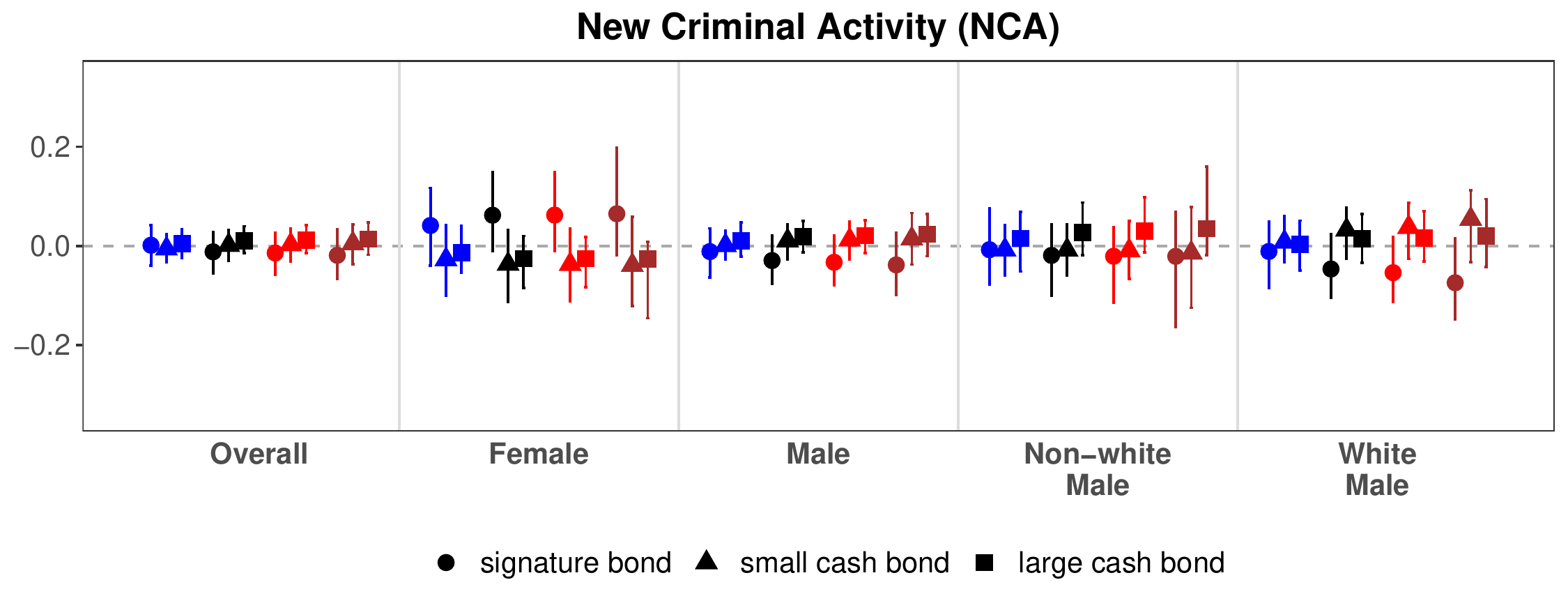}
        \caption{Estimated Average Principal Causal Effects (\ACE) of
          PSA Provision on the Judge's Decision based on Frequentist
          Analysis with Random Effects. Each panel presents the
          overall and subgroup-specific results for a different
          outcome variable. Each column within a panel shows the
          estimated \ACE{} of PSA provision for safe (blue), easily
          preventable (black), preventable (red), and risky (brown)
          cases. For each of these principal strata, we report the
          estimated \ACE{} on the judge's decision to impose a
          signature bond (circles), a small cash bail amount of 1,000
          dollars or less (triangles), and a large cash bail amount of
          greater than 1,000 (squares). The vertical line for each
          estimate represents the $95\%$ credible interval.}
	\label{fig:freq_re}
\end{figure}

\begin{figure}[p!]
	\centering \spacingset{1}
	\includegraphics[width=\textwidth]{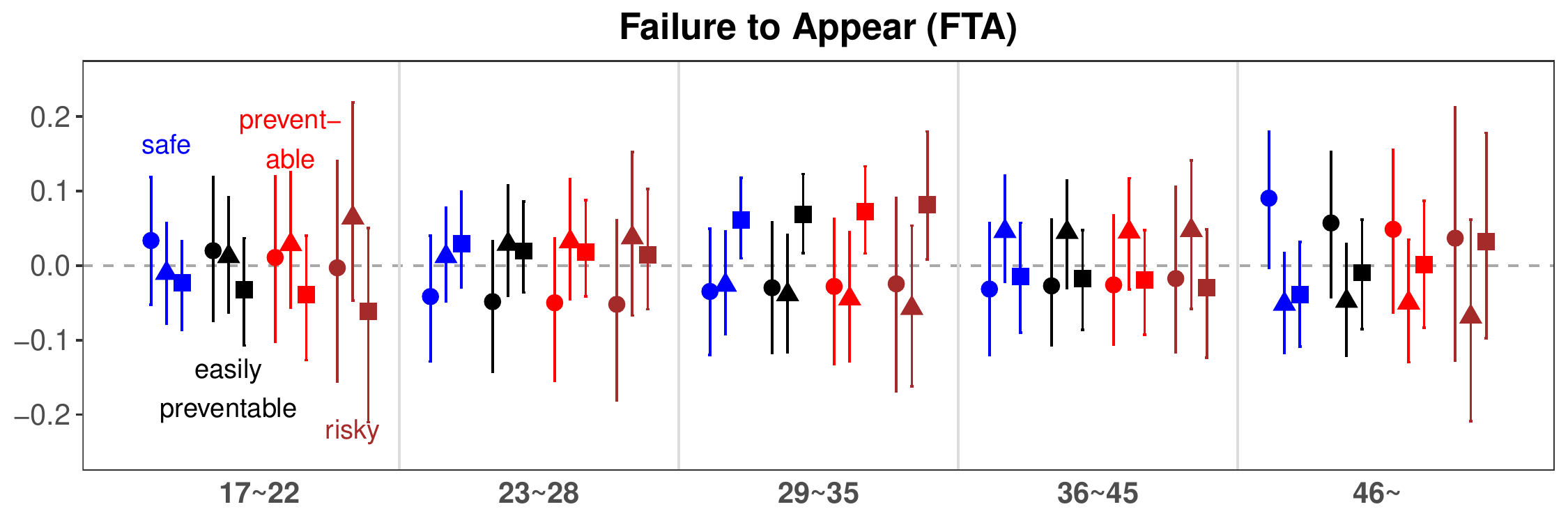}
	\includegraphics[width=\textwidth]{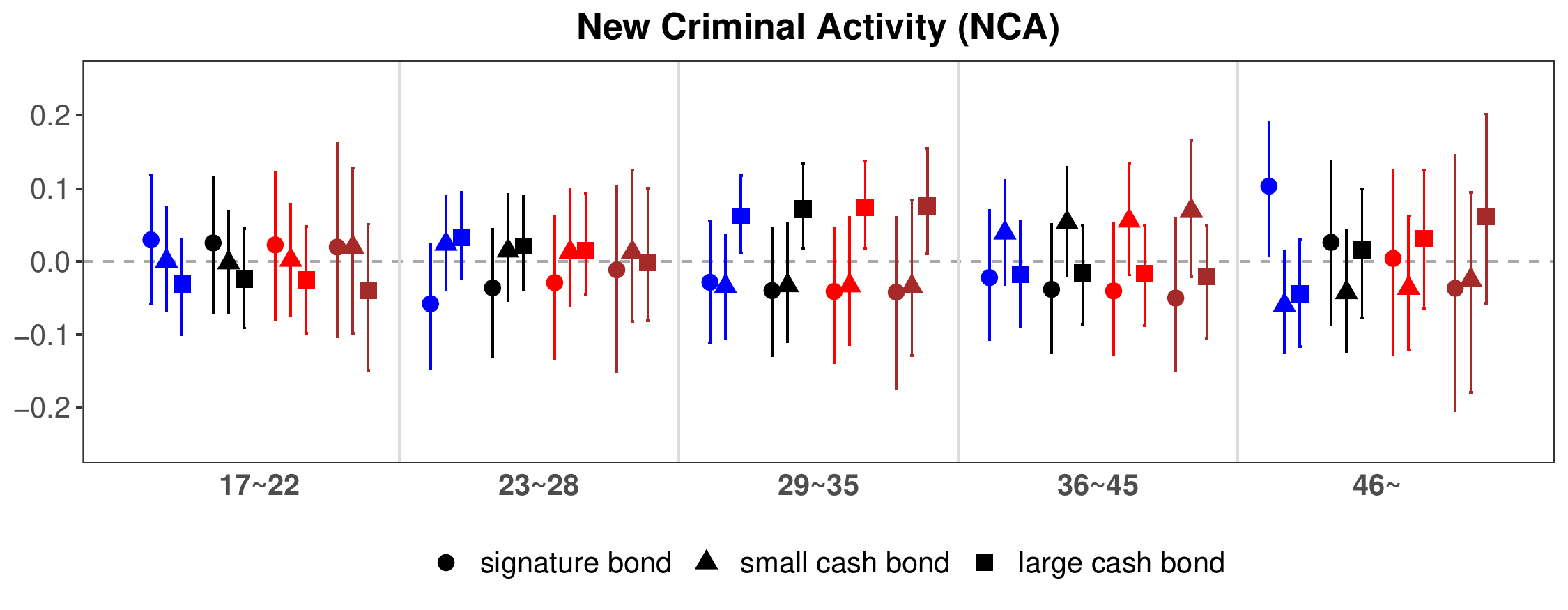}
 	\caption{Estimated Average Principal Causal Effects (\ACE) of
          PSA Provision on the Judge's Decision based on Frequentist
          Analysis. Each panel presents the age group-specific results
          for a different outcome variable. Each column within a panel
          shows the estimated \ACE{} of PSA provision for safe (blue),
          easily preventable (black), preventable (red), and risky
          (brown) cases. For each of these principal strata, we report
          the estimated \ACE{} on the judge's decision to impose a
          signature bond (circles), a small cash bail amount of 1,000
          dollars or less (triangles), and a large cash bail amount of
          greater than 1,000 (squares). The vertical line for each
          estimate represents the $95\%$ credible interval.}
	\label{fig:freq_age}
\end{figure}

Figures~\ref{fig:freq} presents the estimated \ACE{} of PSA provision
on the three ordinal decision categories, separately for FTA and NCA
within each principal stratum. The results for NVCA are not presented
due to the fact that the number of events is too small for an
informative subgroup analysis. The results are largely consistent with
those of the Bayesian analysis presented in the main text. As a
robustness check for the assumption of no interference among the
cases, Figure~\ref{fig:freq_re} presents the estimated \ACE{} of PSA
provision with the model including random effects for the hearing date
of the case, and the results are the same.  Figure~\ref{fig:freq_age}
presents the results for each age group similar to the one in
Appendix~\ref{app:age}.

\clearpage
\section{Nonparametric Sensitivity Analysis}
\label{sec:nonpara_sens}

We consider a nonparametric sensitivity analysis for the ordinal
decision under the monotonicity assumption
(Assumption~\ref{asm::mon-discrete}). We introduce the following
sensitivity parameters, $\xi_{rdz}(\bx)$ for $r,d=0,\ldots,k$ and
$z=0,1$, to characterize the deviation from the unconfoundedness
assumption,
\begin{eqnarray*}
 \xi_{rdz}(\bx) \ = \ \frac{\Pr\{Y_i(r)=1 \mid D_i(z)=d,\bX_i = \bx\}}{\Pr\{Y_i(r)=1 \mid D_i(z)=0,\bX_i = \bx\}},
\end{eqnarray*}
which is equal to 1 for all $(r,d,z)$ and $\bx$ when the
unconfoundedness assumption holds.

We can directly relate the parametric sensitivity parameter $\rho$ to the
parameters of the nonparametric sensitivity analysis. Because $R_i \geq r+1$ is equivalent to $Y_i(r)=1$, we can
obtain the following formula from
Equations~\eqref{eqn::bayes-Dz}~and~\eqref{eqn::bayes-R},
\begin{eqnarray*}
\Pr\{Y_i(r)=1 \mid D_i(z)=d,\bX_i = \bx\} &=& \frac{\Pr(\theta_{zd}< \beta_Z z+ \bx^\top \beta_X +
    z \bx^\top \beta_{ZX} +\epsilon_{i1}\leq \theta_{z,d+1}, \delta_r < \bx^\top \alpha_X+\epsilon_{i2})}{\Pr(\theta_{zd}< \beta_Z z+ \bx^\top \beta_X +
     z \bx^\top \beta_{ZX} +\epsilon_{i1}\leq \theta_{z,d+1})},   
\end{eqnarray*}
where $\theta_{z0}=-\infty$ and $\delta_{k+1} = \infty$. Together with
Proposition~\ref{prop::sensitivity}, we can express the sensitivity
parameters in the nonparametric sensitivity analysis $\xi_{rdz}(\bx)$
in terms of the model parameters given in
Equations~\eqref{eqn::bayes-Dz}~and~\eqref{eqn::bayes-R}. Thus, the
parametric sensitivity analysis, while much simpler, imposes
restrictions on the nonparametric counterpart.

The following proposition gives the identification formulas for
$\Pr\{D_i(z)=d \mid R_i=r\}$ for all $(r,d,z)$ with any given value of
$ \xi_{rdz}(\bx)$.
\begin{proposition}
\label{prop::sensitivity}
Under
Assumptions~\ref{asm::rand},~\ref{asm::ex},~and~\ref{asm::mon-discrete},
if $ \xi_{rdz}(\bx)$ is known for all $(r,d,z)$ and $\bx$, then we
have
\begin{eqnarray*}
\Pr\{D_i(z)=d \mid R_i=r\}
&=& \frac{  \E\left[  \Pr\{Y_i(r-1)=1 \mid D_i(z)=d,\bX_i = \bx\}  \Pr(D_i=d\mid Z_i=z,\bX_i=\bx)\right]}{\E\left[ \Pr\{Y_i(r-1)=1\mid \bX_i=\bx\}-\Pr\{Y_i(r)=1\mid \bX_i=\bx\} \right]}\\
&&- \frac{  \E\left[  \Pr\{Y_i(r)=1 \mid D_i(z)=d,\bX_i = \bx\}  \Pr(D_i=d\mid Z_i=z,\bX_i=\bx)\right]}{\E\left[ \Pr\{Y_i(r-1)=1\mid \bX_i=\bx\}-\Pr\{Y_i(r)=1\mid \bX_i=\bx\} \right]}
\end{eqnarray*}
for $r=1,\ldots,k$ and all $(d,z)$, and 
\begin{eqnarray*}
\Pr\{D_i(z)=d \mid R_i=k+1\}&=&  \frac{  \E\left[  \Pr\{Y_i(k)=1 \mid D_i(z)=d,\bX_i = \bx\}  \Pr(D_i=d\mid Z_i=z,\bX_i=\bx)\right]}{\E\left[ \Pr\{Y_i(k)=1\mid \bX_i=\bx\}\right]},\\
\Pr\{D_i(z)=d \mid R_i=0\}&=&  \frac{  \E\left[  \Pr\{Y_i(0)=0 \mid D_i(z)=d,\bX_i = \bx\}  \Pr(D_i=d\mid Z_i=z,\bX_i=\bx)\right]}{\E\left[ \Pr\{Y_i(0)=0\mid \bX_i=\bx\}\right]}
\end{eqnarray*}
for all $(d,z)$, where
\begin{eqnarray*}
 \Pr\{Y_i(r)=1 \mid D_i(z)=d,\bX_i = \bx\}&=&  \frac{\xi_{rdz}(\bx)}{\xi_{rrz}(\bx)} \cdot \Pr(Y_i=1 \mid Z_i=z,D_i=r,\bX_i = \bx),\\
  \Pr\{Y_i(r)=1 \mid \bX_i = \bx\} &=&   \frac{\sum_{d=0}^k\xi_{rdz}(\bx)\Pr(D_i=d\mid Z_i=z,\bX_i=\bx)}{\xi_{rrz}(\bx)} \\
  && \hspace{.2in} \cdot \Pr(Y_i=1 \mid Z_i=z,D_i=r,\bX_i = \bx).
\end{eqnarray*}
\end{proposition}
{\sc Proof:}
The randomization of treatment assignment (Assumption~\ref{asm::rand})
implies,
\begin{eqnarray*}
 \Pr\{Y_i(r)=1 \mid D_i(z)=r,\bX_i = \bx\}\ =\ \Pr(Y_i=1 \mid Z_i=z,D_i=r,\bX_i = \bx).
\end{eqnarray*}
Therefore, with given values of $\xi_{rdz}(\bx)$, we have,
\begin{eqnarray*}
 \Pr\{Y_i(r)=1 \mid D_i(z)=d,\bX_i = \bx\}&=& \frac{\xi_{rdz}(\bx)}{\xi_{rrz}(\bx)} \cdot \Pr(Y_i=1 \mid Z_i=z,D_i=r,\bX_i = \bx),\\
 \Pr\{Y_i(r)=1 \mid \bX_i = \bx\} &=&  \sum_{d=0}^k  \Pr\{Y_i(r)=1 \mid D(z)=d, \bX_i = \bx\}\Pr\{D(z)=d\mid \bX_i = \bx\}\\
 &=& \frac{\sum_{d=0}^k\xi_{rdz}(\bx)\Pr(D_i=d\mid Z_i=z,\bX_i=\bx)}{\xi_{rrz}(\bx)} \\
 && \hspace{.2in} \cdot \Pr(Y_i=1 \mid Z_i=z,D_i=r,\bX_i = \bx).
\end{eqnarray*}
From the above two terms, we have
\begin{eqnarray*}
&&\Pr\{D_i(z)=d \mid R_i=r\}\\
&=& \frac{ \E\left[ \Pr\{D_i(z)=d, R_i=r\mid \bX_i=\bx\} \right]}{\E \left\{\Pr(R_i=r\mid \bX_i=\bx) \right\}} \\
&=& \frac{ \E\left[ \Pr\{D_i(z)=d, Y_i(r-1)=1\mid \bX_i=\bx\}-\Pr\{D_i(z)=d, Y_i(r)=1\mid \bX_i=\bx\} \right]}{ \E\left[ \Pr\{Y_i(r-1)=1\mid \bX_i=\bx \}-\Pr\{Y_i(r)=1\mid \bX_i=\bx \} \right]} \\
&=& \frac{ \E\left[ \Pr\{Y_i(r-1)=1 \mid D_i(z)=d,\bX_i = \bx\} \Pr(D_i=d\mid Z_i=z,\bX_i=\bx)\right]}{\E\left[ \Pr\{Y_i(r-1)=1\mid \bX_i=\bx\}-\Pr\{Y_i(r)=1\mid \bX_i=\bx\} \right]}\\
&&- \frac{ \E\left[ \Pr\{Y_i(r)=1 \mid D_i(z)=d,\bX_i = \bx\} \Pr(D_i=d\mid Z_i=z,\bX_i=\bx)\right]}{\E\left[ \Pr\{Y_i(r-1)=1\mid \bX_i=\bx\}-\Pr\{Y_i(r)=1\mid \bX_i=\bx\} \right]}
\end{eqnarray*}
for $r=1,\ldots,k$, where the first equality follows from the law of total expectation, and the second equality follows from Assumption~\ref{asm::mon-discrete}.

Similarly, we can obtain 
\begin{eqnarray*}
&&\Pr\{D_i(z)=d \mid R_i=k+1\}\\
&=& \frac{ \E\left[ \Pr\{D_i(z)=d, R_i=k+1\mid \bX_i=\bx\} \right]}{\E \left\{\Pr(R_i=k+1\mid \bX_i=\bx) \right\}} \\
&=& \frac{ \E\left[ \Pr\{D_i(z)=d, Y_i(k)=1\mid \bX_i=\bx\}\right]}{\E\left[ \Pr\{Y_i(k)=1\mid \bX_i=\bx \}\right]} \\
&=& \frac{ \E\left[ \Pr\{Y_i(k)=1 \mid D_i(z)=d,\bX_i = \bx\} \Pr(D_i=d\mid Z_i=z,\bX_i=\bx)\right]}{\E\left[ \Pr\{Y_i(k)=1\mid \bX_i=\bx\}\right]},\\
&&\Pr\{D_i(z)=d \mid R_i=0\}\\
&=& \frac{ \E\left[ \Pr\{D_i(z)=d, R_i=0\mid \bX_i=\bx\} \right]}{\E \left\{\Pr(R_i=0\mid \bX_i=\bx) \right\}} \\
&=& \frac{ \E\left[ \Pr\{D_i(z)=d, Y_i(0)=0\mid \bX_i=\bx\}\right]}{\E\left[ \Pr\{Y_i(0)=0\mid \bX_i=\bx \}\right]} \\
&=& \frac{ \E\left[ \Pr\{Y_i(0)=0 \mid D_i(z)=d,\bX_i = \bx\} \Pr(D_i=d\mid Z_i=z,\bX_i=\bx)\right]}{\E\left[ \Pr\{Y_i(0)=0\mid \bX_i=\bx\}\right]}.
\end{eqnarray*}
\QEDB

Using this result, we can compute the \ACE{} with any given value of
$\xi_{rdz}(\bx)$.  Unfortunately, this nonparametric sensitivity
analysis requires the specification of too many sensitivity
parameters, making it unsuitable for practical use.

\section{Parametric Sensitivity Analysis Results}
\label{sec:sensitivity_results}

\begin{figure}[h!]
	\centering \spacingset{1}
	\includegraphics[width=\textwidth]{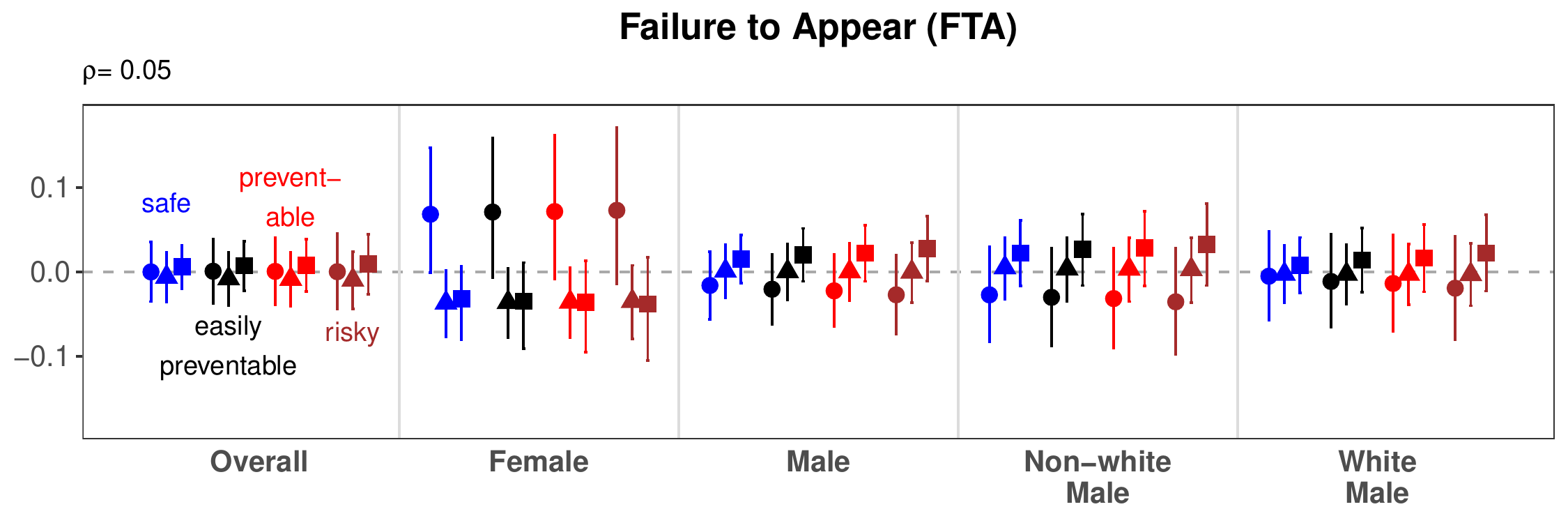}
	\includegraphics[width=\textwidth]{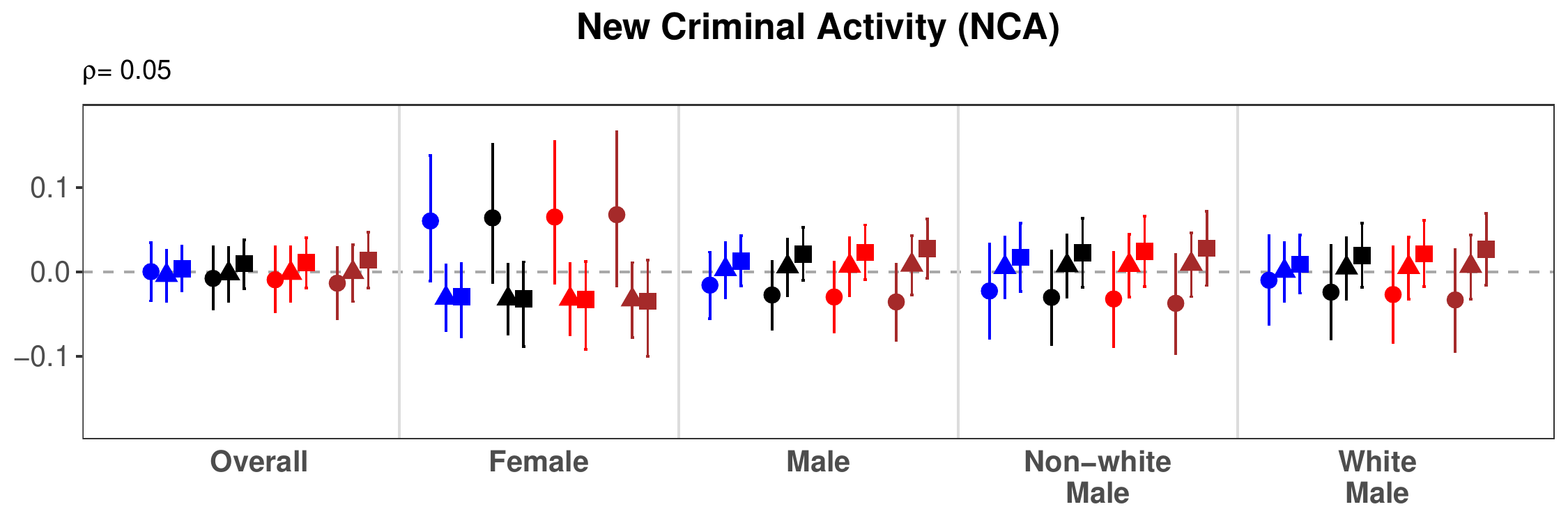}
	\includegraphics[width=\textwidth]{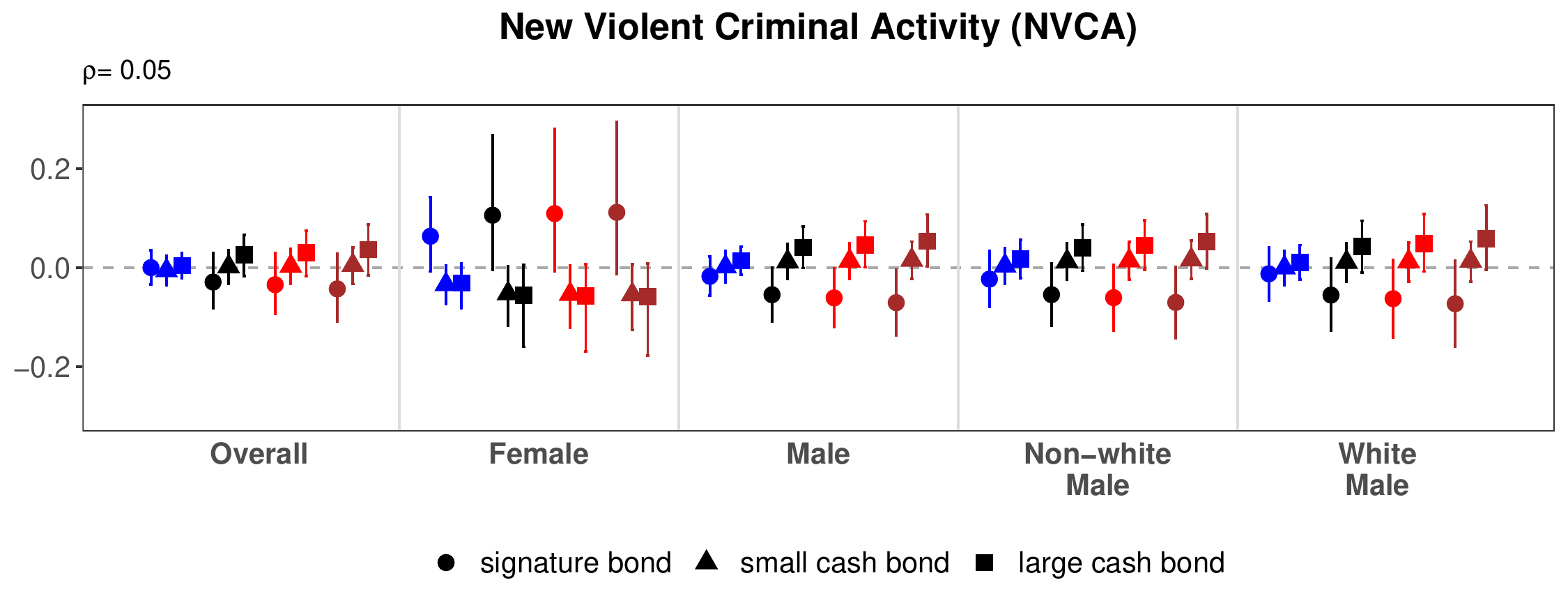}
        \caption{Estimated Average Principal Causal Effects (\ACE) of
          PSA Provision on the Judge's Decision with $\rho=0.05$. Each
          panel presents the overall and subgroup-specific results for
          a different outcome variable. Each column within a panel
          shows the estimated \ACE{} of PSA provision for safe (blue),
          easily preventable (black), preventable (red), and risky
          (brown) cases. For each of these principal strata, we report
          the estimated \ACE{} on the judge's decision to impose a
          signature bond (circles), a small cash bail amount of 1,000
          dollars or less (triangles), and a large cash bail amount of
          greater than 1,000 (squares). The vertical line for each
          estimate represents the Bayesian $95\%$ credible interval.}
	\label{fig:sens1}
\end{figure}

\begin{figure}[p]
	\centering \spacingset{1}
	\includegraphics[width=\textwidth]{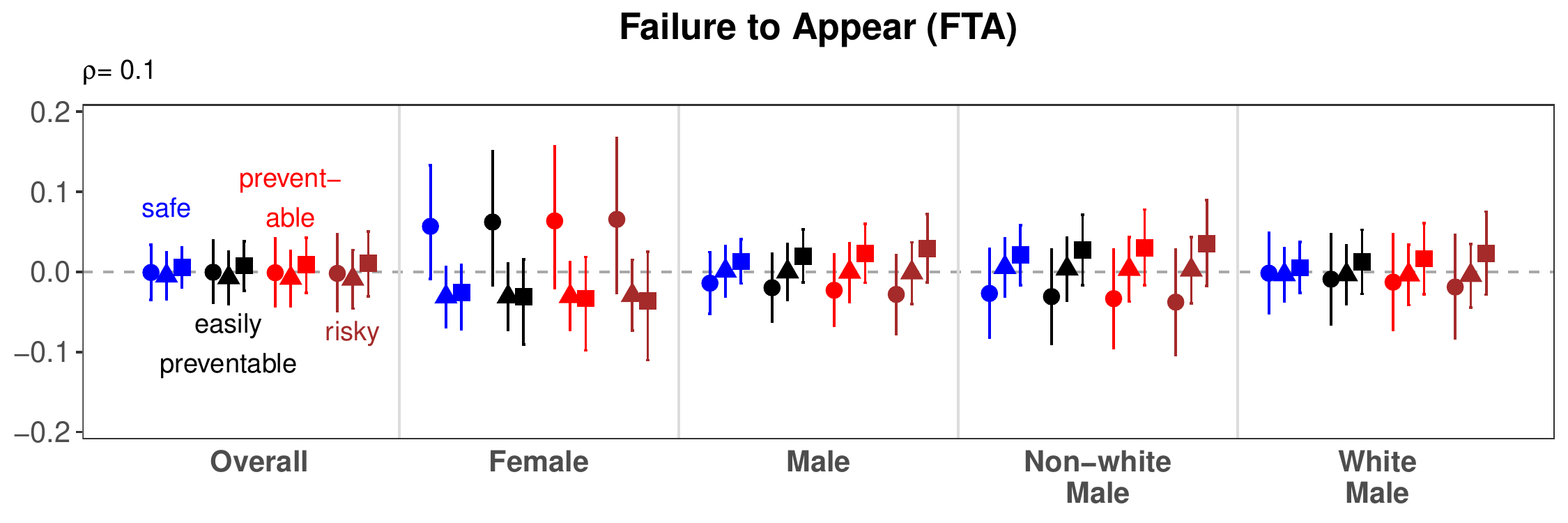}
	\includegraphics[width=\textwidth]{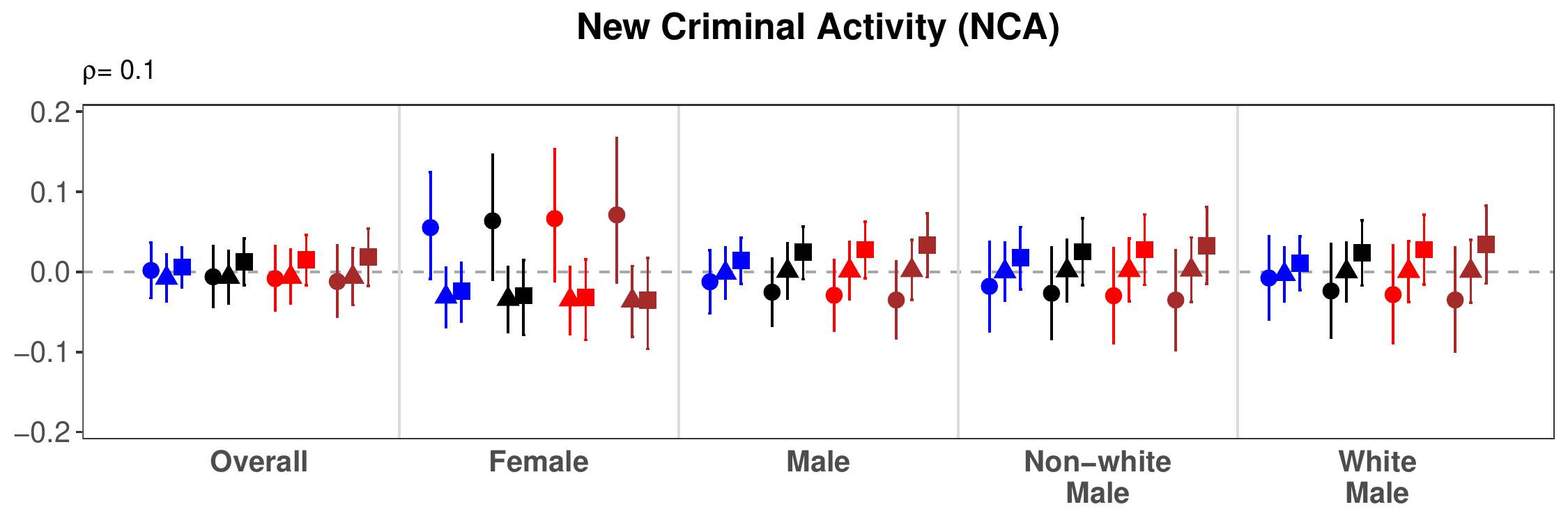}
	\includegraphics[width=\textwidth]{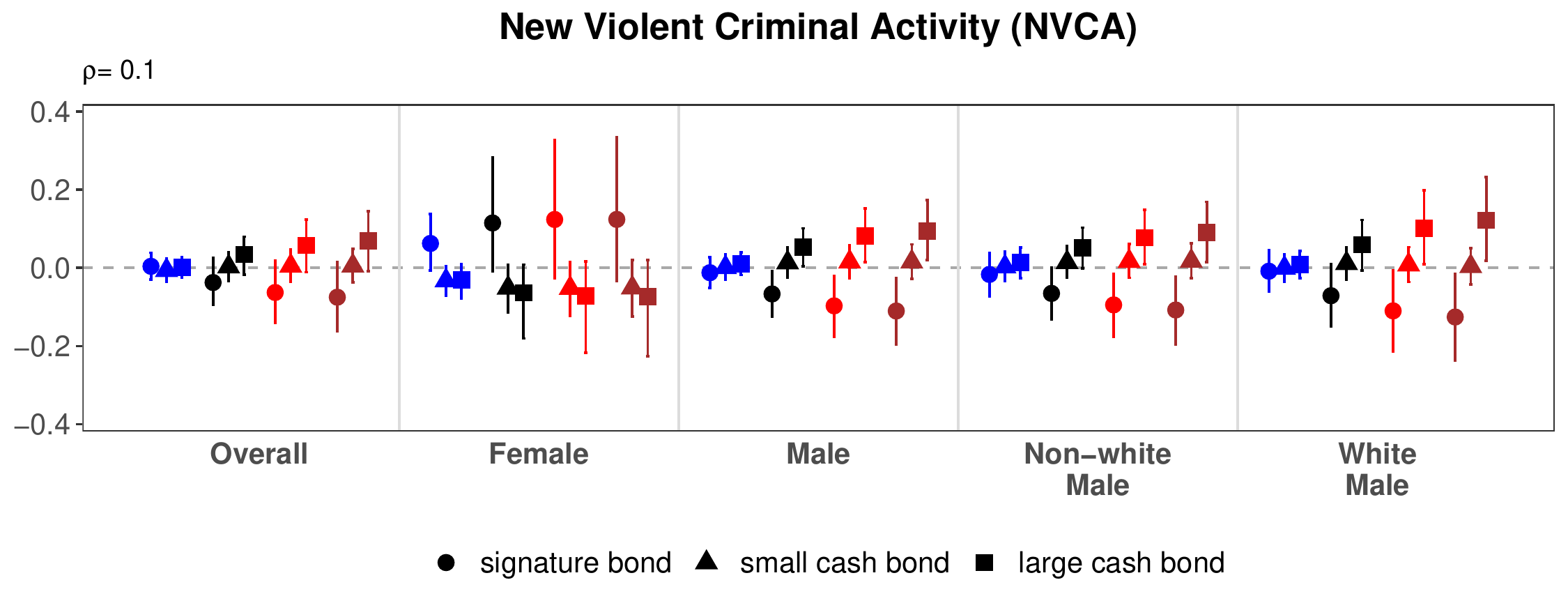}
        \caption{Estimated Average Principal Causal Effects (\ACE) of
          PSA Provision on the Judge's Decision with $\rho=0.1$. Each
          panel presents the overall and subgroup-specific results for
          a different outcome variable. Each column within a panel
          shows the estimated \ACE{} of PSA provision for safe (blue),
          easily preventable (black), preventable (red), and risky
          (brown) cases. For each of these principal strata, we report
          the estimated \ACE{} on the judge's decision to impose a
          signature bond (circles), a small cash bail amount of 1,000
          dollars or less (triangles), and a large cash bail amount of
          greater than 1,000 (squares). The vertical line for each
          estimate represents the Bayesian $95\%$ credible interval.}
	\label{fig:sens2}
\end{figure}

\begin{figure}[p]
	\centering \spacingset{1}
	\includegraphics[width=\textwidth]{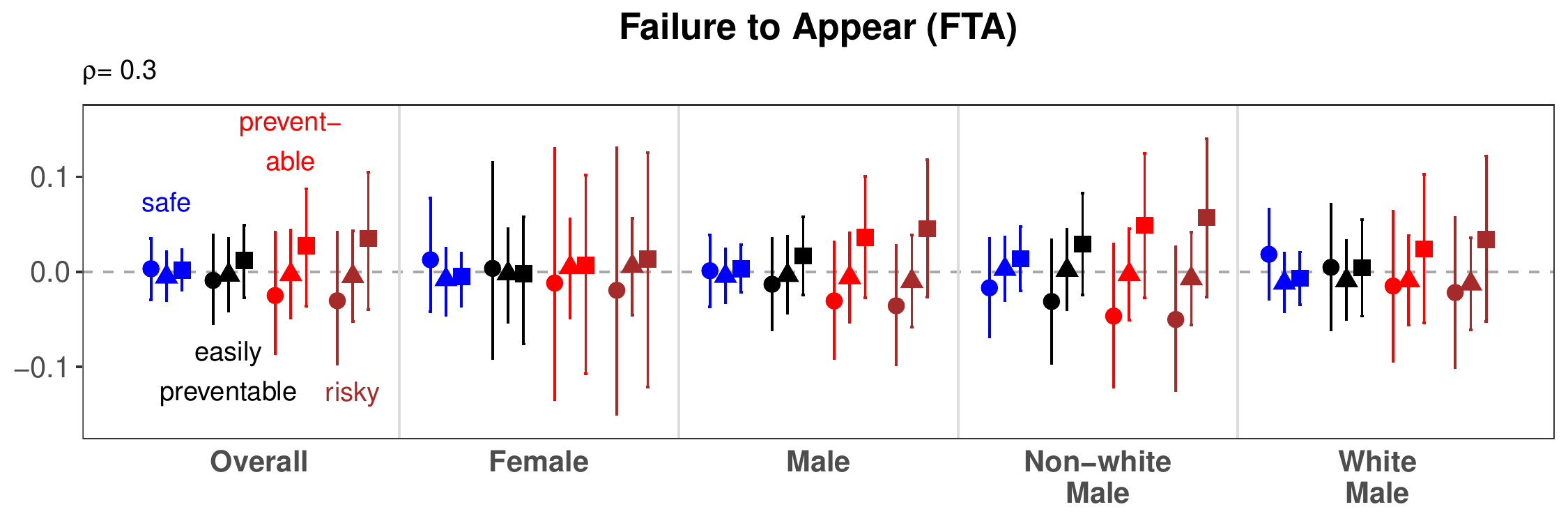}
	\includegraphics[width=\textwidth]{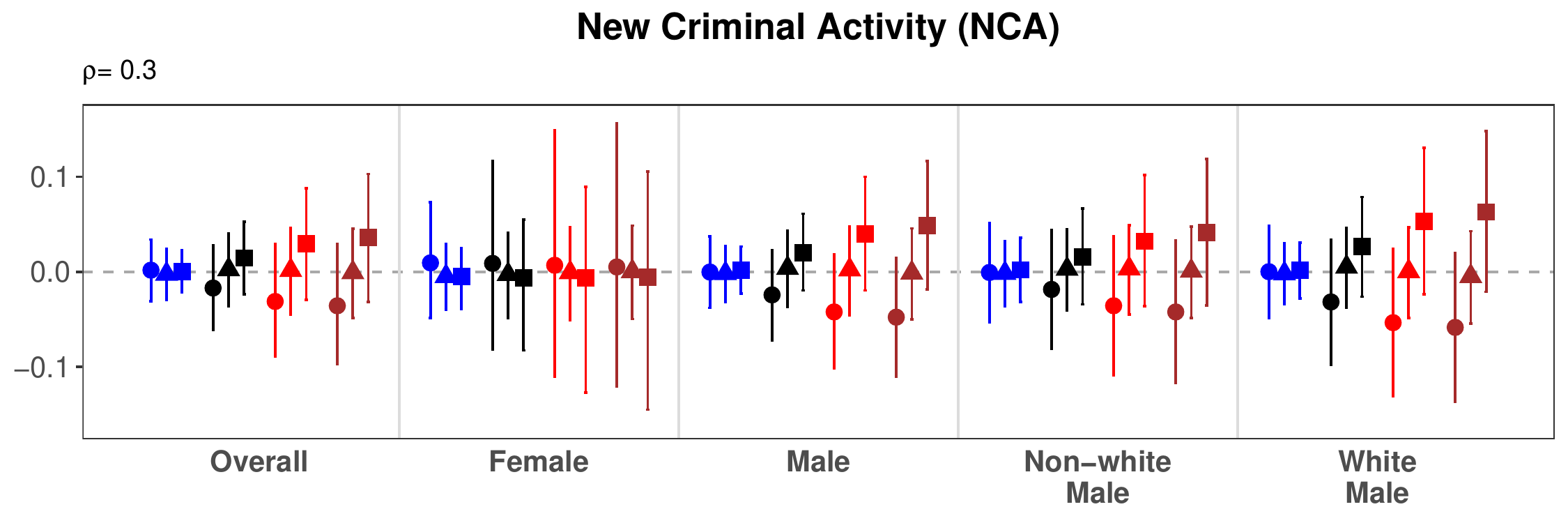}
	\includegraphics[width=\textwidth]{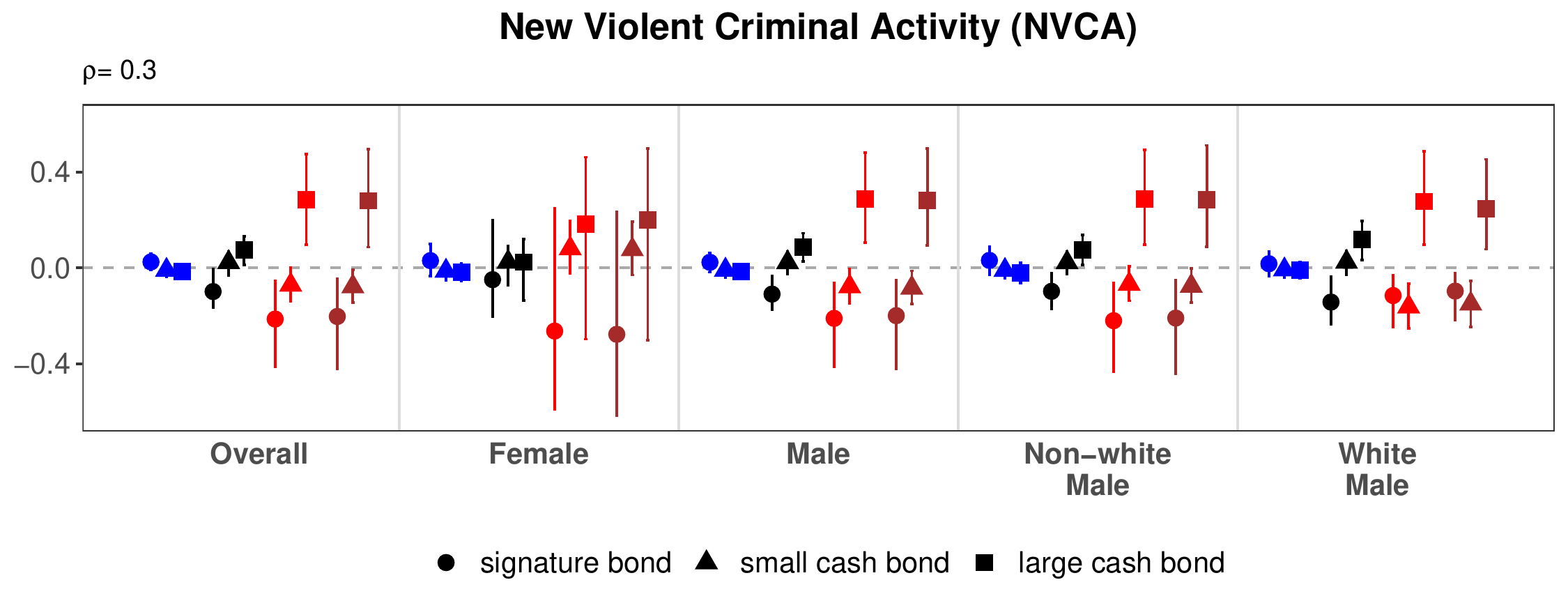}
        \caption{Estimated Average Principal Causal Effects (\ACE) of
          PSA Provision on the Judge's Decision with $\rho=0.3$. Each
          panel presents the overall and subgroup-specific results for
          a different outcome variable. Each column within a panel
          shows the estimated \ACE{} of PSA provision for safe (blue),
          easily preventable (black), preventable (red), and risky
          (brown) cases. For each of these principal strata, we report
          the estimated \ACE{} on the judge's decision to impose a
          signature bond (circles), a small cash bail amount of 1,000
          dollars or less (triangles), and a large cash bail amount of
          greater than 1,000 (squares). The vertical line for each
          estimate represents the Bayesian $95\%$ credible interval.}
	\label{fig:sens3}
      \end{figure}

In this appendix, we implement sensitivity analysis for
unconfoundedness assumption (Assumption~\ref{asm::indep}) and present
the results. 
Figures~\ref{fig:sens1},~\ref{fig:sens2},~and~\ref{fig:sens3} show the
results for the parametric sensitivity analysis.
The patterns of the estimated \ACE s of PSA provision with different
sets of sensitivity parameters are generally consistent with those in
the case where the unconfoundedness assumption holds.

\clearpage

\section{Additional Results for Optimal Decision}
\label{app:optimalresults}

\begin{figure}[h!]
	\centering \spacingset{1}
 \begin{subfigure}[t]{\textwidth}
	\subcaption{The cases whose DMF recommendation is a signature bond}
	\includegraphics[width = \textwidth]{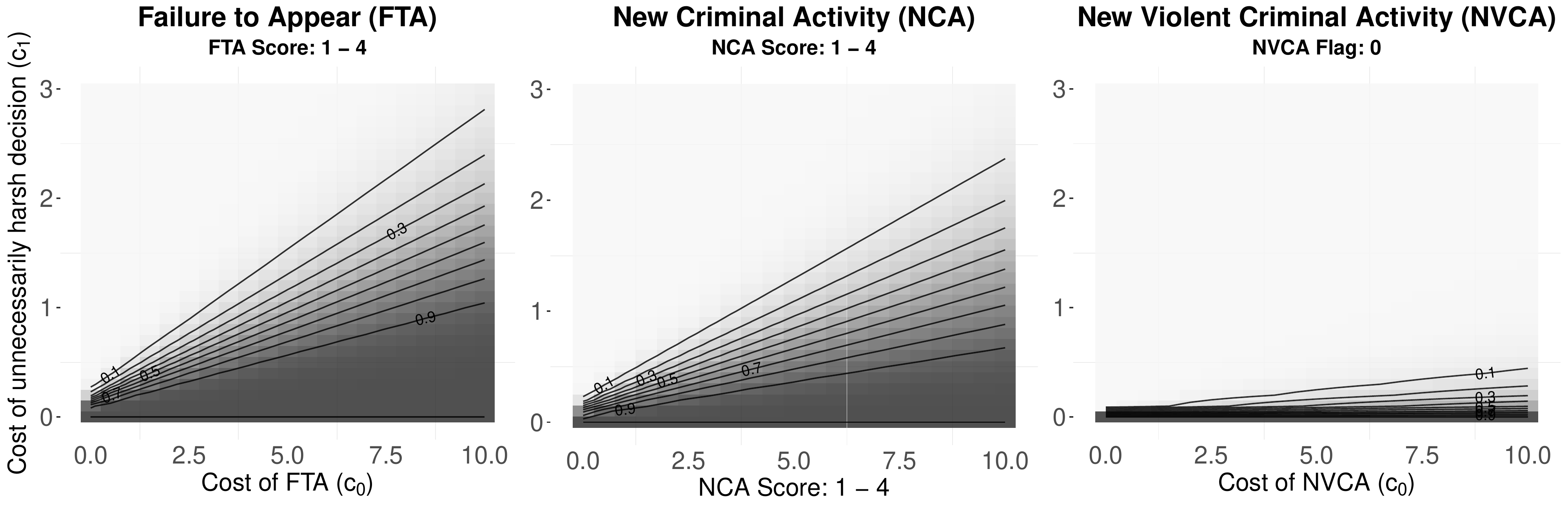}
\end{subfigure} \\
\vspace{.25in}
\begin{subfigure}[t]{\textwidth}
	\subcaption{The cases whose DMF recommendation is a cash bond}
	\includegraphics[width = \textwidth]{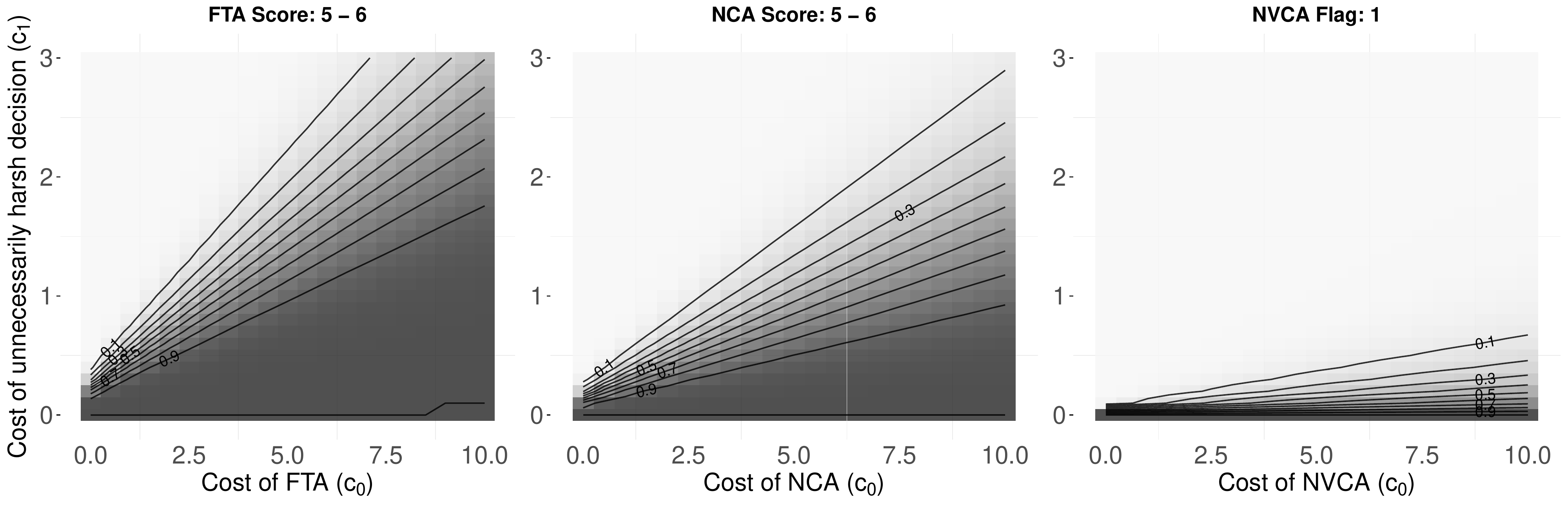}
\end{subfigure}
\caption{Estimated Proportion of Cases for Which Cash Bond is
  Optimal. Each column represents the results based on one of the
  three outcomes (FTA, NCA, and NVCA). The top (bottom) panel shows
  the results for the cases whose DMF recommendation is a signature
  (cash) bond. Unlike Figure~\ref{fig:opt_comb}, which uses the
  overall DMF recommendation, the results are based on the separate
  DMF recommendation for each outcome. In each plot, the contour lines
  represents the estimated proportion of cases, for which a cash bond
  is optimal, given the cost of an unnecessarily harsh decision
  ($c_1$; $y$-axis) and that of a negative outcome ($c_0$;
  $x$-axis). A grey area represents a greater proportion of such
  cases. }
	 \label{fig:opt_sep}
\end{figure}

\newpage
\section{Additional Results for the Comparison between Judge's Decisions and DMF Recommendations}
\label{app:utility}

\begin{figure}[h!]
	\centering \spacingset{1}
	\begin{subfigure}[t]{\textwidth}
		\subcaption{Treatment Group}
		\includegraphics[width = \textwidth]{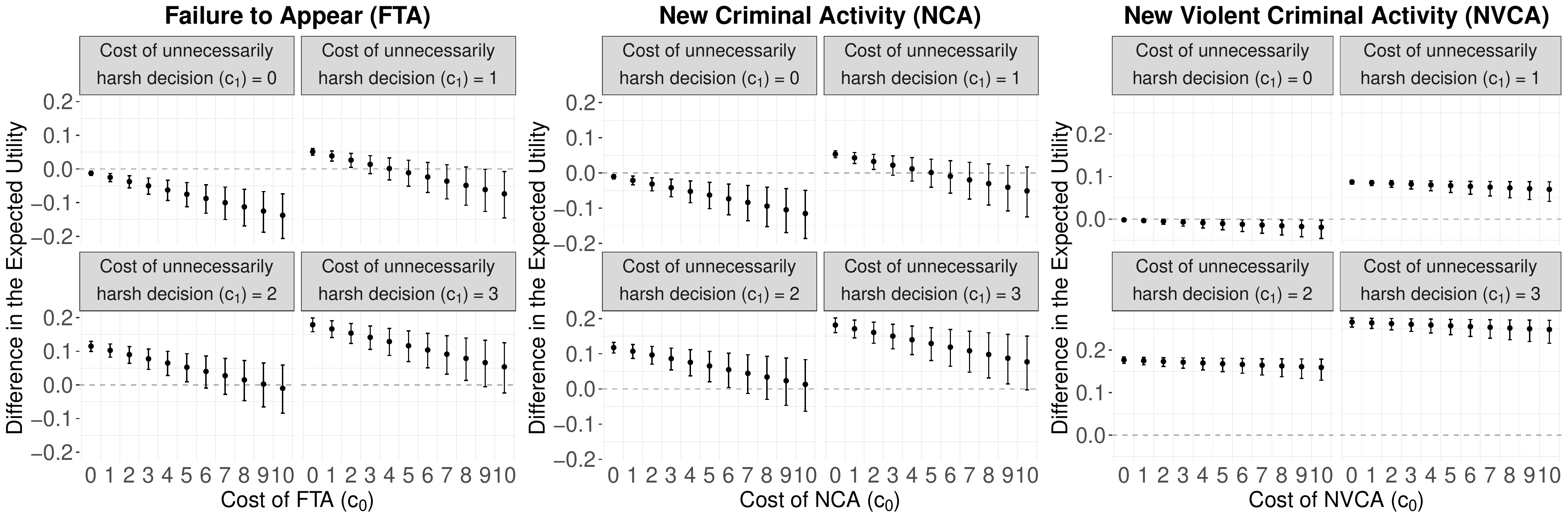}
	\end{subfigure} \\
	\vspace{.25in}
	\begin{subfigure}[t]{\textwidth}
		\subcaption{Control Group}
		\includegraphics[width = \textwidth]{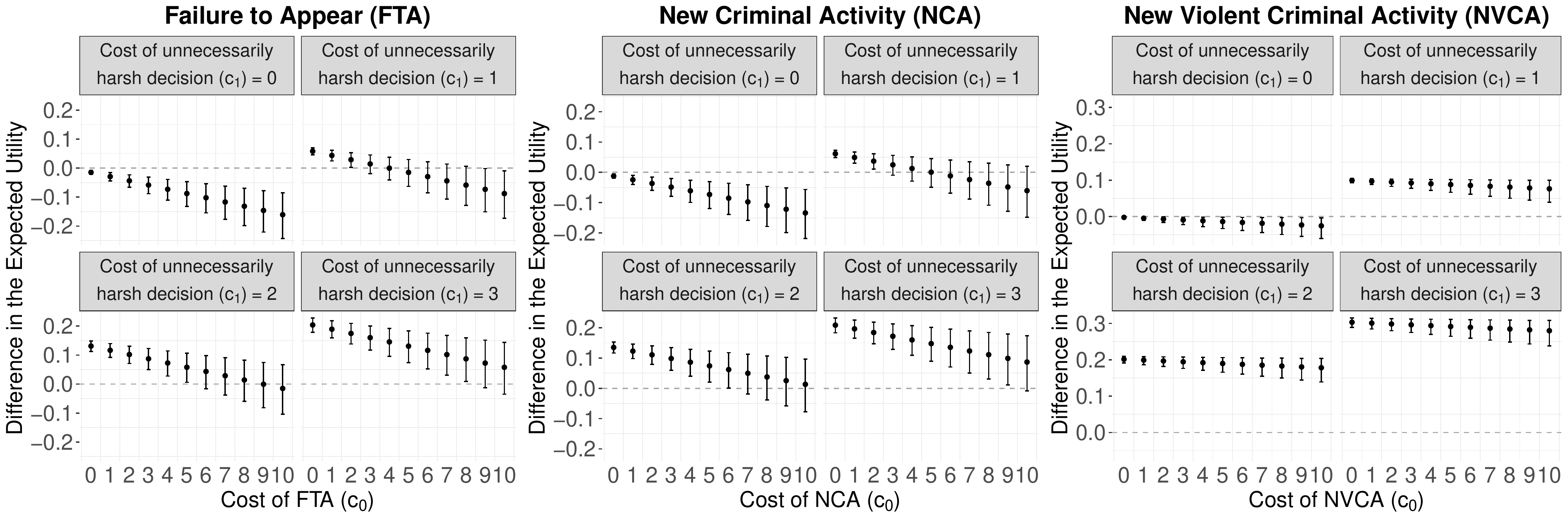}
	\end{subfigure}
	\caption{Estimated Difference in the Expected Utility under
          Selected Values of Cost Parameters between Judge's Decisions
          and DMF Recommendations for the Treatment (top row) and
          Control (bottom row) Group. Each column represents the
          results base on one of the three outcomes, given the cost of
          an unnecessarily harsh decision ($c_1$; each panel) and that
          of a negative outcome ($c_0$; $x$-axis). A positive value
          implies that the judge's decision yields a higher expected
          utility (i.e., more optimal) than the corresponding DMF
          recommendation. The vertical line for each estimate
          represents the Bayesian $95\%$ credible interval.}
	\label{fig:utility_ci}
\end{figure}

\end{document}